\documentclass[twocolumn]{aastex63}
\usepackage{graphicx}
\usepackage{float}
\usepackage{amsfonts,amsmath,amssymb}
\usepackage{natbib}
\usepackage{url}
\usepackage{xcolor}
\usepackage{hyperref}
\usepackage[numbered]{bookmark}
\usepackage{calculator}
\usepackage{placeins} 
\let\splitbox\relax
\usepackage{adjustbox} 
\usepackage{enumitem} 
\usepackage{xifthen}
\usepackage{chngcntr} 

\usepackage{bm}

\usepackage{tabularx}

\hypersetup{
    colorlinks,
    linkcolor={magenta!80!white},
    citecolor={blue!80!black},
    urlcolor={magenta!80!black},
}

\usepackage{listings} 
\usepackage{color}
\definecolor{lightgray}{gray}{0.9}

\usepackage{lmodern}
\usepackage{slantsc}

\usepackage{multirow}

\DeclareRobustCommand{\disambiguate}[3]{#3}

\newcommand{\TODO}[2][]{%
\ifthenelse{\isempty{#1}}%
{\textbf{\textcolor{red}{TODO: #2}}}%
{\textbf{\textcolor{red}{TODO (#1) #2}}}%
}

\def\gtsima{$\; \buildrel > \over \sim \;$}
\def\ltsima{$\; \buildrel < \over \sim \;$}
\def\prosima{$\; \buildrel \propto \over \sim \;$}
\def\gsim{\lower.5ex\hbox{\gtsima}}
\def\lsim{\lower.5ex\hbox{\ltsima}}
\def\simgt{\lower.5ex\hbox{\gtsima}}
\def\simlt{\lower.5ex\hbox{\ltsima}}
\def\simpr{\lower.5ex\hbox{\prosima}}

\def\h1{$h^{-1}$}
\def\beq{\begin{equation}}
\def\eeq{\end{equation}}
\def\8mu{8\,$\mu{\rm m}$}
\def\16mu{16\,$\mu{\rm m}$}
\def\24mu{24\,$\mu{\rm m}$}
\def\70mu{70\,$\mu{\rm m}$}

\def\Umean{\left<U\right>}
\def\Umin{U_{\mathrm{min}}}
\def\Umax{U_{\mathrm{max}}}
\def\SNR{\mathrm{S/N}}

\def\deltaGDR{\delta_{\mathrm{GDR}}}

\def\Tkin{T_{\mathrm{kin}}}
\def\nmean{\left<n_{\mathrm{H}_2}\right>}
\def\nthresh{n_{\mathrm{H_2,thresh}}}
\def\nH2{n_{\mathrm{H}_2}}
\def\NH2_{N_{\mathrm{H}_2}}
\def\percmcubic{\mathrm{cm}^{-3}}

\def\dvddr{\mathrm{d}v/\mathrm{d}r}
\def\lgNH2_{\mathrm{log}_{10}\,(N_{\mathrm{H}_2} / \mathrm{cm^{-2}})}
\def\lgnmean{\mathrm{log}_{10}\,\left<n_{\mathrm{H_2}}\right>}
\def\lgnthresh{\mathrm{log}_{10}\,(n_{\mathrm{H_2,thresh}})}
\def\alphavir{\alpha_{\mathrm{vir}}}
\def\H2{\mathrm{H}_2}
\def\HIrom{\mathrm{H}\textnormal{\textup{\uppercase\expandafter{\romannumeral 1}}}}
\def\HIIrom{\mathrm{H}\textnormal{\textup{\uppercase\expandafter{\romannumeral 2}}}}

\def\Msun{\mathrm{M}_{\odot}}
\def\Lsun{\mathrm{L}_{\odot}}
\def\Kkmspc2{\mathrm{K\,km\,s^{-1}\,pc^{2}}}

\def\SFR{\mathrm{SFR}}
\def\LIR{L_{\mathrm{IR}}}
\def\Mstar{M_{\star}}
\def\Mdust{M_{\mathrm{dust}}}
\def\Tdust{T_{\mathrm{dust}}}
\def\Mgas{M_{\mathrm{gas}}}

\def\R52{R_{52}}
\def\fPDR{f_{\mathrm{PDR}}}

\def\fH2_{f_{\mathrm{H_2}}}
\def\MH2_{M_{\mathrm{H_2}}}

\def\Jykms{\mathrm{Jy\,km/s}}
\def\Kkms_{\mathrm{K\,km/s}}

\def\michi2{\textsc{MiChi2}}

\newcommand{\J}[2][]{%
	\ifthenelse{\isempty{#1}}%
	{\begingroup%
		\def\JlowerN{}%
		\ADD{#2}{-1}{\JlowerN}%
		J=#2{\to}\JlowerN%
		\endgroup%
	}%
	{J=#1{\to}#2}%
}

\newcommand{\rom}[1]{{{\uppercase\expandafter{\romannumeral #1}}}}
\newcommand{\romup}[1]{{\textup{\uppercase\expandafter{\romannumeral #1}}}}

\newcommand{\incode}[1]{{\raggedright\lstinline|#1|}}
\newcommand{\incodep}[1]{{\raggedright\lstinline|"#1"|}}

\defcitealias{Liudz2015}{L15}
\defcitealias{DL07}{DL07}
\defcitealias{V20}{V20}

\renewcommand{\figurename}{Fig.}

\usepackage{twoopt}
\usepackage[normalem]{ulem}
\newcommandtwoopt{\REVISED}[3][][]{%
\ifthenelse{\isempty{#3}}%
{%
\ifthenelse{\isempty{#2}}%
{
\textbf{\textcolor{black}{#1}}%
}%
{
\textcolor{gray}{\sout{#2}}%
}%
}%
{%
\ifthenelse{\isempty{#2}}%
{
\textbf{\textcolor{black}{#3}}%
}%
{
\textcolor{gray}{\sout{#2}}%
\textbf{\textcolor{black}{#3}}%
}%
}%
}

\newcommand{\DONE}[2][]{%
\ifthenelse{\isempty{#1}}%
{%
\textit{\textcolor{green!55!black}{DONE: #2}}}%
{%
\IfSubStr{#1}{ES}{%
\textit{\textcolor{magenta!95!black}{DONE (#1) #2}}%
}{%
\IfSubStr{#1}{SKL}{%
\textit{\textcolor{orange!95!black}{DONE (#1) #2}}%
}{%
\IfSubStr{#1}{PL}{%
\textit{\textcolor{cyan!95!black}{DONE (#1) #2}}%
}{%
\textit{\textcolor{green!55!black}{DONE (#1) #2}}%
}%
}%
}%
}%
}

\received{September 29, 2020}
\revised{December 22, 2020}
\accepted{January 1, 2021}

\shorttitle{CO Excitation}
\shortauthors{D. Liu et al.}

\begin{document}

\title{CO excitation, molecular gas density and interstellar radiation field in local and high-redshift galaxies}

\author{Daizhong Liu}
\affiliation{Max Planck Institute for Astronomy, K\"{o}nigstuhl 17, D-69117 Heidelberg, Germany}
\email{dzliu@mpia.de}

\author{Emanuele Daddi} 
\affiliation{CEA, Irfu, DAp, AIM, Universit\`e Paris-Saclay, Universit\`e de Paris, CNRS, F-91191 Gif-sur-Yvette, France}

\author{Eva Schinnerer} 
\affiliation{Max Planck Institute for Astronomy, K\"{o}nigstuhl 17, D-69117 Heidelberg, Germany}

\author{Toshiki Saito} 
\affiliation{Max Planck Institute for Astronomy, K\"{o}nigstuhl 17, D-69117 Heidelberg, Germany}

\author{Adam Leroy} 
\affiliation{18 Department of Astronomy, The Ohio State University, 140 West 18th Ave, Columbus, OH 43210, USA}

\author{John Silverman} 
\affiliation{Kavli Institute for the Physics and Mathematics of the Universe, The University of Tokyo (Kavli IPMU, WPI), Kashiwa 277-8583, Japan}
\affiliation{Department of Astronomy, School of Science, The University of Tokyo, 7-3-1 Hongo, Bunkyo, Tokyo 113-0033, Japan}

\author{Francesco Valentino} 
\affiliation{Cosmic Dawn Center (DAWN), Copenhagen, Denmark}
\affiliation{Niels Bohr Institute, University of Copenhagen, Lyngbyvej 2, DK-2100 Copenhagen {\O}, Denmark}

\author{Georgios Magdis} 
\affiliation{Cosmic Dawn Center (DAWN), Copenhagen, Denmark}
\affiliation{DTU-Space, Technical University of Denmark, Elektrovej 327, DK-2800 Kgs. Lyngby, Denmark}
\affiliation{Niels Bohr Institute, University of Copenhagen, Lyngbyvej 2, DK-2100 Copenhagen {\O}, Denmark}
\affiliation{Institute for Astronomy, Astrophysics, Space Applications and Remote Sensing, National Observatory of Athens, GR-15236 Athens, Greece}

\author{Yu Gao} 
\affiliation{Department of Astronomy, Xiamen University, Xiamen, Fujian 361005, People's Republic of China}
\affiliation{Purple Mountain Observatory \& Key Lab of Radio Astronomy, Chinese Academy of Sciences (CAS), Nanjing 210033, People's Republic of China}

\author{Shuowen Jin} 
\affiliation{Instituto de Astrof\'{i}sica de Canarias (IAC), E-38205 La Laguna, Tenerife, Spain}
\affiliation{Universidad de La Laguna, Dpto. Astrof\'{i}sica, E-38206 La Laguna, Tenerife, Spain}

\author{Annagrazia Puglisi} 
\affiliation{Center for Extragalactic Astronomy, Durham University, South Road, Durham DH13LE, United Kingdom}

\author{Brent Groves} 
\affiliation{Research School of Astronomy and Astrophysics, Australian National University, Canberra ACT, 2611, Australia}
\affiliation{International Centre for Radio Astronomy Research, University of Western Australia, Crawley, Perth, Western Australia, 6009, Australia}

\begin{abstract}
We study the Carbon Monoxide (CO) excitation, mean molecular gas density and interstellar radiation field (ISRF) intensity in a comprehensive sample of 76 galaxies from local to high redshift ($z\sim0-6$), selected based on detections of their CO transitions $J=2\to1$ and $5\to4$ and their optical/infrared/(sub-)millimeter spectral energy distributions (SEDs). 
We confirm the existence of a tight correlation between CO excitation as traced by the CO(5-4)$/$(2-1) line ratio $\R52$, and the mean ISRF intensity $\Umean$ as derived from infrared SED fitting using dust SED templates. 
By modeling the molecular gas density probability distribution function (PDF) in galaxies and predicting CO line ratios with large velocity gradient radiative transfer calculations, we present a framework linking global CO line ratios to the mean molecular hydrogen gas density $\nmean$ and kinetic temperature $\Tkin$. 
Mapping in this way observed $\R52$ ratios to $\nmean$ and $\Tkin$ probability distributions, we obtain positive $\Umean$--$\nmean$ and $\Umean$--$\Tkin$ correlations, which imply a scenario in which the ISRF in galaxies is mainly regulated by $\Tkin$ and (non-linearly) by $\nmean$. A small fraction of starburst galaxies showing enhanced $\nmean$ could be due to merger-driven compaction. 
Our work demonstrates that ISRF and CO excitation are tightly coupled, and that density-PDF modeling is a promising tool for probing detailed ISM properties inside galaxies. 
\end{abstract}

\keywords{galaxies: ISM --- galaxies: star formation --- ISM: general --- ISM: evolution}

\section{Introduction}
\label{Section_Introduction}

Star formation in galaxies is regulated by their reservoir of molecular gas. 
Globally, the star formation rate (SFR) correlates with the total amount of molecular gas mass via the Kennicutt-Schmidt law (\citealt{Schmidt1959,Kennicutt1998ApJ}). 
Meanwhile, physical properties like density and temperature of the molecular gas also play an important role. 
For example, observations of different carbon monoxide (CO) rotational transition ($J$) lines reveal a relatively denser ($\nH2 \sim 10^{3-4} \; \percmcubic$), highly-excited phase of molecular gas in addition to a more diffuse ($\nH2 \sim 10^{2-3} \; \percmcubic$), less-excited phase (e.g., \citealt{Harris1991}; \citealt{Wild1992M82}; \citealt{Guesten1993}; \citealt{Mao2000M82}; \citealt{Weiss2001M82,Weiss2005M82}; \citealt{Israel2002,Israel2003,Israel2005,Israel2006,Israel2009a,Israel2009b,Israel2014,Israel2015}; \citealt{Bradford2003}; \citealt{Bayet2004,Bayet2006}; \citealt{Papadopoulos2007b,Papadopoulos2010a,Papadopoulos2010b,Papadopoulos2012}; \citealt{Kamenetzky2011,Kamenetzky2012,Kamenetzky2014,Kamenetzky2016,Kamenetzky2017,Kamenetzky2018}; \citealt{ZhangZhiyu2014}; \citealt{Liudz2015} [hereafter \citetalias{Liudz2015}]; \citealt{Daddi2015}; \citealt{Saito2017}). 
While observations of rotational transition lines of high dipole moment molecules like hydrogen cyanide (HCN) reveal the densest phase of the gas ($\nH2 \gtrsim 10^{3-4} \; \percmcubic$; e.g., \citealt{Downes1992,Brouillet1993,Gao2004,Gao2004ApJS,Papadopoulos2007a,Shirley2015}). 

In turbulent star formation theory, 
variations of molecular gas properties are naturally created by turbulence which is ubiquitous in galaxies (e.g., \citealt{Nordlund1999}, \citealt{Ostriker1999}, \citealt{Padoan2002,Padoan2011,Padoan2012}; \citealt{Krumholz2005,Krumholz2007}; \citealt{Feldmann2011}; \citealt{Hennebelle2011}; \citealt{Salim2015}; \citealt{Leroy2017}; \citealt{Elmegreen2018}). 
Turbulence generates certain gas density probability distribution functions (PDFs). At each gas density, CO molecules have different excitation conditions. By solving radiative transfer equations with the large velocity gradient (LVG) assumption (e.g., \citealt{Goldreich1974}), CO line fluxes can be calculated for each given state of gas volume density, column density, CO abundance, and LVG velocity gradient, etc. 
The integrated CO line fluxes from all gas states give the total CO spectral line energy distribution (SLED) as observed. 
Therefore, CO SLED could be a powerful tracer of turbulence and of molecular gas properties.

Meanwhile, dust grains are also important ingredients of the interstellar medium (ISM), mixed with gas. They are exposed to and heated by the interstellar radiation field (ISRF), and their thermal emission dominates the (far-)infrared/(sub-)millimeter part of galaxies' spectral energy distributions (SEDs). Like molecular gas, dust grains do not physically have a single state. Although observational studies sometimes approximate galaxies' dust SEDs by one- or two-components in modified-blackbody fitting, physical models based on assuming PDFs for the ISRF have been proposed and calculated by \citet{Dale2001}, \citet{Dale2002}, \citet{Li2002} and \citet[][hereafter \citetalias{DL07}]{DL07}. See also subsequent applications in \citet{Draine2007SINGS,Draine2014}, \citet{Aniano2012,Aniano2020}, \citet{Magdis2012SED}, \citet{Daddi2015}, \citet{Dale2017} and \citet{Schreiber2018}. 

Through the study of both CO excitation and dust SED traced mean ISRF intensity ($\Umean$) in about 20 galaxies, \cite{Daddi2015} found that the CO(5-4)/(2-1) line ratio, $\R52$, is tightly correlated with $\Umean$. This indicates that CO excitation, or its related ISM properties, is indeed sensitive to the ISRF. However, how the underlying gas density and temperature correlate with ISRF, and how this relates to other known correlations like the Kennicutt-Schmidt law is still unclear. 

In this work, we study the CO excitation, molecular gas density and ISRF in a large sample of 76 (unlensed) galaxies from local to high redshift. 
The sample is selected from a large compilation of local and high-redshift CO observations from the literature, where we require galaxies to have both CO(2-1) and CO(5-4) detections together with well-sampled dust SEDs. 
This also includes CO(5-4) observations newly presented here, from the Institute de Radioastronomie Millim\'etrique (IRAM) Plateau de Bure Interferometer (PdBI; now upgraded to the NOrthern Extended Millimeter Array [NOEMA]) for six starburst-type galaxies at $z\sim1.6$ in the COSMOS field, which have Atacama Large Millimeter/Submillimeter Array (ALMA) CO(2-1) from \cite{Silverman2015b}.

To estimate gas density and temperature from observed line ratios, we model gas density PDFs following \citet{Leroy2017} but with a new approach incorporating assumptions based on the observed correlations between the gas volume density, column density and velocity dispersion. 
We propose a conversion method from the line ratio to the mean molecular hydrogen gas density $\nmean$ and kinetic temperature $\Tkin$ for galaxies at global scale\,\footnote{A \textsc{Python} package (\incode{co-excitation-gas-modeling}) is provided with this paper for the calculation: \url{https://pypi.org/project/co-excitation-gas-modeling}. It fits an input line ratio with error to our model grid and determines the probable ranges of $\nmean$ and $\Tkin$.}. 
Our model-predicated $J_{\mathrm{u}}<10$ CO SLEDs also show good agreement with the current data. 

The structure of this paper is as follows. 
Sect.~\ref{Section_Sample} describes the sample and data. 
Sect.~\ref{Section_SED_Fitting} describes the SED fitting technique for $\Umean$ and other galaxy properties. 
In Sect.~\ref{Section_Umean_R52}, we present correlations between $\R52$ and various galaxy properties. 
Then, in Sect.~\ref{Section_gas_modeling}, we describe details of our gas modeling and the conversion from $\R52$ to $\nmean$ and $\Tkin$, while the resulting correlations between $\nmean$, $\Tkin$ and $\Umean$ are presented in Sect.~\ref{Section_Umean_nH2}. 
We discuss the physical meaning of $\Umean$, the connection from the $\Umean$-- and $\Tkin$--$\nmean$ correlations to the Kennicutt-Schmidt Law and the limitations and outlook of our study in Sect.~\ref{Section_Discussion} . 
Finally, we summarize in Sect.~\ref{Section_Summary}. 

Throughout this paper, line ratios for CO are expressed as flux-flux ratio, where fluxes are in units of $\mathrm{Jy\,km\,s^{-1}}$. 
We adopt a flat $\Lambda$CDM cosmology with $H_0=73\;\mathrm{km\,s^{-1}\,Mpc^{-1}}$, $\Omega_M=0.27$, and a \cite{Chabrier2003} initial mass function (IMF).

\vspace{0.5truecm}

\section{Sample and Data}
\label{Section_Sample}

We search the literature for CO observations of local and high-redshift galaxies, and seek for galaxies which have multiple CO line detections. This is not a complete search, but we have included 132 papers presenting CO observations from 1975 to 2020\,\footnote{A \incode{MySQL/MariaDB} database is available for interested readers by request.}. 
We require a galaxy to have one low-$J$ CO line, CO(2-1), and one mid/high-$J$ CO line, CO(5-4), for this work. This approach is chosen to maximize the sample size while covering most high-redshift main-sequence (MS)\,\footnote{MS is defined as a sequence between galaxies' stellar mass and SFR at each redshift, see, \cite{Noeske2007}, \cite{Elbaz2007}, \cite{Daddi2007}. In this work we use the \cite{Speagle2014} MS equation.} 
galaxies' CO observations. 

We also require multi-wavelength coverage including optical, near-IR, far-infrared and (sub-)millimeter, in order to fit their panchromatic SEDs and obtain stellar and dust properties.

In this way, we build up a sample of 76 galaxies. They are divided into the following subsamples:
\begin{itemize}
\item 22 ``local (U)LIRG'': local (ultra-)luminous infrared galaxies with IR luminosity $\LIR \ge 10^{11} \, \Lsun$. Their high-$J$ CO data are from the HerCULES (\citealt{Rosenberg2015}) and GOALS (\citealt{Armus2009,LuNanyao2014,LuNanyao2015,LuNanyao2017}) surveys using the Spectral and Photometric Imaging Receiver (SPIRE; \citealt{Griffin2010}) Fourier Transform Spectrometer (FTS; \citealt{Naylor2010}) on board the \textit{Herschel Space Observatory} (\citealt{Pilbratt2010}) and analyzed by \citetalias{Liudz2015}. Their low-$J$ CO data are from ground-based observations in the literature (see references in Table~\ref{Table_All_in_one}). 
\item 16 ``local SFG'': local star-forming galaxies (SFGs), most of which have high-$J$ CO from the KINGFISH (\citealt{Kennicutt2011}) and VNGS (PI: C. Wilson) surveys using \textit{Herschel} SPIRE FTS, also analyzed by \citetalias{Liudz2015}. Many of them have low-$J$ CO mapping from the ground-based HERACLES survey (\citealt{Leroy2009}), while others have CO(2-1) single pointing observations in the literature. 
\item 6 ``high-$z$ SB FMOS'': redshift $z\sim1.5$ starburst (SB)\,\footnote{%
A SB galaxy is defined by its SFR being $4\times$ greater than the MS SFR (e.g., the \citealt{Speagle2014} equation) given its redshift and stellar mass. Vice versa, a MS galaxy is defined by its SFR being within $4\times$ the MS SFR.
}
galaxies from the FMOS-COSMOS survey (\citealt{Silverman2015a}), where CO(2-1) data are from \citet{Silverman2015b} and CO(5-4) data are newly presented in this work. 
\item 4 ``high-$z$ MS BzK'': $z\sim1.5$ MS galaxies from \cite{Daddi2008BzK,Daddi2010BzK,Daddi2015}, selected using $BzK$ color criterion (\citealt{Daddi2004}) and representing high-redshift massive star-forming disks. 
\item 4 ``high-$z$ SB SMG'': $z\sim2-6$ starbursty, (sub)millimeter-selected galaxies (SMGs), including GN20 (\citealt{Daddi2009GN20,Carilli2010GN20}), AzTEC-3 (\citealt{Riechers2010AzTEC3}), COSBO11 (\citealt{Aravena2008COSBO11}), and HFLS3 (\citealt{Riechers2013HFLS3}). 
\item 8 ``high-$z$ MS V20'': high-redshift MS galaxies from \citet{Valentino2020arXiv200612521V}, with SFR within $4\times$ the MS SFR. 
\item 16 ``high-$z$ SB V20'': high-redshift SB galaxies from \citet{Valentino2020arXiv200612521V}, with SFR greater than $4\times$ the MS SFR. 
\end{itemize}

Our sample is shown in Table~\ref{Table_All_in_one}, where references for the CO(2-1) and CO(5-4) observations are provided. We note that there are also additional interesting galaxies observed in these CO lines: for example, strongly lensed galaxies (e.g., \citealt{Yang2017,Harrington2018}), or galaxies that have observations of different CO lines (e.g., \citealt{Boogaard2019}). As the sample we compiled in this work already covers a large variety of galaxy types (e.g., MS/SB, local/high-redshift), we chose not to further include these data for simplicity and consistency. Applying our method to an extended sample of galaxies could be the subject of a future study. 

In the following, we present more details about the CO and multi-wavelength photometry data for subsamples.

\vspace{0.25truecm}

\subsection{Local (U)LIRGs and SFGs}
\label{Section_Sample_Local}

For local galaxies, all high-$J$ ($J_{\mathrm{u}}\sim4$ to $13$) CO observations are taken with \textit{Herschel} SPIRE FTS. 
\citetalias{Liudz2015} explored the full public \textit{Herschel} Science Archive and reduced the spectra for almost all (167) FTS-observed local galaxies\,\footnote{Their catalog is available at \url{https://zenodo.org/record/3632388}}. 
Based on their sample, we select galaxies with CO(5-4) $\SNR>3$ and cross-match them with low-$J$ ($J_{\mathrm{u}}\sim1$ and $2$) observations in the literature (i.e., 132 papers). 
There are about 40 galaxies which meet our criterion. 

The FTS's spatial pixel (``spaxel'') has a beam size of about 20--40$''$ across its frequency range of 447--1568~GHz. As we attempt to recover the total flux from the finite beam size as reliably as possible, a few interacting galaxies (e.g., NGC~4038/39; Arp~299~A/B/C) and very nearby, large galaxies (e.g., Cen~A, NGC~891, M~83) have been excluded. 
This gives us a sample of 38 galaxies with both CO(2-1) and CO(5-4) detections, of which 22 are local (U)LIRGs whose CO(5-4) transitions were mainly observed by the HerCULES and GOALS surveys, while ground-based CO(2-1) was provided by various works in the literature (see Table~\ref{Table_All_in_one}). Meanwhile, 16 are local star-forming spiral galaxies, whose CO(5-4) data are mostly taken by the KINGFISH and VNGS surveys, and 12 of which have CO(2-1) mapping from the HERACLES survey (\citealt{Leroy2009})\,\footnote{Their data are available at \url{http://www.mpia.de/HERACLES/Overview.html}}. 

We provide some notes about galaxies which have multiple, possibly inconsistent CO measurements in the literature in Appendix~\ref{Section_Appendix_CO}. In some cases these early observations do not fully agree with each other, even after accounting for the effect of different beam sizes. This could be due to absolute flux calibration or single-dish baseline issues. Thus it is likely that the uncertainty in these CO fluxes could be quite high, e.g., a factor of two.

To correct for the fact that FTS spaxel beam sizes are smaller than entire galaxies, \citetalias{Liudz2015} measured the \textit{Herschel} PACS 70--160$\,\mu$m aperture photometries within each FTS CO line beam size as well as for the entire galaxy, and calculated the ratio between the beam-aperture photometry and the entire galaxy photometry, namely ``BeamFrac'', as listed in the full Table~\ref{Table_All_in_one} (online version). This BeamFrac is then used to scale the measured CO line flux in the FTS central spaxel to the entire galaxy scale. This method is based on the assumption that PACS 70--160$\,\mu$m luminosity linearly traces CO(5-4) luminosity, and is also adopted by other works, e.g., \citet{Kamenetzky2014,Kamenetzky2016,Kamenetzky2017} and \citet{LuNanyao2017}. 

For nearby galaxies which have CO(2-1) maps from HERACLES, we measure their CO(2-1) integrated fluxes using our own photometry method, as some of them do not have published line fluxes in \citet{Leroy2009}. Because the signal-to-noise ratio is relatively poor when reaching galaxies' outer disks in the HERACLES data, aperture photometry can be strongly affected by the choice of aperture size. We thus perform a signal masking of the HERACLES moment-0 maps to distinguish pure noise pixels from signal pixels. The mask is iteratively generated, median-filtered and binary-dilated based on pixels above 1-$\sigma$, where $\sigma$ is the rms noise iteratively determined on the pixels outside the signal mask. In this way, we obtain a Gaussian-distributed pixel value histogram outside the mask, and a total CO(2-1) line flux from the sum of pixels within the mask. 
We compared our CO(2-1) line fluxes with those published in \cite{Leroy2009} for available galaxies, finding relative differences to be as small as 5--10\%. 

To study the dust SED and ISRF of these galaxies, we further collected multi-wavelength photometry data in the literature. In our sample, 22, 15, 7 and 6 galaxies have \textit{Herschel} far-IR photometry from \cite{Chu2017}, \cite{Dale2017}, \cite{Clark2018} and \cite{Clements2018}, respectively. Eight have SCUBA2 850\,$\mu$m photometry from \cite{Lisenfeld2000}. 
Note that \cite{Dale2017} provide the full UV/optical-to-infrared/sub-mm SEDs\,\footnote{Including \textit{GALEX} far-UV, near-UV, $B$, $V$, $R$, $I$, $u$, $g$, $r$, $i$, $z$, $J$, $H$, $K$, \textit{Spitzer}/IRAC 3.6, 4.5, 5.8, 8.0~$\mu$m, WISE 12~$\mu$m, \textit{Spitzer}/MIPS 24~$\mu$m, \textit{Herschel}/PACS 70, 100, 160~$\mu$m, \textit{Herschel}/SPIRE 250, 350, 500~$\mu$m, and JCMT/SCUBA 850~$\mu$m. See their Table~2.}. 
All of these local galaxies have \textit{Herschel} PACS 70 or 100~$\mu$m and 160~$\mu$m photometry from \citetalias{Liudz2015}. Fluxes are consistent among these works. For example, comparing \citetalias{Liudz2015} with \cite{Chu2017}, we find 13 galaxies in common, and their median flux ratio in logarithm is -0.01\,dex, with a scatter of 0.04\,dex. 
For our SED fitting, we average all available fluxes for each band. 

In addition, we cross-matched with \cite{Brown2014}, \cite{Jarrett2003}, \cite{Brauher2008}, and the NASA Extra-galactic Database (NED) for missing optical to near-/mid-infrared photometry. 
All local galaxies have 2MASS near-IR photometry from \cite{Jarrett2003} except for NGC~2369 and NGC~3256. 
For 9 galaxies which do not have any optical photometry from \cite{Dale2017} and \cite{Brown2014}, we use the optical/near-/mid-IR photometry from NED\,\footnote{%
	These are: 
	Mrk231,
	NGC0253,
	NGC1365,
	NGC2369,
	NGC3256,
	NGC4945,
	NGC5135,
	NGC7469 and 
	NGC7582. 
	Note that we carefully selected photometric data with large enough aperture to cover entire galaxies.
}.

\vspace{0.25truecm}

\subsection{High-$z$ SB FMOS galaxies with new PdBI observations}
\label{Section_Sample_FMOS_COSMOS}

We observed the CO(5-4) line emission in six $z\sim1.6$ starburst galaxies from the FMOS-COSMOS survey (\citealt{Silverman2015a}) with IRAM PdBI in the winter of 2014 (program ID W14DS). 
These galaxies have ALMA CO(2-1) observations presented in \citet{Silverman2015b}. 
Our PdBI observations are at 1.3~mm. Phase centers are set to the ALMA CO(2-1) emission peak position for each galaxy, and the on-source integration time is 1.5 to 3.1~hrs per source. Sensitivity is 0.6--0.7~$\mathrm{mJy/beam}$ over the expected line widths of 200--600~MHz, depending on the ALMA CO(2-1) line properties of each source. 
With robust weighting (robust factor 1), the cleaned images have synthesized beam FWHM of 2.0--3.3$''$.

As the ALMA CO(2-1) data have much higher $\SNR$ than the PdBI CO(5-4) data, we extract the CO(5-4) line fluxes in the $uv$ plane by Gaussian source fitting with fixed CO(2-1) positions and line widths (from \citealt{Silverman2015b}), using the {GILDAS}\,\footnote{\url{http://gildas.iram.fr}} \incode{MAPPING} \incode{UV_FIT} task. 
The achieved line flux $\SNR$ are 1.8--5.4 within the subsample. 
For two sources, PACS-819 and PACS-830, which are spatially resolved in ALMA CO(2-1) data, we also fix their CO(5-4) sizes to the measured CO(2-1) sizes ($\sim$0.3 to 1.0$''$, respectively) in the \incode{UV_FIT} fitting, so as to account that they are marginally resolved in the PdBI data. For other galaxies with smaller ALMA CO(2-1) sizes, we consider them unresolved by the PdBI beam. 

Furthermore, we partially observed their CO(1-0) line emission with the Very Large Array (VLA; project code 17A-233). The observing program is incomplete, and none have full integration (PACS-867, 299 and 164 each have about 90 minutes on-source integration), but we provide face-value measurements obtained as for CO(2-1).
We list the new CO(5-4) and CO(1-0) line fluxes and upper limits together with the \cite{Silverman2015b} CO(2-1) line fluxes in Table~\ref{Table_Appendix_A} in Appendix~\ref{Section_Appendix_PdBI_Observation}. 

Multi-wavelength photometry is available from \cite{Laigle2016} and \cite{Jin2018}, thanks to the rich observational data in the COSMOS deep field (see also \citealt{McCracken2012,Muzzin2013,Ilbert2013,Liudz2019a}).

\vspace{0.25truecm}

\subsection{High-$z$ MS BzK galaxies}
\label{Section_Sample_BzKs_GOODSN}

We include 4 $BzK$ color selected MS galaxies from \citet{Daddi2015} in our sample. They represent typical high-redshift star-forming MS galaxies and are consistent with having a disk-like morphology. 
Their CO(5-4) observations were taken with IRAM PdBI in 2009 and 2011 by \citet{Daddi2015}, and CO(2-1) in 2007--2009 by \citet{Daddi2010BzK}. 

These galaxies have optical-to-near-IR photometry from \cite{Skelton2014}, far-IR to (sub-)mm and radio photometry from \cite{Liudz2018} based on the \textit{Herschel} PEP (\citealt{Lutz2011}), HerMES (\citealt{Roseboom2010}) and GOODS-\textit{Herschel} surveys (\citealt{Elbaz2011}), ground-based SCUBA2 S2CLS (\citealt{Geach2017}) and AzTEC+MAMBO surveys (\citealt{Perera2008,Greve2008,Penner2011}). 

\cite{Daddi2015} presented a similar panchromatic SED fitting as in this work with the full \citetalias{DL07} dust models (see Sect.~\ref{Section_SED_Fitting}) to estimate ISRF $\Umean$ and other SED properties, but without including an AGN component in the modeling. Our SED fitting allows for the inclusion of a mid-IR AGN component, but we confirm that such an AGN component is not required, based on the chi-square statistics. Thus we obtain similar results in terms of $\Umean$ as \cite{Daddi2015}.

\vspace{0.25truecm}

\subsection{High-$z$ SB SMGs}
\label{Section_Sample_SMGs}

We include 4 sub-mm selected high-redshift galaxies in our study: 
GN20 (\citealt{Daddi2009GN20,Carilli2010GN20,Tan2014GN20}), AzTEC-3 (\citealt{Riechers2010AzTEC3}), COSBO11 (\citealt{Aravena2008COSBO11}), and HFLS3 (\citealt{Riechers2013HFLS3,Cooray2014HFLS3,Laporte2015HFLS3}). 
Due to their sub-mm selection, they usually have very high SFRs compared to MS galaxies with similar stellar masses, therefore we consider them as SB. 
We note that there are now more than one hundred sub-mm selected high-redshift ($z\gtrsim1$) galaxies that have CO detections, but only a few tens have both CO(5-4) and (2-1) detections. 
We further excluded strongly-lensed galaxies lacking optical/near-IR SEDs, for example those from \cite{Cox2011}, \cite{Yang2017}, \cite{Bothwell2017}, \cite{Canameras2018}, and \cite{Harrington2018,Harrington2019}, despite
the fairly good sampling of their CO SLEDs. Their strong magnification ($\gtrsim 10$) largely reduces the observing time ($\times 1/100$) for CO observations compared to unlensed targets. Yet their optical to mid-IR SEDs are usually not well sampled. \cite{Harrington2020} present a study of CO excitation and far-IR/(sub-)mm dust SED modeling in strongly lensed galaxies, based on a similar gas density PDF modeling. 

Among our SMG subsample, GN20 is in the GOODS-North field, and AzTEC-3 and COSBO11 are in the COSMOS field. They have rich multi-wavelength photometry as mentioned earlier. 
\citet{Tan2014GN20} fitted the GN20 SED with \citetalias{DL07} templates without an AGN component, and our new fitting to the same photometry data shows that a mid-IR AGN component is indistinguishable from the warm dust component in the \citetalias{DL07} models. The inclusion of the AGN component in this work, however, leads to more realistic uncertainties in the derived $\Umean$ parameter.

\subsection{High-$z$ MS and SB galaxies from V20}
\label{Section_Sample_V20}

We further include 8 MS and 16 SB galaxies from \cite{Valentino2020arXiv200612521V} which have both CO(2-1) and CO(5-4) $\SNR>3$ detections and far-IR photometric data. \citet{Valentino2018,Valentino2020arXiv200612521V,Valentino2020ApJ...890...24V} surveyed 123, 75 and 15 galaxies with ALMA through Cycle 3, 4, and 7, respectively. Cycle 3 and 4 observations targeted CO(5-4) and CO(2-1), respectively. Their sample is selected from the COSMOS field at $z \approx 1.1 - 1.7$ based on predicted CO line luminosities, which are further based on the CO--IR luminosity correlation (\citealt{Daddi2015}). By this selection, this sample contains both MS and SB galaxies. We divide MS and SB galaxies into two subsamples for illustration in the later sections. 

These galaxies have multi-wavelength photometry similarly to the other COSMOS galaxies mentioned above, and most of them also have one or more ALMA dust continuum measurements from the public ALMA archive, reduced by \citet{Liudz2019a,Liudz2019b} and from line-free channels of CO observations in \cite{Valentino2020arXiv200612521V}. \cite{Valentino2020arXiv200612521V} did multi-component SED fitting including stellar, AGN and \citetalias{DL07} warm and cold dust components following \cite{Magdis2012SED,Magdis2017}. They adopt a slightly different definition of ISRF $\Umean_{\mathrm{V20}} = 1/125 \times \LIR / \Mdust$, where their $\LIR$ also includes the AGN contribution. 
In this work, we assembled all available ALMA photometry and re-fitted their SEDs with our own code. To be consistent within this work, we still use the $\Umean$ definition according to \citetalias{DL07} (their Eq.~33), and use only the star-forming dust components without the contribution of AGN torus. Because of the different definition and treatment of the AGN component, there are some noticeable differences in $\Umean$ between \cite{Valentino2020arXiv200612521V} and our study. However, if we were to adopt the same $\Umean_{\mathrm{V20}}$ definition, the $\Umean$ derivations would become fully consistent.

\startlongtable
\begin{deluxetable*}{l @{\hspace{1.0\tabcolsep}} l @{\hspace{1.0\tabcolsep}} c @{\hspace{0.5\tabcolsep}} c c c c c @{\hspace{0.5\tabcolsep}} C @{\hspace{0.5\tabcolsep}} C}
\tabletypesize{\scriptsize}
\tablecaption{%
Sample of galaxies used in this work with measured and derived physical properties. 
\label{Table_All_in_one}}
\tablehead{
    \colhead{ Source } &
    \colhead{ Subsample } &
    \colhead{ z } &
    \colhead{ R52 } &
    \colhead{ $ \log \nmean $ } &
    \colhead{ $ \Umean $ } &
    \colhead{ $ \log L_{\mathrm{IR}} $ } &
    \colhead{ $ \log M_{\star} $ } &
    \colhead{ \multirow{2}{3.5em}{\centering Ref. CO54} } &
    \colhead{ \multirow{2}{3.5em}{\centering Ref. CO21} }\\[-2.5ex]
}
\startdata
Arp193          & local (U)LIRG   & $     0.023 $ & $     1.8 \pm     0.3 $ & $     2.4 \pm     0.4 $ & $   17.0 _{  +0.0} ^{  +2.4} $ & $   11.6 _{  -0.0} ^{  +0.0} $ & $   10.3 _{  -0.0} ^{  +0.0} $ & L15        & P14                \\
Arp220          & local (U)LIRG   & $     0.018 $ & $     2.4 \pm     0.5 $ & $     2.7 \pm     0.7 $ & $   20.6 _{  -0.1} ^{  +0.4} $ & $   12.2 _{  -0.0} ^{  +0.0} $ & $   11.0 _{  -0.0} ^{  +0.0} $ & L15        & K14/G09            \\
IRASF17207-0014 & local (U)LIRG   & $     0.043 $ & $     3.9 \pm     1.0 $ & $     3.5 \pm     1.3 $ & $   35.0 _{  -0.4} ^{  +0.2} $ & $   12.4 _{  -0.0} ^{  +0.0} $ & $   11.0 _{  -0.0} ^{  +0.1} $ & L15        & K14/B08/W08/P12    \\
IRASF18293-3413 & local (U)LIRG   & $     0.018 $ & $     2.2 \pm     0.1 $ & $     2.6 \pm     0.4 $ & $   11.1 _{  -2.9} ^{  +0.9} $ & $   11.7 _{  -0.1} ^{  +0.0} $ & $    9.6 _{  -0.2} ^{  +0.2} $ & L15        & G93                \\
M82             & local SFG       & $     0.001 $ & $     1.7 \pm     0.2 $ & $     2.4 \pm     0.3 $ & $   25.4 _{  -0.5} ^{  +0.2} $ & $   10.6 _{  -0.0} ^{  +0.0} $ & $    9.8 _{  -0.0} ^{  +0.0} $ & L15        & L09                \\
Mrk231          & local (U)LIRG   & $     0.042 $ & $     3.0 \pm     0.9 $ & $     2.9 \pm     1.3 $ & $   50.0 _{  -2.5} ^{  +0.0} $ & $   12.3 _{  -0.0} ^{  +0.0} $ & $   10.9 _{  -0.0} ^{  +0.0} $ & L15        & K14/P12/A07        \\
Mrk273          & local (U)LIRG   & $     0.038 $ & $     3.4 \pm     0.9 $ & $     3.2 \pm     1.3 $ & $   37.7 _{  +0.0} ^{  +5.6} $ & $   12.0 _{  -0.0} ^{  +0.0} $ & $   10.6 _{  -0.2} ^{  +0.0} $ & L15        & K14/P12            \\
NGC0253         & local SFG       & $     0.001 $ & $     2.9 \pm     0.7 $ & $     3.0 \pm     1.0 $ & $    6.3 _{  -0.5} ^{  +0.0} $ & $   10.6 _{  -0.0} ^{  +0.0} $ & $   10.9 _{  -0.1} ^{  +0.0} $ & L15        & K14(43.5)/H99      \\
NGC0828         & local (U)LIRG   & $     0.018 $ & $     0.9 \pm     0.4 $ & $     2.1 \pm     0.4 $ & $    3.5 _{  -0.0} ^{  +0.5} $ & $   11.3 _{  -0.0} ^{  +0.0} $ & $   11.2 _{  -0.0} ^{  +0.0} $ & L15        & P12(22)            \\
NGC1068         & local (U)LIRG   & $     0.004 $ & $     0.8 \pm     0.2 $ & $     2.1 \pm     0.2 $ & $    5.8 _{  -0.7} ^{  +0.2} $ & $   11.2 _{  -0.0} ^{  +0.0} $ & $   10.5 _{  -0.0} ^{  +0.0} $ & L15        & K14(43.5)/K11/B08  \\
NGC1266         & local SFG       & $   0.00724 $ & $     2.6 \pm     0.7 $ & $     2.8 \pm     1.0 $ & $   13.3 _{  -2.0} ^{  +3.3} $ & $   10.3 _{  -0.0} ^{  +0.1} $ & $   10.5 _{  -0.0} ^{  +0.0} $ & L15        & K14(43.5)/A11/Y11  \\
NGC1365         & local (U)LIRG   & $     0.005 $ & $     1.5 \pm     0.4 $ & $     2.3 \pm     0.5 $ & $    3.5 _{  -0.0} ^{  +0.6} $ & $   11.2 _{  -0.0} ^{  +0.0} $ & $   10.9 _{  -0.0} ^{  +0.0} $ & L15        & K14(43.5)/S95      \\
NGC1614         & local (U)LIRG   & $     0.016 $ & $     1.4 \pm     0.3 $ & $     2.3 \pm     0.3 $ & $   31.2 _{  +0.0} ^{ +28.6} $ & $   11.6 _{  -0.0} ^{  +0.1} $ & $   10.5 _{  -0.2} ^{  +0.0} $ & L15        & A95(22)            \\
NGC2369         & local (U)LIRG   & $     0.011 $ & $     2.0 \pm     0.4 $ & $     2.5 \pm     0.5 $ & $    5.8 _{  -1.9} ^{  +0.2} $ & $   11.1 _{  -0.1} ^{  +0.0} $ & $   10.5 _{  -0.0} ^{  +0.0} $ & L15        & A95(22)/B08        \\
NGC2623         & local (U)LIRG   & $     0.018 $ & $     4.1 \pm     0.8 $ & $     3.6 \pm     1.1 $ & $   19.1 _{  -2.0} ^{  +4.3} $ & $   11.4 _{  -0.0} ^{  +0.0} $ & $   10.3 _{  -0.0} ^{  +0.0} $ & L15        & P12/W08            \\
NGC2798         & local SFG       & $   0.00576 $ & $     1.9 \pm     0.3 $ & $     2.5 \pm     0.4 $ & $   13.3 _{  -2.6} ^{  +2.8} $ & $   10.5 _{  -0.0} ^{  +0.0} $ & $    9.7 _{  -0.0} ^{  +0.0} $ & L15        & L09                \\
NGC3256         & local (U)LIRG   & $     0.009 $ & $     1.7 \pm     0.2 $ & $     2.4 \pm     0.3 $ & $   23.9 _{  -9.5} ^{  +0.5} $ & $   11.6 _{  -0.1} ^{  +0.0} $ & $   10.4 _{  -0.0} ^{  +0.0} $ & L15        & A95(24)/B08/G93    \\
NGC3351         & local SFG       & $    0.0026 $ & $     0.9 \pm     0.2 $ & $     2.1 \pm     0.2 $ & $    2.0 _{  -0.3} ^{  +0.4} $ & $    9.8 _{  -0.0} ^{  +0.0} $ & $    9.8 _{  -0.0} ^{  +0.0} $ & L15        & L09                \\
NGC3627         & local SFG       & $   0.00243 $ & $     1.8 \pm     0.4 $ & $     2.4 \pm     0.6 $ & $    3.6 _{  -1.0} ^{  +0.8} $ & $   10.3 _{  -0.0} ^{  +0.0} $ & $   10.1 _{  -0.0} ^{  +0.2} $ & L15        & L09                \\
NGC4321         & local SFG       & $   0.00524 $ & $     0.8 \pm     0.1 $ & $     2.1 \pm     0.2 $ & $    1.8 _{  +0.0} ^{  +0.7} $ & $   10.4 _{  -0.0} ^{  +0.0} $ & $   10.5 _{  -0.0} ^{  +0.0} $ & L15        & L09                \\
NGC4536         & local SFG       & $   0.00603 $ & $     1.7 \pm     0.3 $ & $     2.4 \pm     0.4 $ & $    3.9 _{  -0.1} ^{  +0.6} $ & $   10.3 _{  -0.0} ^{  +0.0} $ & $   10.1 _{  -0.0} ^{  +0.0} $ & L15        & L09                \\
NGC4569         & local SFG       & $  -0.00078 $ & $     1.1 \pm     0.1 $ & $     2.2 \pm     0.2 $ & $    1.9 _{  -0.3} ^{  +0.2} $ & $    9.6 _{  -0.0} ^{  +0.0} $ & $   10.5 _{  -0.0} ^{  +0.0} $ & L15        & L09                \\
NGC4631         & local SFG       & $   0.00202 $ & $     0.9 \pm     0.2 $ & $     2.1 \pm     0.3 $ & $    2.8 _{  -0.6} ^{  +0.4} $ & $   10.2 _{  -0.0} ^{  +0.0} $ & $    9.4 _{  -0.0} ^{  +0.0} $ & L15        & L09                \\
NGC4736         & local SFG       & $   0.00103 $ & $     0.6 \pm     0.1 $ & $     2.0 \pm     0.1 $ & $    4.1 _{  -0.4} ^{  +1.5} $ & $    9.6 _{  -0.0} ^{  +0.0} $ & $    9.8 _{  -0.0} ^{  +0.0} $ & L15        & L09                \\
NGC4826         & local SFG       & $   0.00136 $ & $     1.6 \pm     0.3 $ & $     2.4 \pm     0.4 $ & $    3.6 _{  -0.6} ^{  +0.8} $ & $    9.5 _{  -0.0} ^{  +0.0} $ & $   10.4 _{  -0.0} ^{  +0.0} $ & L15        & A95(28)            \\
NGC4945         & local (U)LIRG   & $     0.002 $ & $     4.0 \pm     0.8 $ & $     3.6 \pm     1.2 $ & $    7.0 _{  -1.0} ^{  +0.3} $ & $   11.1 _{  -0.0} ^{  +0.0} $ & $    9.7 _{  -0.0} ^{  +1.1} $ & L15        & W04/B08(22)        \\
NGC5135         & local (U)LIRG   & $     0.014 $ & $     1.6 \pm     0.4 $ & $     2.4 \pm     0.4 $ & $    8.2 _{  -1.2} ^{  +2.5} $ & $   11.2 _{  -0.1} ^{  +0.0} $ & $   11.1 _{  -0.7} ^{  +0.0} $ & L15        & P12(22)            \\
NGC5194         & local SFG       & $     0.002 $ & $     0.7 \pm     0.1 $ & $     2.1 \pm     0.1 $ & $    3.0 _{  -0.2} ^{  +0.7} $ & $   10.2 _{  -0.0} ^{  +0.0} $ & $    9.6 _{  -0.0} ^{  +0.0} $ & L15        & L09                \\
NGC5713         & local SFG       & $   0.00633 $ & $     1.0 \pm     0.2 $ & $     2.1 \pm     0.2 $ & $    5.2 _{  -1.1} ^{  +0.6} $ & $   10.4 _{  -0.0} ^{  +0.0} $ & $   10.1 _{  -0.0} ^{  +0.0} $ & L15        & L09                \\
NGC6240         & local (U)LIRG   & $     0.024 $ & $     2.8 \pm     0.8 $ & $     2.9 \pm     0.9 $ & $   20.0 _{  -0.5} ^{  +0.2} $ & $   11.7 _{  -0.0} ^{  +0.0} $ & $   10.8 _{  -0.0} ^{  +0.0} $ & L15        & G09                \\
NGC6946         & local SFG       & $   0.00013 $ & $     1.1 \pm     0.1 $ & $     2.2 \pm     0.2 $ & $    4.2 _{  -1.1} ^{  +0.4} $ & $   10.4 _{  -0.1} ^{  +0.0} $ & $   10.3 _{  -0.0} ^{  +0.3} $ & L15        & L09                \\
NGC7469         & local (U)LIRG   & $     0.016 $ & $     1.1 \pm     0.3 $ & $     2.1 \pm     0.2 $ & $   13.1 _{  +0.0} ^{  +5.1} $ & $   11.6 _{  -0.0} ^{  +0.0} $ & $   10.0 _{  -0.0} ^{  +0.3} $ & L15        & P12                \\
NGC7552         & local (U)LIRG   & $     0.005 $ & $     2.4 \pm     0.5 $ & $     2.7 \pm     0.7 $ & $   14.0 _{  -0.5} ^{  +0.2} $ & $   11.1 _{  -0.0} ^{  +0.0} $ & $   10.2 _{  -0.0} ^{  +0.0} $ & L15        & A95                \\
NGC7582         & local SFG       & $     0.005 $ & $     1.5 \pm     0.3 $ & $     2.3 \pm     0.4 $ & $   11.7 _{  -0.3} ^{  +0.2} $ & $   10.9 _{  -0.0} ^{  +0.0} $ & $   10.9 _{  -0.0} ^{  +0.0} $ & L15        & A95                \\
MCG+12-02-001   & local (U)LIRG   & $     0.016 $ & $     1.6 \pm     0.3 $ & $     2.3 \pm     0.3 $ & $   17.4 _{  -2.1} ^{  +3.6} $ & $   11.5 _{  -0.0} ^{  +0.0} $ & $   11.9 _{  -1.1} ^{  +0.6} $ & L15        & K16(43.5)          \\
Mrk331          & local (U)LIRG   & $     0.018 $ & $     2.4 \pm     0.4 $ & $     2.7 \pm     0.7 $ & $   14.7 _{  -2.1} ^{  +0.4} $ & $   11.4 _{  -0.0} ^{  +0.0} $ & $   10.9 _{  -1.5} ^{  +0.0} $ & L15        & K16(43.5)          \\
NGC7771         & local (U)LIRG   & $     0.014 $ & $     1.3 \pm     0.2 $ & $     2.2 \pm     0.3 $ & $    7.0 _{  -0.5} ^{  +0.1} $ & $   11.3 _{  -0.0} ^{  +0.0} $ & $   11.4 _{  -0.0} ^{  +0.0} $ & L15        & K16(43.5)          \\
IC1623          & local (U)LIRG   & $      0.02 $ & $     1.8 \pm     0.3 $ & $     2.4 \pm     0.5 $ & $   13.2 _{  -2.1} ^{  +0.5} $ & $   11.6 _{  -0.0} ^{  +0.0} $ & $    9.1 _{  -0.0} ^{  +0.0} $ & L15        & K16(43.5)          \\
BzK16000        & high-z MS BzK   & $      1.52 $ & $     1.5 \pm     0.3 $ & $     2.3 \pm     0.4 $ & $   15.2 _{ -13.3} ^{ +33.2} $ & $   11.8 _{  -0.0} ^{  +0.0} $ & $   11.0 _{  -0.2} ^{  +0.0} $ & D15        & D15/M12            \\
BzK17999        & high-z MS BzK   & $      1.41 $ & $     2.2 \pm     0.2 $ & $     2.6 \pm     0.4 $ & $   14.4 _{  -7.7} ^{ +13.7} $ & $   12.0 _{  -0.0} ^{  +0.0} $ & $   10.7 _{  -0.0} ^{  +0.2} $ & D15        & D15/M12            \\
BzK21000        & high-z MS BzK   & $      1.52 $ & $     2.3 \pm     0.2 $ & $     2.6 \pm     0.4 $ & $   25.2 _{ -12.2} ^{  +5.2} $ & $   12.3 _{  -0.0} ^{  +0.0} $ & $   11.0 _{  -0.2} ^{  +0.1} $ & D15        & D15/M12            \\
BzK4171         & high-z MS BzK   & $      1.47 $ & $     1.8 \pm     0.2 $ & $     2.4 \pm     0.4 $ & $   16.5 _{  -8.1} ^{  +4.5} $ & $   12.0 _{  -0.0} ^{  +0.0} $ & $   10.7 _{  -0.1} ^{  +0.1} $ & D15        & D15/M12            \\
GN20            & high-z SB SMG   & $      4.06 $ & $     3.4 \pm     0.4 $ & $     3.2 \pm     0.7 $ & $   35.4 _{  -8.8} ^{  +4.9} $ & $   13.3 _{  -0.1} ^{  +0.0} $ & $   11.2 _{  -0.1} ^{  +0.0} $ & C10        & D09                \\
AzTEC-3         & high-z SB SMG   & $       5.3 $ & $     4.0 \pm     0.2 $ & $     3.6 \pm     0.5 $ & $  120.2 _{ -84.9} ^{  +8.9} $ & $   13.3 _{  -0.1} ^{  +0.0} $ & $   10.8 _{  -0.0} ^{  +0.2} $ & R10        & R10                \\
COSBO11         & high-z SB SMG   & $      1.83 $ & $     3.7 \pm     0.1 $ & $     3.3 \pm     0.5 $ & $   20.0 _{  -3.0} ^{  +0.4} $ & $   12.9 _{  -0.0} ^{  +0.0} $ & $   10.8 _{  -0.0} ^{  +0.0} $ & A08        & A08                \\
HFLS3           & high-z SB SMG   & $      6.34 $ & $     5.9 \pm     0.3 $ & $     5.0 \pm     0.4 $ & $   68.3 _{  -0.0} ^{ +10.5} $ & $   13.7 _{  -0.0} ^{  +0.1} $ & $   10.5 _{  -0.4} ^{  +0.5} $ & R13        & R13                \\
PACS-819        & high-z SB FMOS  & $      1.45 $ & $     3.5 \pm     0.2 $ & $     3.2 \pm     0.5 $ & $   27.7 _{  -0.1} ^{  +5.6} $ & $   12.5 _{  -0.0} ^{  +0.1} $ & $   10.7 _{  -0.1} ^{  +0.1} $ & THIS       & S15                \\
PACS-830        & high-z SB FMOS  & $      1.46 $ & $     1.6 \pm     0.2 $ & $     2.4 \pm     0.4 $ & $   24.3 _{  -2.3} ^{  +6.7} $ & $   12.4 _{  -0.0} ^{  +0.0} $ & $   11.0 _{  -0.0} ^{  +0.0} $ & THIS       & S15                \\
PACS-867        & high-z SB FMOS  & $      1.57 $ & $     1.6 \pm     0.3 $ & $     2.4 \pm     0.4 $ & $    2.8 _{  -2.3} ^{ +18.6} $ & $   12.0 _{  -0.0} ^{  +0.0} $ & $   10.8 _{  -0.1} ^{  +0.1} $ & THIS       & S15                \\
PACS-299        & high-z SB FMOS  & $      1.65 $ & $     2.6 \pm     0.2 $ & $     2.7 \pm     0.4 $ & $   28.3 _{ -19.1} ^{ +38.5} $ & $   12.4 _{  -0.0} ^{  +0.0} $ & $   10.1 _{  -0.0} ^{  +0.4} $ & THIS       & S15                \\
PACS-325        & high-z SB FMOS  & $      1.65 $ & $     0.0 \pm     3.4 $ &       \nodata           & $    1.2 _{  -0.7} ^{ +14.6} $ & $   11.8 _{  -0.1} ^{  +0.1} $ & $   10.4 _{  -0.1} ^{  +0.0} $ & THIS       & S15                \\
PACS-164        & high-z SB FMOS  & $      1.65 $ & $     1.9 \pm     0.4 $ & $     2.4 \pm     0.7 $ & $   18.1 _{ -15.4} ^{ +35.0} $ & $   12.5 _{  -0.0} ^{  +0.0} $ & $   10.2 _{  -0.2} ^{  +0.3} $ & THIS       & S15                \\
V20-ID41458     & high-z SB V20   & $      1.29 $ & $     1.8 \pm     0.2 $ & $     2.4 \pm     0.3 $ & $   33.5 _{  +0.0} ^{ +11.0} $ & $   12.5 _{  -0.0} ^{  +0.0} $ & $   11.1 _{  -0.0} ^{  +0.0} $ & V20        & V20                \\
V20-ID21060     & high-z SB V20   & $      1.28 $ & $     3.6 \pm     0.9 $ & $     3.4 \pm     1.3 $ & $   51.5 _{  -1.5} ^{  +0.5} $ & $   12.3 _{  -0.0} ^{  +0.0} $ & $   10.0 _{  -0.0} ^{  +0.1} $ & V20        & V20                \\
V20-ID51599     & high-z SB V20   & $      1.17 $ & $     2.1 \pm     0.2 $ & $     2.5 \pm     0.4 $ & $   14.4 _{  -1.9} ^{  +3.7} $ & $   12.5 _{  -0.0} ^{  +0.0} $ & $   11.1 _{  -0.0} ^{  +0.1} $ & V20        & V20                \\
V20-ID30694     & high-z MS V20   & $      1.16 $ & $     1.2 \pm     0.2 $ & $     2.2 \pm     0.3 $ & $   15.0 _{  -2.9} ^{  +5.5} $ & $   12.0 _{  -0.0} ^{  +0.1} $ & $   10.9 _{  -0.0} ^{  +0.2} $ & V20        & V20                \\
V20-ID38053     & high-z SB V20   & $      1.15 $ & $     1.3 \pm     0.4 $ & $     2.3 \pm     0.4 $ & $   18.9 _{  -0.1} ^{ +11.6} $ & $   12.0 _{  -0.0} ^{  +0.0} $ & $   10.5 _{  -0.0} ^{  +0.0} $ & V20        & V20                \\
V20-ID48881     & high-z SB V20   & $      1.16 $ & $     1.9 \pm     0.5 $ & $     2.5 \pm     0.5 $ & $   42.9 _{  -0.2} ^{  +0.5} $ & $   12.3 _{  -0.0} ^{  +0.0} $ & $   10.6 _{  -0.0} ^{  +0.0} $ & V20        & V20                \\
V20-ID37250     & high-z SB V20   & $      1.15 $ & $     1.1 \pm     0.1 $ & $     2.2 \pm     0.2 $ & $    9.9 _{  -1.1} ^{  +3.7} $ & $   12.2 _{  -0.0} ^{  +0.0} $ & $   11.0 _{  -0.2} ^{  +0.0} $ & V20        & V20                \\
V20-ID44641     & high-z MS V20   & $      1.15 $ & $     1.0 \pm     0.3 $ & $     2.2 \pm     0.5 $ & $    9.4 _{  -2.8} ^{  +2.4} $ & $   12.0 _{  -0.1} ^{  +0.0} $ & $   11.2 _{  -0.3} ^{  +0.0} $ & V20        & V20                \\
V20-ID51936     & high-z SB V20   & $       1.4 $ & $     1.9 \pm     0.3 $ & $     2.4 \pm     0.3 $ & $    5.3 _{  -0.1} ^{  +2.1} $ & $   12.0 _{  -0.0} ^{  +0.0} $ & $   10.5 _{  -0.0} ^{  +0.0} $ & V20        & V20                \\
V20-ID31880     & high-z SB V20   & $       1.4 $ & $     2.2 \pm     0.4 $ & $     2.6 \pm     0.5 $ & $   20.5 _{  +0.0} ^{  +2.8} $ & $   12.3 _{  -0.0} ^{  +0.0} $ & $   11.0 _{  -0.0} ^{  +0.0} $ & V20        & V20                \\
V20-ID2299      & high-z SB V20   & $      1.39 $ & $     3.4 \pm     0.3 $ & $     3.2 \pm     0.5 $ & $   13.8 _{  -0.4} ^{  +0.4} $ & $   12.7 _{  -0.0} ^{  +0.0} $ & $   11.1 _{  -0.1} ^{  +0.0} $ & V20        & V20                \\
V20-ID21820     & high-z MS V20   & $      1.38 $ & $     2.1 \pm     0.4 $ & $     2.5 \pm     0.5 $ & $   15.7 _{  -2.3} ^{  +7.6} $ & $   12.2 _{  -0.0} ^{  +0.0} $ & $   11.0 _{  -0.1} ^{  +0.0} $ & V20        & V20                \\
V20-ID13205     & high-z SB V20   & $      1.27 $ & $     3.0 \pm     0.7 $ & $     3.1 \pm     1.2 $ & $   49.8 _{ -10.8} ^{ +16.6} $ & $   12.3 _{  -0.0} ^{  +0.0} $ & $   11.1 _{  -0.2} ^{  +0.0} $ & V20        & V20                \\
V20-ID13854     & high-z MS V20   & $      1.27 $ & $     1.8 \pm     0.3 $ & $     2.5 \pm     0.5 $ & $   20.0 _{  -3.0} ^{  +0.4} $ & $   12.2 _{  -0.0} ^{  +0.0} $ & $   11.1 _{  -0.0} ^{  +0.0} $ & V20        & V20                \\
V20-ID19021     & high-z SB V20   & $      1.26 $ & $     1.9 \pm     0.3 $ & $     2.4 \pm     0.4 $ & $   25.0 _{  +0.0} ^{  +5.4} $ & $   12.3 _{  -0.0} ^{  +0.0} $ & $   10.4 _{  -0.0} ^{  +0.0} $ & V20        & V20                \\
V20-ID35349     & high-z MS V20   & $      1.26 $ & $     0.8 \pm     0.2 $ & $     2.1 \pm     0.2 $ & $    8.2 _{  -0.6} ^{  +4.0} $ & $   12.0 _{  -0.0} ^{  +0.0} $ & $   11.2 _{  -0.1} ^{  +0.0} $ & V20        & V20                \\
V20-ID42925     & high-z SB V20   & $       1.6 $ & $     2.1 \pm     0.4 $ & $     2.5 \pm     0.5 $ & $   59.9 _{ -18.5} ^{  +0.4} $ & $   12.7 _{  -0.0} ^{  +0.0} $ & $   11.0 _{  -0.0} ^{  +0.0} $ & V20        & V20                \\
V20-ID38986     & high-z MS V20   & $      1.61 $ & $     2.8 \pm     0.9 $ & $     2.8 \pm     1.4 $ & $   19.5 _{ -16.1} ^{+155.1} $ & $   12.0 _{  -0.1} ^{  +0.0} $ & $   11.1 _{  -0.0} ^{  +0.0} $ & V20        & V20                \\
V20-ID30122     & high-z MS V20   & $      1.46 $ & $     2.0 \pm     0.4 $ & $     2.6 \pm     0.6 $ & $   13.4 _{  -4.2} ^{  +1.3} $ & $   12.2 _{  -0.0} ^{  +0.0} $ & $   10.9 _{  -0.0} ^{  +0.1} $ & V20        & V20                \\
V20-ID41210     & high-z SB V20   & $      1.31 $ & $     2.1 \pm     0.2 $ & $     2.5 \pm     0.4 $ & $   25.0 _{  -9.6} ^{  +0.4} $ & $   12.3 _{  -0.1} ^{  +0.0} $ & $   10.6 _{  -0.0} ^{  +0.0} $ & V20        & V20                \\
V20-ID2993      & high-z SB V20   & $      1.19 $ & $     1.4 \pm     0.3 $ & $     2.3 \pm     0.4 $ & $   13.1 _{  -3.0} ^{  +7.4} $ & $   12.2 _{  -0.0} ^{  +0.0} $ & $   11.0 _{  -0.2} ^{  +0.1} $ & V20        & V20                \\
V20-ID48136     & high-z MS V20   & $      1.18 $ & $     1.5 \pm     0.2 $ & $     2.3 \pm     0.3 $ & $   14.9 _{  -3.3} ^{  +3.0} $ & $   12.3 _{  -0.1} ^{  +0.0} $ & $   11.1 _{  -0.0} ^{  +0.1} $ & V20        & V20                \\
V20-ID51650     & high-z SB V20   & $      1.34 $ & $     2.8 \pm     0.4 $ & $     2.9 \pm     0.6 $ & $   21.1 _{  -5.0} ^{  +9.4} $ & $   12.2 _{  -0.0} ^{  +0.1} $ & $   10.9 _{  -0.0} ^{  +0.0} $ & V20        & V20                \\
V20-ID15069     & high-z SB V20   & $      1.21 $ & $     1.7 \pm     0.6 $ & $     2.4 \pm     1.0 $ & $    6.1 _{  -1.2} ^{  +2.3} $ & $   12.0 _{  -0.0} ^{  +0.0} $ & $   10.8 _{  -0.3} ^{  +0.1} $ & V20        & V20                \\
\enddata
\tablerefs{%
THIS = This work (see Appendix~\ref{Section_Appendix_PdBI_Observation}); 
L15 = \cite{Liudz2015}; 
P14 = \cite{Papadopoulos2014ApJ...788..153P}; 
K14 = \cite{Kamenetzky2014}; 
G09 = \cite{Greve2009ApJ...692.1432G}; 
E90 = \cite{Eckart1990ApJ...363..451E}; 
I14 = \cite{Israel2014}; 
B08 = \cite{Baan2008}; 
W08 = \cite{Wilson2008}; 
P12 = \cite{Papadopoulos2012}; 
G93 = \cite{Garay1993AaA...277..405G}; 
L09 = \cite{Leroy2009}; 
B06 = \cite{Bayet2006}; 
A07 = \cite{Albrecht2007AaA...462..575A}; 
H99 = \cite{Harrison1999}; 
B92 = \cite{Braine1992}; 
K11 = \cite{Kamenetzky2011}; 
A11 = \cite{Alatalo2011ApJ...735...88A}; 
Y11 = \cite{Young2011MNRAS.414..940Y}; 
S95 = \cite{Sandqvist1995}; 
A95 = \cite{Aalto1995}; 
W04 = \cite{Wang2004NGC4945}; 
K16 = \cite{Kamenetzky2016}; 
D15 = \cite{Daddi2015}; 
M12 = \cite{Magnelli2012AaA...548A..22M}; 
D09 = \cite{Daddi2009GN20}; 
W12 = \cite{Walter2012Natur.486..233W}; 
C10 = \cite{Carilli2010GN20}; 
R10 = \cite{Riechers2010AzTEC3}; 
A08 = \cite{Aravena2008COSBO11}; 
R13 = \cite{Riechers2013HFLS3}; 
S15 = \cite{Silverman2015b}; 
V20 = \cite{Valentino2020arXiv200612521V}; 
}
\tablecomments{%
Only a few selected key columns are shown here. The full sample table has more columns including galaxy properties of \citetalias{DL07} warm and cold dust luminosities, AGN luminosities, offset from the MS, which are used in Fig.~\ref{Plot_R52_vs_params}. 
The full machine-readable table is available at \url{https://doi.org/10.5281/zenodo.3958271}. 
}
\end{deluxetable*}

\begin{figure*}[htb]
\centering
\includegraphics[width=0.490\textwidth]{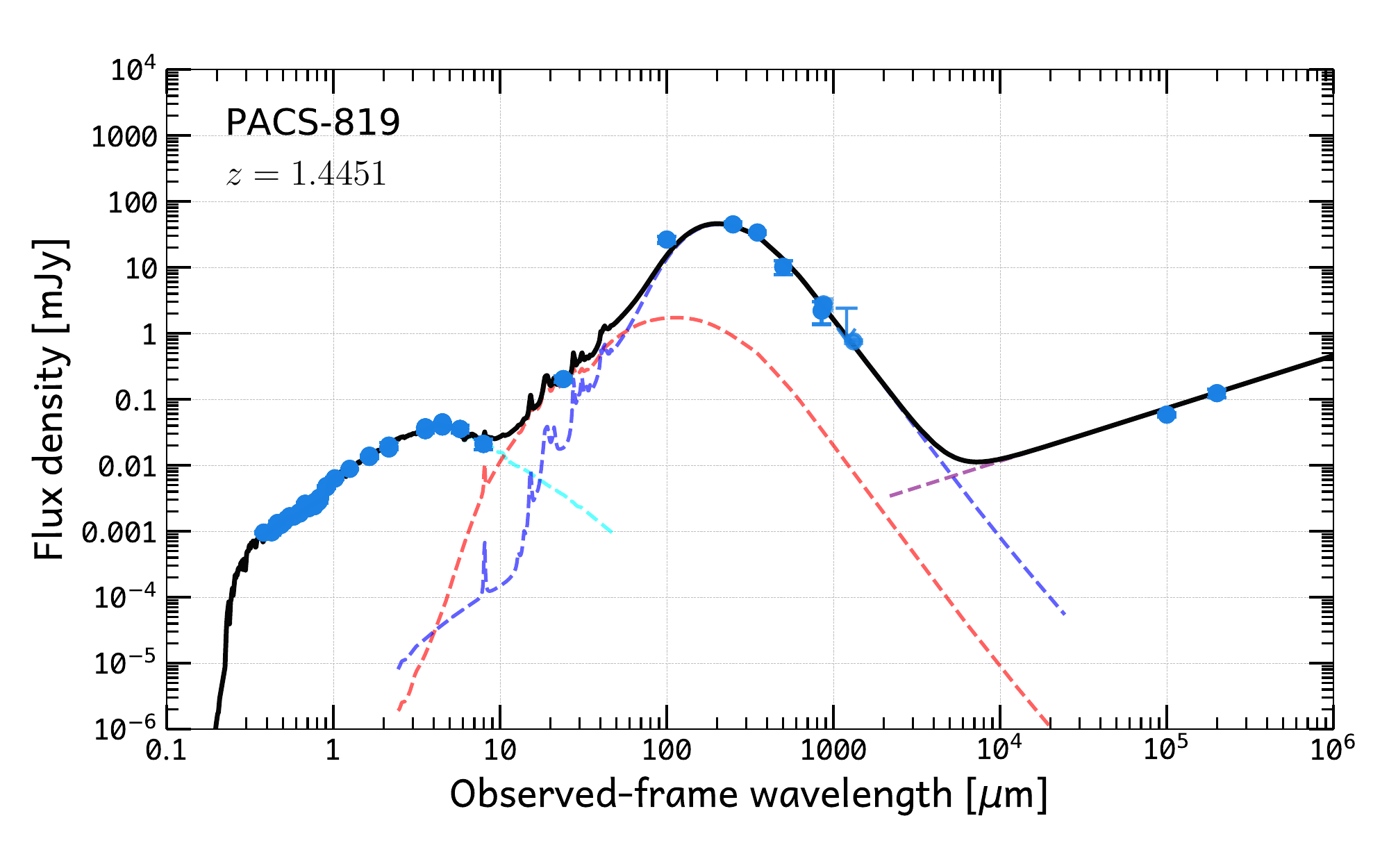}
\includegraphics[width=0.490\textwidth]{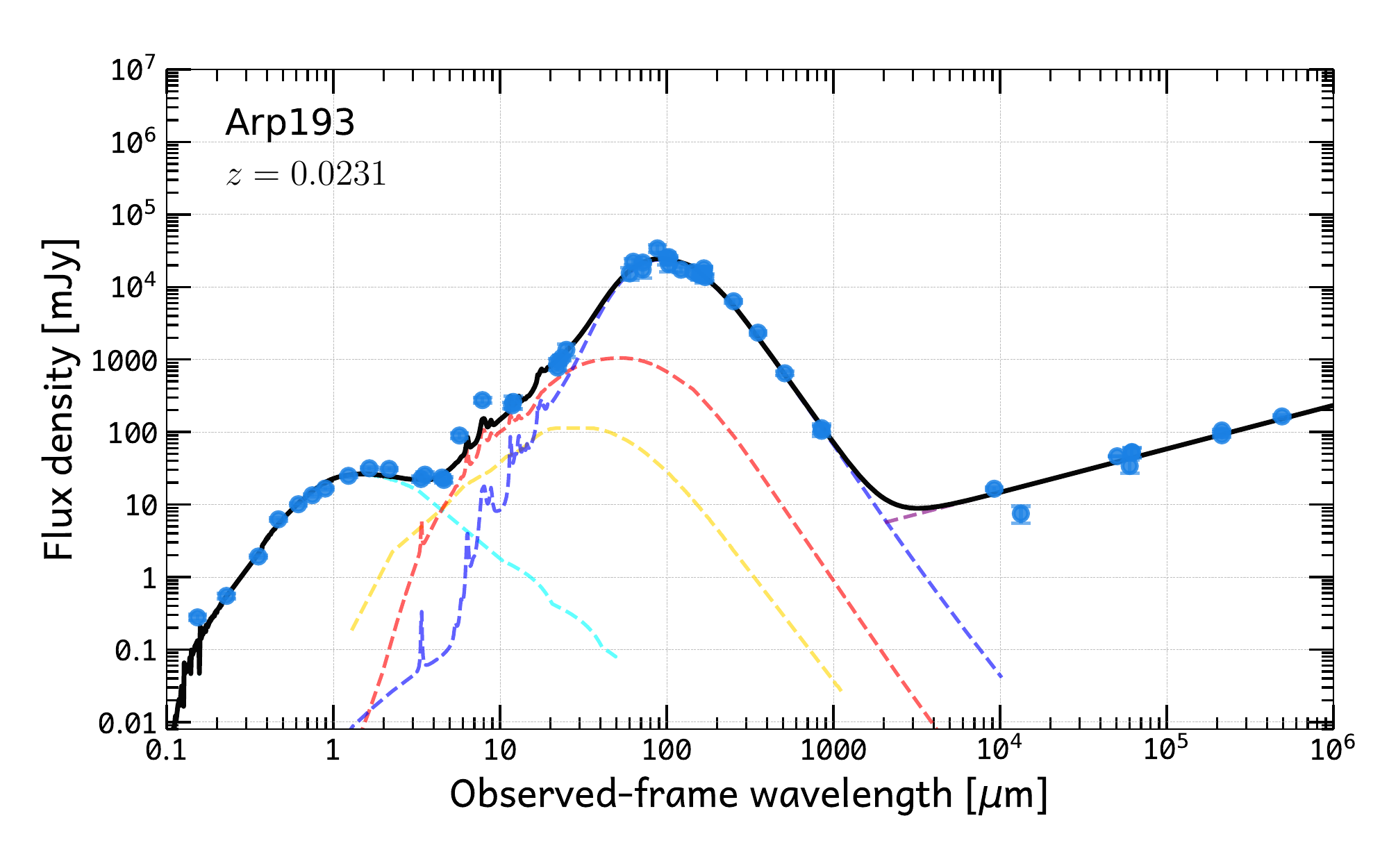}
\includegraphics[width=0.490\textwidth, trim=0 -6mm 0 12mm]{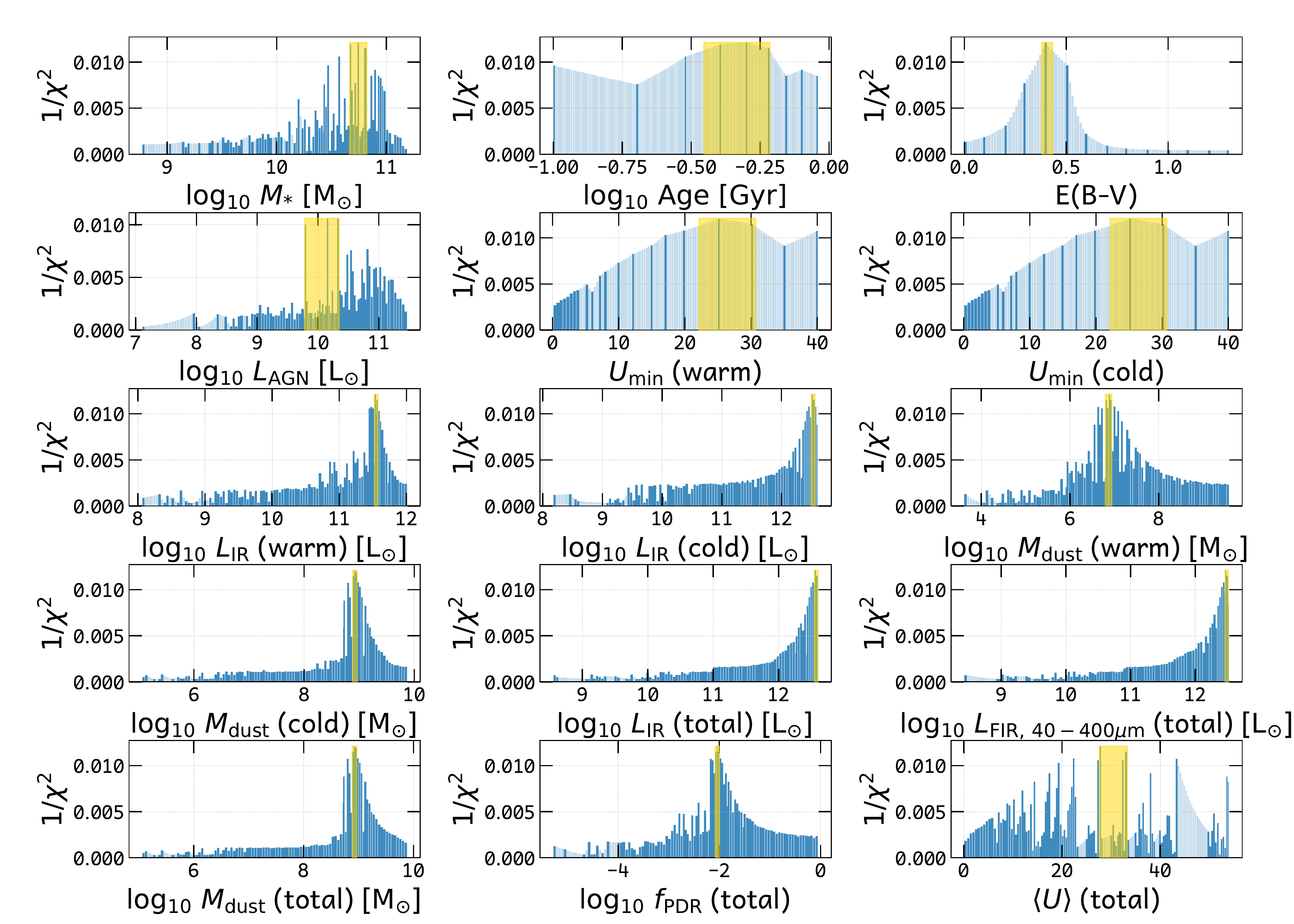}
\includegraphics[width=0.490\textwidth, trim=0 -6mm 0 12mm]{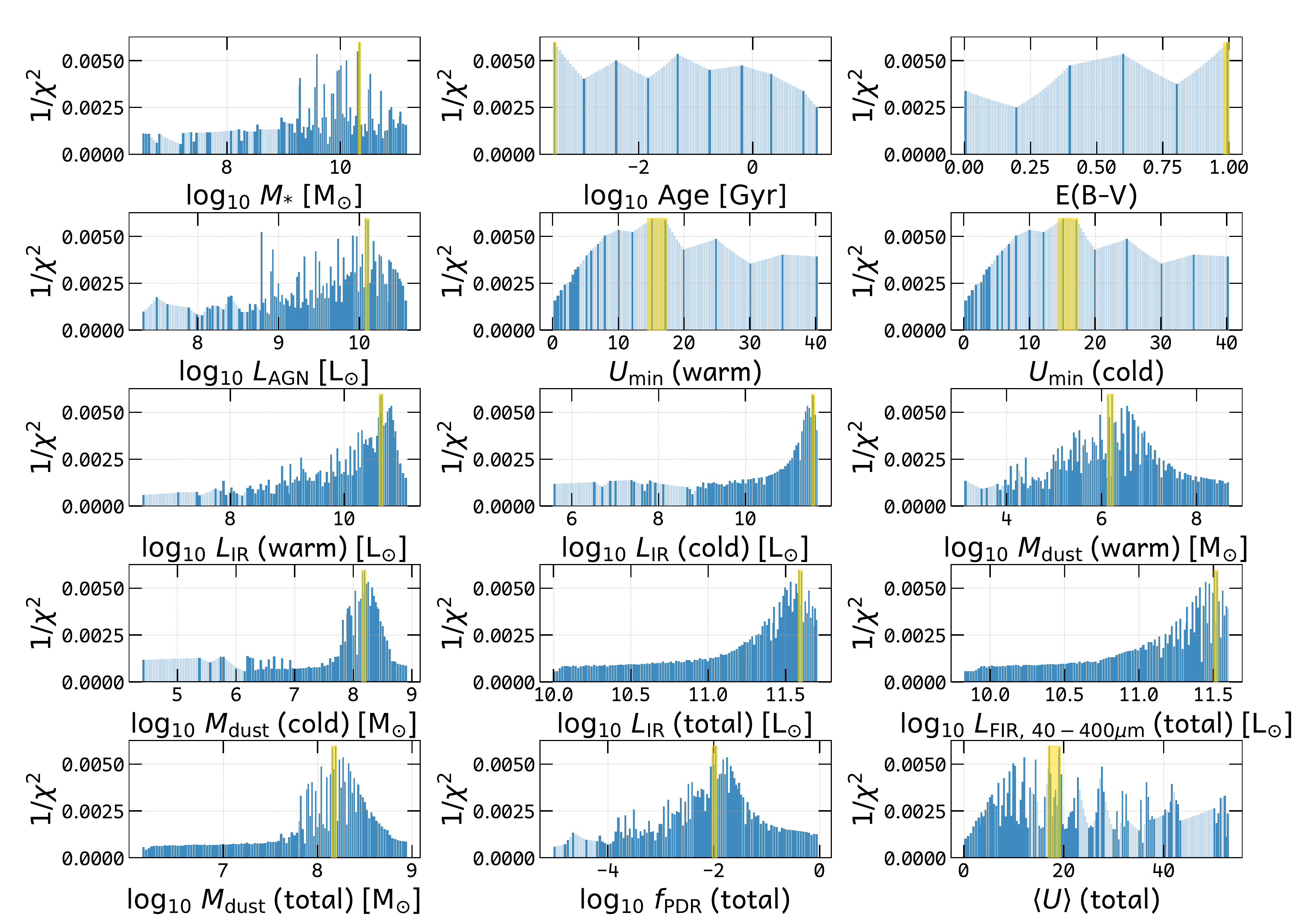}
\caption{%
Two examples of our SED fitting for PACS-819 \textit{(left)} and Arp~193 \textit{(right)} with our \michi2{} code as described in Sect.~\ref{Section_SED_Fitting}. 
Upper panels show the best-fit SED (black line) and SED components, which are stellar (cyan dashed line), mid-IR AGN (yellow dashed line, optional if AGN is present), PDR dust (red dashed line) and cold/ambient dust (blue dashed line). Photometry data are shown by circles with errorbars or downward arrows for upper limits if $\SNR<3$. 
Lower panels show $1/\chi^2$ distributions for several galaxy properties from our SED fitting. In each sub-panel, the height of histogram indicates the highest $1/\chi^2$ in each bin of the $x$-axis galaxy property. A higher $1/\chi^2$ means a better fit. The 68\% confidence level for our five SED component fitting is indicated by the yellow shading. 
{\it (Figures for all sources are available at \url{https://doi.org/10.5281/zenodo.3958271}.)}
\label{Figure_SED_fitting}
}
\end{figure*}

\vspace{0.5truecm}

\section{Spectral Energy Distribution (SED) Fitting: the \michi2{} code}
\label{Section_SED_Fitting}

The well-sampled SEDs from optical to far-IR/mm allow us to obtain accurate dust properties by fitting them with SED templates. Particularly, since dust grains do not have a single temperature in a galaxy, the mean ISRF intensity, $\Umean$, has been considered to be a more physical proxy of dust emission properties (\citetalias{DL07}). It represents the 0--13.6~eV intensity of interstellar UV radiation in units of the \cite{Mathis1983} ISRF intensity (see \citealt{Draine2007SINGS}).

The $\Umean$ parameter has advantages in describing mixture states of ISRF over using a single or several dust temperatures values to describe galaxy dust SEDs. In \citetalias{DL07} dust models, 
the majority of dust grains are exposed to a minimum ambient ISRF with intensity $\Umin$, while the rest are exposed to the photon-dominated region (PDR) ISRF, with intensities ranging from $\Umin$ to $\Umax$ in a power-law PDF (in mass). The mass fraction of the latter dust grain population (``warm dust'' or ``PDR dust'') is expressed as $\fPDR$ in this work and is a free parameter in the fit. $\Umin$ is another free parameter, while $\Umax$ is empirically fixed, as well as the power-law index (see more detailed introduction in \citealt{Draine2007SINGS,Draine2014}; \citealt{Aniano2012,Aniano2020}). 
As pointed out by \citet{Dale2002}, such a physically driven dust model actually fits the mass distribution of molecular clouds (\citealt{Stutzki2001}; \citealt{Shirley2002}; \citealt{Elmegreen2002}). 
Based on this model, \citetalias{DL07} generated SED templates which can then be used for fitting by other works using their own SED fitting code.

In this work, we use our own-developed SED fitting code, \michi2{}\,\footnote{\url{https://github.com/1054/Crab.Toolkit.michi2}}, providing us the flexibility in combining multiple SED components and choosing SED templates for each component. Comparing with popular panchromatic (UV-to-mm/radio) SED fitting codes, e.g., \incode{MAGPHYS} (\citealt{daCunha2008,daCunha2015}), \incode{LePhare} (\citealt{Arnouts1999,Ilbert2006}), \incode{CIGALE} (\citealt{Noll2009,Ciesla2015,Boquien2019}), our code fits SEDs well and produces similar best-fitting results 
(see Appendix~\ref{Section_Appendix_SED})%
. 
Our code also performs $\chi^2$-based posterior probability distribution analysis and estimates reasonable (asymmetric) uncertainties for each free or derived parameter (e.g., Fig.~\ref{Figure_SED_fitting}). 

Our code can also handle an arbitrary number of 
SED libraries as the components of the whole SED. For example, we use five SED libraries/components representing stellar, AGN, \citetalias{DL07} warm dust, \citetalias{DL07} cold dust, and radio emissions (see below). Our code samples their combinations in the five-dimensional space, then generates a composite SED (after multiplying the model with the filter curves), then fits to the observed photometric data and obtains $\chi^2$ statistics. The post-processing of the $\chi^2$ distribution provides the best-fit and probability range of each physical parameter in the SED libraries (following \citealt{Press1992}, chapter 15.6).

Details of the five SED libraries/components are: 
\begin{itemize}[itemsep=1pt,topsep=2pt]
	\item stellar component: for high-redshift ($z>1$) star-forming galaxies, we use the \cite{BC03} code to generate solar metallicity, constant star formation history, \cite{Chabrier2003}-IMF SED templates, then apply the \cite{Calzetti2000} attenuation law with a range of $\mathrm{E(B-V)} = 0.0$ to $1.0$ to construct our SED library. For local galaxies, we use the FSPS (\citealt{Conroy2009,Conroy2010,Conroy2010ascl}) code to generate solar metallicity, $\tau$-declining star formation history, \citealt{Chabrier2003}-IMF SED templates (also with \citealt{Calzetti2000} attenuation law), as this generates a larger variety of SED templates which fit local galaxies better. 
	
	\item mid-IR AGN component: we use the observationally calibrated AGN torus SED templates from \cite{Mullaney2011}. They cover $6-100\,\mathrm{\mu m}$ in wavelengths, and can fit both Type~1, Type~2 and intermediate-type AGNs as demonstrated by \cite{Mullaney2011}. 
	
	\item \citetalias{DL07} warm dust component for dust grains exposed to the PDR ISRF with intensity ranging from $\Umin$ to $\Umax=10^{7}$ in a power-law PDF with an index of $-2$ (updated version; see \citealt{Draine2014,Aniano2020}). The fraction of dust mass in Polycyclic Aromatic Hydrocarbons (PAHs) is described by $q_{\mathrm{PAH}}$. The contribution of such warm dust to total ISM dust in mass is described by $\fPDR$ in this work (i.e., the $\gamma$ in \citetalias{DL07}). Free parameters are $\Umin$, $q_{\mathrm{PAH}}$ and $\fPDR$. 
	
	\item \citetalias{DL07} cold dust component for dust grains exposed to the ambient ISRF with intensity of $\Umin$. The $\Umin$ and $q_{\mathrm{PAH}}$ of the cold dust are fixed to be the same as the warm dust in our fitting. 
	
	\item radio component: a simple power-law with index $-0.8$ is assumed. Our code has the option to fix the normalization of the radio component at rest-frame 1.4~GHz to the total IR luminosity $L_{\mathrm{IR}\,(8-1000{\mu}{\mathrm{m}})}$ (integrating warm and cold dust components only) via assumptions about the IR-radio correlation (e.g., \citealt{Condon1991}; \citealt{Yun2001}; \citealt{Ivison2010}; \citealt{Magnelli2015}) when galaxies lack sufficient IR photometric data and display no obvious radio excess due to AGN (e.g., \citealt{Liudz2018}). 
	As radio is not the focus of this work, we only use the simple power-law assumption for illustration purposes. 
	
\end{itemize}

Note that we do not balance the dust attenuated stellar light with the total dust emission. This has the advantage of allowing for optically thick dust emission that is only seen in the infrared. 
Our fitting then outputs $\chi^2$ distributions for the following parameters of interest (see bottom panels in Fig.~\ref{Figure_SED_fitting}):
\begin{itemize}[itemsep=1pt,topsep=2pt]
	\item Stellar properties, including stellar mass $\Mstar$, dust attenuation $\mathrm{E(B-V)}$, and light-weighted stellar age. 
	
	\item AGN luminosity $L_{\mathrm{AGN}}$, integrated over the AGN SED component. 
	
	\item IR luminosities for cold dust ($L_{\mathrm{IR,\,cold\,dust}}$), warm dust ($L_{\mathrm{IR,\,PDR\,dust}}$) and their sum ($L_{\mathrm{IR,\,total\,dust}}$). 
	
	\item Mean ISRF intensity $\Umean$, minimum ISRF intensity $\Umin$, and the mass fraction of warm/PDR-like dust in the \citetalias{DL07} model $\fPDR$. 
	
\end{itemize}

In Fig.~\ref{Figure_SED_fitting} we show two examples of our SED fitting. 
Best-fit parameters and their errors are also listed in our full sample table (Table~\ref{Table_All_in_one} online version). 

To verify our SED fitting, we also fit our high-$z$ galaxies' SEDs with \textsc{MAGPHYS} and \textsc{CIGALE} 
(see more details in Appendix~\ref{Section_Appendix_SED}). 
We find that for most high-$z$ galaxies the stellar masses and IR luminosities are agreed within 
$\sim 0.2 - 0.3$~dex. 
The IR luminosities are more consistent than stellar masses among the results of three fitting codes, with a scatter of $\sim 0.2$\,dex. 
In several outlier cases, our code produces more reasonable fitting to the data (e.g., AzTEC-3, Arp220, NGC0253), which is likely because we do not have an energy balance constraint in the code. Our code has no systematic bias against \textsc{CIGALE}, but there is a noticeable trend that \textsc{MAGPHYS} fits slightly larger stellar masses than the other two. A possible reason is the use of the \cite{Charlot2000} double attenuation law in \textsc{MAGPHYS} (see \citealt{LoFaro2017}) rather than the \cite{Calzetti2000} attenuation law in our \michi2{} and \textsc{CIGALE} fitting.

Given the general agreement between our code and \textsc{CIGALE}/\textsc{MAGPHYS}, and
to be consistent within this paper, we fit all SEDs with our \michi2{} SED fitting code with the five SED libraries as mentioned above.

\vspace{0.5truecm}

\section{Interstellar Radiation Field Traces CO Excitation: The $\Umean$--$\R52$ Correlation}
\label{Section_Umean_R52}

\begin{figure*}[htb]
\centering
\includegraphics[width=\textwidth]{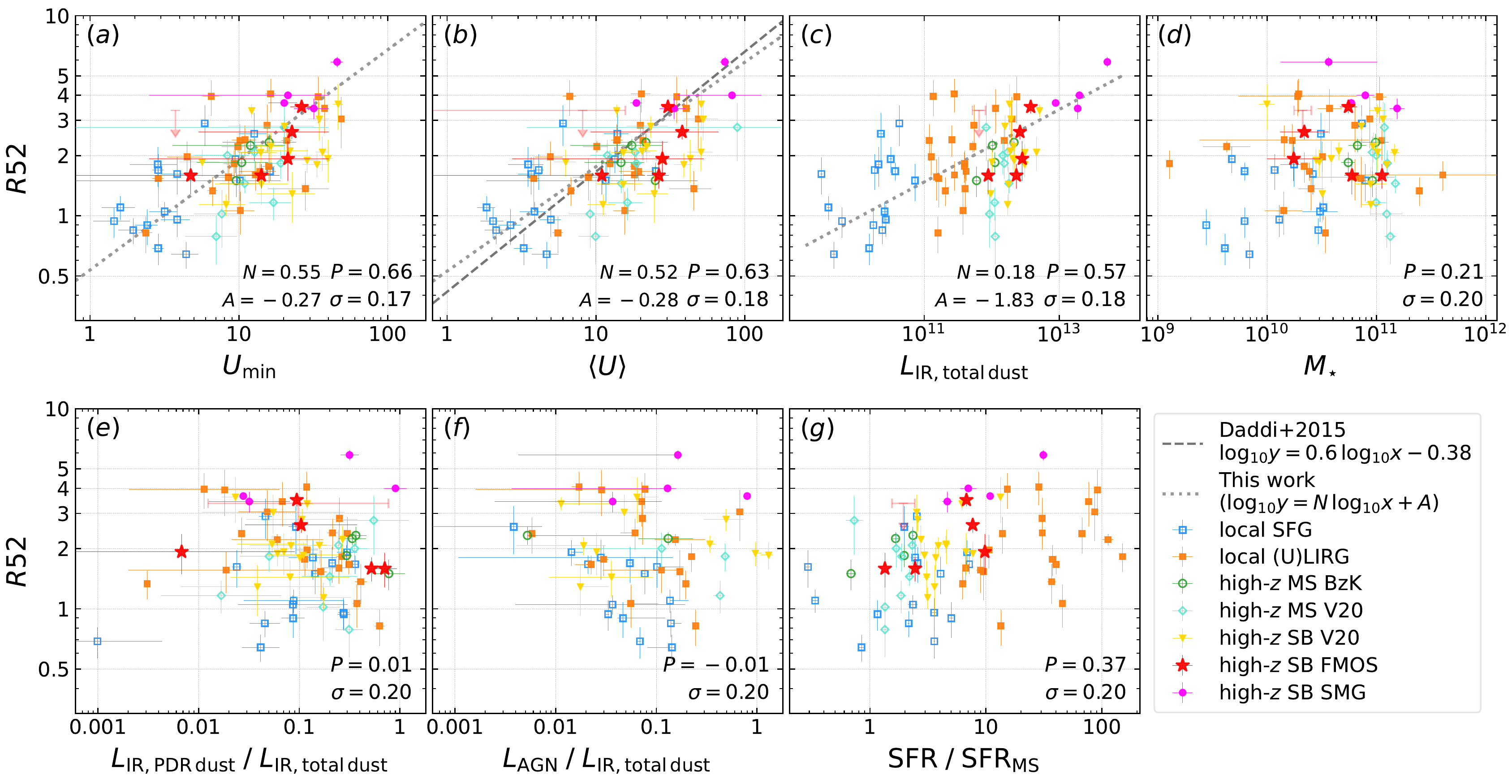}
\vspace{-1ex}
\caption{%
CO(5-4) to(2-1) line ratio $\R52$ versus various galaxy properties: 
(\textit{a}) ambient ISRF intensity ($\Umin$); 
(\textit{b}) mean ISRF intensity ($\Umean$); 
(\textit{c}) dust IR luminosity; 
(\textit{d}) stellar mass; 
(\textit{e}) luminosity fraction of dust exposed to warm/PDR-like ISRF to total dust in ISM (does not include AGN torus); 
(\textit{f}) luminosity ratio between mid-IR AGN and total ISM dust (AGN luminosity is integrated over for all available wavelengths while dust luminosity is integrated over rest-frame 8-1000\,$\mu$m); 
and (\textit{g}) the offset to the MS in terms of SFR. 
The Pearson coefficient $P$ and scatter $\sigma$ for each correlation are shown at bottom right. We performed orthogonal distance regression (ODR) linear regression fitting to the data points and their $x$ and $y$ errors in panels (a), (b) and (c), where $P>0.5$. Dotted lines are the best-fits from this work, with slope $N$ and intercept $A$ shown at the bottom. 
The dashed line in panel (b) is the best-fit linear regression from \cite{Daddi2015}. 
\label{Plot_R52_vs_params}
}
\end{figure*}

We use our SED fitting results and the compiled CO data to study the empirical correlation between the CO(5-4)/CO(2-1) line ratio $\R52$ and the mean ISRF intensity $\Umean$. This correlation physically links molecular gas and dust properties together, supporting the idea that gas and dust are generally mixed together at large scales and exposed to the same local ISRF. 

In Fig.~\ref{Plot_R52_vs_params} we correlate $\R52$ with various galaxy properties derived from our SED fitting. Panel \textit{(a)} shows a tight correlation between $\R52$ and the ambient ISRF intensity $\Umin$, and panel \textit{(b)} confirms the tight correlation between $\R52$ and $\Umean$ which was first reported by \cite{Daddi2015}. 
Panels \textit{(c)} and \textit{(d)} show that CO excitation is also well correlated with galaxies' dust luminosities, but not with their stellar masses. In panels \textit{(e)}~to~\textit{(g)}, we show that $\R52$ exhibits no correlation with $\fPDR$ and mid-IR AGN fraction, while a very weak correlation seems to exist between $\R52$ and the offset to the MS SFR, $\SFR/\SFR_{\mathrm{MS}}$. In each panel, the Pearson correlation coefficient $P$ is computed and shown at bottom right. These correlations, or lack there-of, demonstrate that $\R52$ or mid-$J$ CO excitation is indeed mostly driven by dust-related quantities, i.e., $\LIR$, $\Umean$ and $\Umin$. 

Our best fitting $\R52$--$\Umean$ correlation is close to the one found by \cite{Daddi2015}, yet somewhat shallower than that. \cite{Valentino2020arXiv200612521V} also reported a shallower slope of the $\R52$--$\Umean$ correlation, given that the high-$z$ V20 sample is used in both their and this work. Indeed, sub-samples behave slightly differently in Fig.~\ref{Plot_R52_vs_params}. While local SFGs and local (U)LIRGs are scattered well around the average $\R52$--$\Umean$ correlation line, high-$z$ MS and SB galaxies from the FMOS and V20 subsamples tend to lie below it. Given the varied $\SNR$ of IR data as reflected by the $\Umean$ error bars, the majority of those high-$z$ galaxies do not have a high-quality constraint on $\Umean$. High-$z$ sample selections for CO observations are usually also biased to high-$z$ IR-bright galaxies. Therefore, it is difficult to draw a conclusion about any redshift evolution of the $\R52$--$\Umean$ correlation with the current dataset.

From panel \textit{(f)} of Fig.~\ref{Plot_R52_vs_params}, we can see that there are several galaxies showing a high AGN to ISM dust luminosity ratio (note that the AGN luminosity is integrated over all wavelengths, while the IR luminosity is only \citetalias{DL07} warm+cold dust integrated over 8--1000$\mu$m). The three galaxies with $L_{\mathrm{AGN},\,\mathrm{all}\,\lambda} / L_{\mathrm{IR,\,8\textnormal{--}1000{\mu}m}} \gtrsim 0.9$ are V20-ID38986, V20-ID51936 and V20-ID19021, from high to low respectively. They all clearly show power-law shape SEDs from the near-IR IRAC bands to mid-IR MIPS 24\,$\mu$m and PACS 100\,$\mu$m\,\footnote{Their SEDs figures are accessible at the link mentioned in the caption of Fig.~\ref{Figure_SED_fitting}. With high $\SNR$ IRAC to MIPS~24\,$\mu$m data, their mid-IR AGN and any PAH feature if present can be well distinguished by our SED fitting. Yet, we note that for galaxies with low $\SNR$ IRAC to MIPS~24\,$\mu$m data the uncertainty in AGN component identification could be high.}. 
However, their $\R52$ do not tend to be higher. This likely supports that these mid-$J$ ($J_{\mathrm{u}}\sim5$) CO lines are not overwhelmingly affected by AGN. 

We note that the correlations in Fig.~\ref{Plot_R52_vs_params} are not the only ones worth exploring. $\R52$ also correlates with dust mass in a way similar to $\Umean$ but with larger scatter, and $\Umean$ can be considered as the ratio of $\LIR/\Mdust$, therefore here we omit the correlation with $\Mdust$. \cite{Daddi2015} also investigated how SFR surface density ($\Sigma_{\SFR}$), star formation efficiency ($\SFR/\Mgas$), gas-to-dust ratio ($\deltaGDR$) and massive star-forming clumps affect $\Umean$ and $\R52$. Their results support the idea that a larger fraction of massive star-forming clumps with denser molecular gas compared to the diffuse, low density molecular gas is the key for a high CO excitation (as proposed by the simulation work of \citealt{Bournaud2015}). Therefore, to understand the key physical drivers of CO excitation, information on molecular gas density distributions is likely the most urgently required.

\vspace{0.5truecm}

\section{Modeling of Molecular Gas Density Distribution in Galaxies}
\label{Section_gas_modeling}

CO line emission in galaxies arises mainly from the cold molecular gas, and CO line ratios/SLEDs are sensitive to local molecular gas physical conditions, i.e., volume density $\nH2$, column density $\NH2_$, and kinetic temperature $\Tkin$. These properties typically vary by one to three orders of magnitude within a galaxy, e.g., as seen in observations as reviewed by \citet{Young1991Review}, \citet{Solomon2005Review}, \citet{Carilli2013Review}, \citet{Combes2018Review} and references therein, 
and also in modeling and simulations, e.g., by \citet{Krumholz2007}, \citet{Glover2012b}, \citet{Smith2014b}, \citet{Narayanan2014}, \citet{Bournaud2015}, \citet{Glover2015}, \citet{Glover2016}, \citet{Popping2016,Popping2019}, \citet{Renaud2019a,Renaud2019b} and \citet{Tress2020}. 

In practice, studies of the CO SLED at a global galaxy scale or at sub-kpc scales usually require the presence of a relative dense gas component ($\nH2 \sim 10^{3\textnormal{--}5}\mathrm{cm^{-3}}$; $\Tkin \gtrsim 50\textnormal{--}100 \,\mathrm{K}$) in addition to a relatively diffuse gas ($\nH2 \sim 10^{2\textnormal{--}3}\mathrm{cm^{-3}}$; $\Tkin \sim 20\textnormal{--}100 \,\mathrm{K}$), via non-local thermodynamic equilibrium (non-LTE) LVG radiative transfer modeling (e.g., \citealt{Israel1995,Israel2001,Israel2002,Israel2003,Israel2015,Mao2000M82,Weiss2001M82,Weiss2005M82,Bradford2003,Zhu2003,Bayet2004,Bayet2006,Bayet2009,Papadopoulos2007b,Papadopoulos2008,Papadopoulos2010a,Papadopoulos2010b,Papadopoulos2012,HaileyDunsheath2008,HaileyDunsheath2012,Panuzzo2010,Rangwala2011,Papadopoulos2012,Spinoglio2012,Meijerink2013,PereiraSantaella2013,Rigopoulou2013,Rosenberg2014,Rosenberg2014Arp299,LuNanyao2014,LuNanyao2017,ZhangZhiyu2014,Greve2014,Liudz2015,Kamenetzky2014,Kamenetzky2016,Kamenetzky2017,Schirm2014,Schirm2017,Mashian2015,WuRonin2015,Yang2017,Valentino2020arXiv200612521V}). A third state which is mostly responsible for $J_{\mathrm{u}} \gtrsim 10$ CO lines is also found in the case of AGN (e.g., \citealt{vanderWerf2010,Rangwala2011,Spinoglio2012}) or mechanical heating (e.g., \citealt{Rosenberg2014}). 
Therefore, a mid-to-low-$J$ CO line ratio like $\R52$ not only reflects the excitation condition of a single gas state, but also the relative amount of the denser, warmer to the more diffuse gas component.

\cite{Leroy2017} have conducted pioneer modeling of the sub-beam gas density PDF to understand line ratios of CO isotopologue and dense gas tracers. The method includes constructing a series of one-zone clouds, performing non-LTE LVG calculation, and compositing line fluxes by the gas density PDF. 
They demonstrated that such modeling can successfully reproduce observed isotopologue or dense gas tracers to CO line ratios. 
Inspired by this work, we present in this section similar sub-beam density-PDF gas modeling to study the CO excitation, and propose a useful conversion from $\R52$ observations to $\nmean$ and $\Tkin$ for galaxies at global scales.

\begin{figure*}[htb]
	\centering%
	\includegraphics[width=\textwidth]{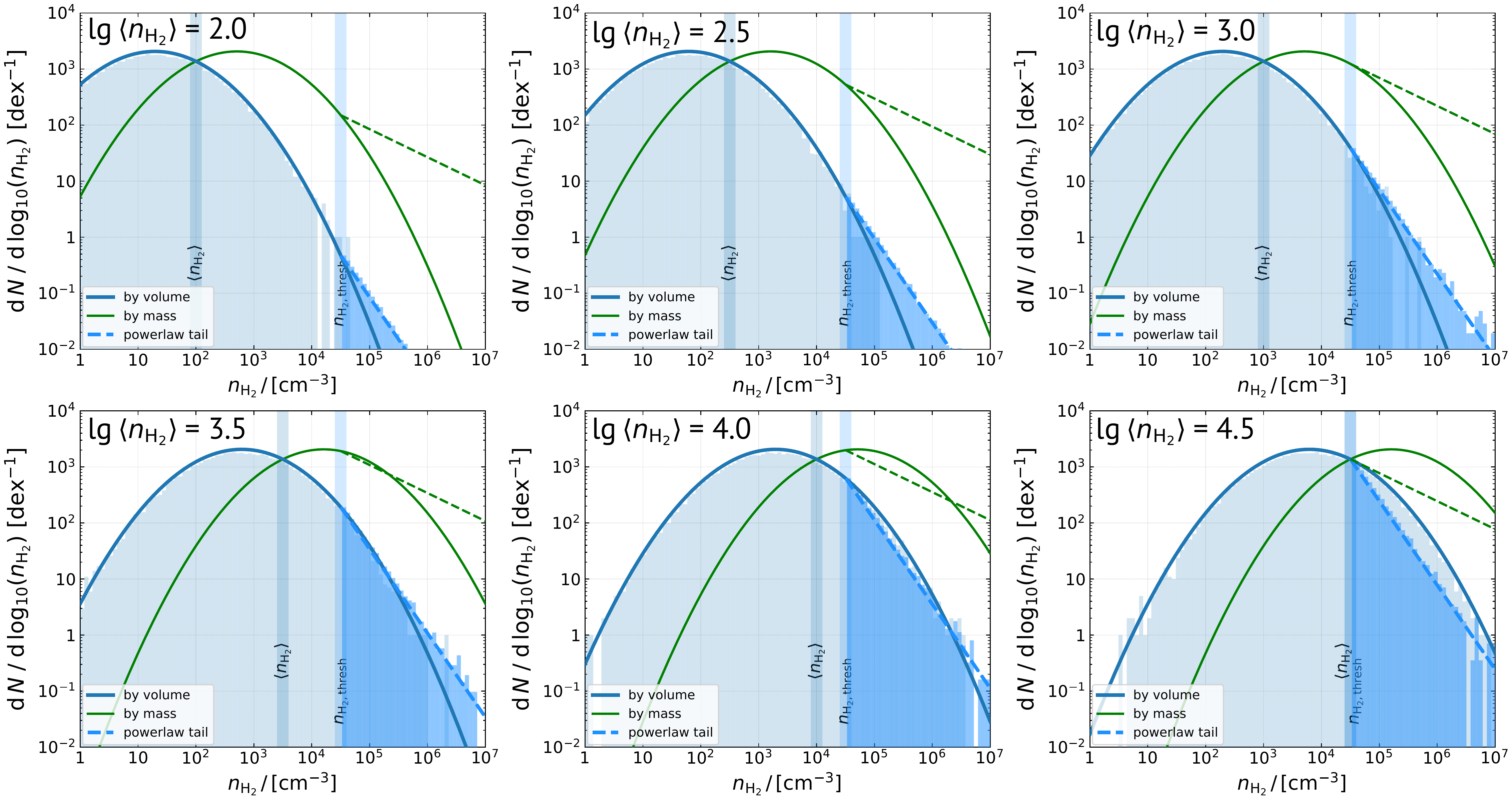}
	\vspace{0.5ex}
	\caption{%
		Example of composite gas density PDFs in our modeling with varied log-normal PDF's mean gas density $\lgnmean$ (from 2.0 to 4.5 in panels from left to right and top to bottom) and a fixed power-law tail threshold gas density $\lgnthresh=4.5$. The $\lgnmean$ and $\lgnthresh=4.5$ are indicated by the vertical transparent bars and labels in each panel. The thick blue and thin green solid (dashed) lines represent the volume- and mass-weighted PDFs of the log-normal (power-law tail) gas component, respectively. 
		\label{Plot_generated_PDF}
	}
\end{figure*}

\subsection{Observational evidences of gas density PDF}
\label{Section_gas_density_PDF_observations}

Observation of gas density PDF at molecular cloud scale requires high angular resolution (e.g., sub-hundred-pc scales) and full spatial information, therefore could only be obtained either with sensitive single-dish mapping in the Galaxy and nearest large galaxies, or with sensitive interferometric plus total power observations. 
For external galaxies, the MAGMA survey by \citet{Pineda2009LMC}, \cite{Hughes2010LMC} and \citet{Wong2011LMC} mapped CO(1-0) in the LMC at 11~pc resolution with the Mopra~22m single-dish telescope. \cite{Gardan2007M33}, \cite{Gratier2010M33} and \cite{Druard2014M33} mapped M~33 CO(2-1) emission at 50~pc scale with the IRAM~30m single-dish telescope. The PAWS survey provides M~51 CO maps at 40~pc obtained with the IRAM PdBI and with IRAM~30m data (\citealt{Schinnerer2013,Schinnerer2017,Pety2013,Hughes2013a,Leroy2016}). The on-going PHANGS-ALMA survey\,\footnote{\url{http://phangs.org}} maps CO(2-1) at $\sim$60--100~pc scales in more than 70 nearby galaxies using ALMA with total power (Leroy et al., submitted; see also \citealt{SunJiayi2018,SunJiayi2020,Kreckel2018,Schinnerer2019,Chevance2020}). 
Meanwhile, higher physical resolution observations are also available for Galactic clouds and filaments, e.g., \cite{Kainulainen2013}, \cite{Lombardi2014,Lombardi2015}, \cite{Kainulainen2017}, \cite{ZhangMiaomiao2019}. 

These observations at large scales reveal a smooth gas density PDF which can be described by a log-normal distribution plus a high-density power-law tail (e.g., \citealt{Wong2011LMC,Hughes2013a,Druard2014M33}). The width of the log-normal PDF and the slope of the power-law tail do slightly vary among galaxies, but the most prominent difference is seen for the mean of the log-normal PDF (hereafter $\nmean$), which changes by more than one order of magnitude (for a relatively small sample of $<10$ spiral galaxies; see Fig.~7 of \citealt{Leroy2016}). 

Interestingly, such a log-normal PDF is consistently predicted by isothermal homogeneous supersonic turbulent theories or diverse cloud models (e.g., \citealt{Ostriker2001,VazquezSemadeni2001,Padoan2002,Padoan2004a,Padoan2004b,Padoan2011,Padoan2012}; \citealt{Tassis2010}; \citealt{Kritsuk2017}; \citealt{Raskutti2017}; see also references in \citealt{Raskutti2017}), and the additional power-law PDF is also expected, e.g., for a multi-phase ISM and/or due to the cloud evolution/star formation at late times (e.g., \citealt{Klessen2000a}; \citealt{Tassis2010}; \citealt{Kritsuk2017}; \citealt{Raskutti2017} and references therein). Therefore, modeling gas density PDFs assuming a log-normal distribution plus a power-law tail appears to be a very reasonable approach.

\subsection{Sub-beam gas density PDF modeling}
\label{Section_gas_density_modeling}

We thus assume that the line-of-sight volume density of molecular gas in a galaxy follows a log-normal PDF, with a small portion of line-of-sights following a power-law PDF at the high-density tail. 
Representative PDFs are shown in Fig.~\ref{Plot_generated_PDF}. 
Each PDF samples the $\nH2$ from $1$ to $10^{7}\;\mathrm{cm^{-3}}$ in 100 bins in logarithmic space. 
For each $\nH2$ bin, the height of the PDF is thus proportional to the number of sight lines with a density of $\nH2$. We assume that the CO line emission surface brightness from each line-of-sight can be computed from an equivalent ``one-zone'' cloud with a single $\nH2$, $\NH2_$, $\Tkin$, velocity gradient and CO abundance. Thus the total CO line emission surface brightness is the sum of all sight lines in the PDF. 

The shape of the gas density PDF is described by the following parameters: 
the mean gas density of the log-normal PDF $\nmean$, the threshold density of the power-law tail $\nthresh$, 
the width of the log-normal PDF, 
and the slope of the power-law tail. 
We model a series of PDFs by varying the $\nmean$ from $10^{2.0}$ to $10^{5.0}\;\mathrm{cm^{-3}}$ in steps of 0.25~dex, and $\nthresh$ from $10^{4.0}$ to $10^{5.25}\;\mathrm{cm^{-3}}$ in steps of 0.25~dex, to build our model grid which can cover most situations observed in galaxies. 
The slope of the power-law tail is fixed to $-1.5$, which is an intermediate value as indicated by simulations (\citealt{Federrath2013}), also previously adopted by \cite{Leroy2017}. 
The width of the log-normal PDF is physically characterized by the Mach number of the supersonic turbulent ISM (see \citealt{Padoan2002,Padoan2011} and Eq.~5 of \citealt{Leroy2017}): $\sigma \approx 0.43 \sqrt{\ln (1 + 0.25 \mathcal{M}^2)}$, which ranges typically within 4 to 20 in star-forming regions as shown by simulations (e.g., \citealt{Krumholz2005}; \citealt{Padoan2011}). Here we adopt a fiducial Mach number of 10, as done previously by \cite{Leroy2017}. 
Note that a high Mach number $\sim80$ is also found in merger systems and starburst galaxies (e.g., \citealt{Leroy2016}). It corresponds to a log-normal PDF width $1.56\times$ our fiducial value, and marginally affects the CO excitation in a similar way as a higher $\nmean$. Thus, for simplicity in this work we fix the Mach number and allow $\nmean$ to vary.

\begin{figure*}[htb]
\centering%
\includegraphics[width=0.95\textwidth]{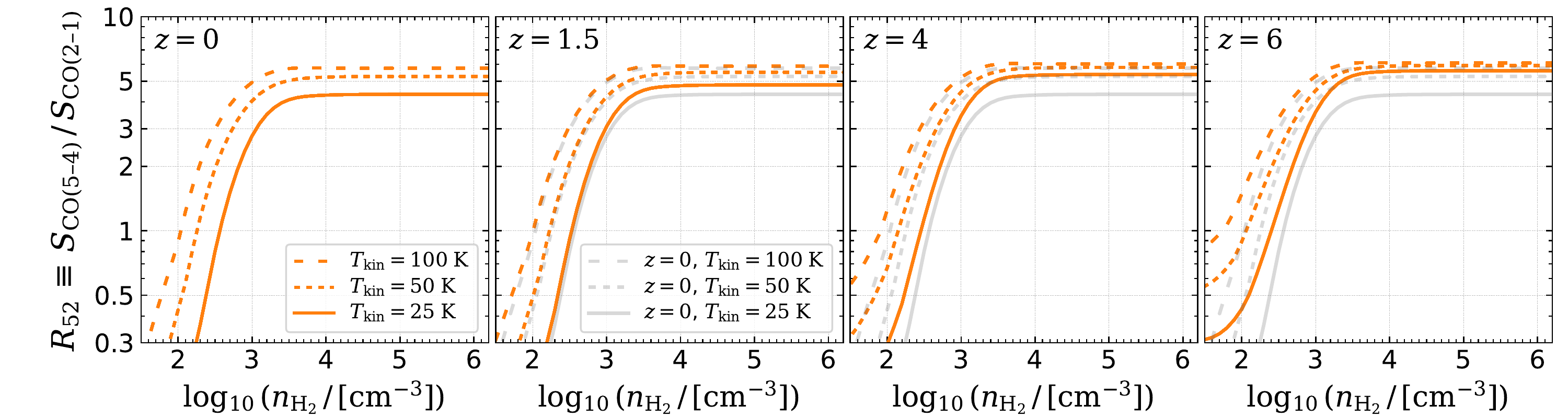}
\vspace{1.5ex}
\caption{%
CO(5-4)/CO(2-1) line ratio ($\R52$) from single one-zone LVG calculation. The four panels show the calculations at four representative redshifts $z=0$, 1.5, 4 and 6, from left to right, respectively. Solid, dashed and long-dashed lines are for gas kinetic temperature $\Tkin=25$, $50$ and $100\;\mathrm{K}$, respectively. 
The grey lines in the second, third and fourth panels are the corresponding $z=0$ lines. 
\label{Plot_single_LVG_R52}
}
\end{figure*}

\begin{figure*}[htb]
\centering%
\includegraphics[width=0.95\textwidth, trim=0 0 0 1.8cm]{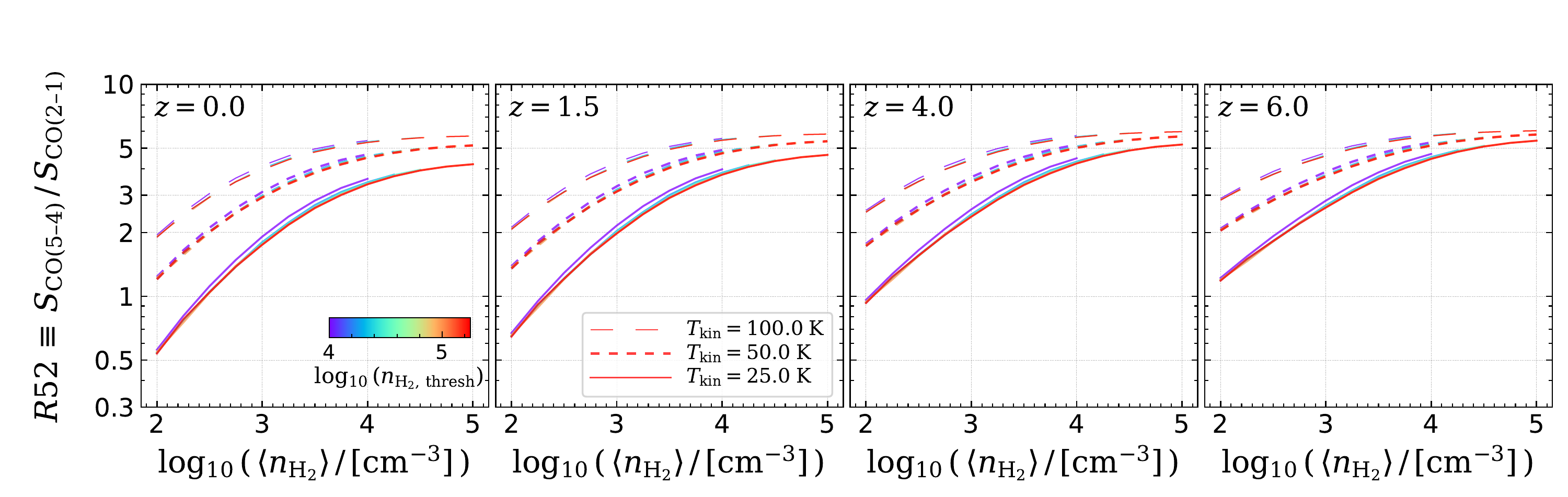}
\vspace{0.5ex}
\caption{%
$\R52$ as functions of the mean gas density ($\lgnmean$) as predicted from our composite gas modeling. The four panels show the models at four different representative redshifts. In each panel, color indicates the threshold density of the power-law tail ($\lgnthresh$; which alters the line ratio only slightly), and three line styles are models at three representative kinetic temperatures ($\Tkin=25$, $50$ and $100\,\mathrm{K}$ for solid, dashed and long-dashed lines, respectively). 
\label{Plot_modeled_volume_density_nH2_R52}
}\vspace{2ex}
\end{figure*}

\subsection{One-zone gas cloud calculation}
\label{Section_gas_modeling_assumptions}

For a given gas density PDF, each $\nH2$ bin is composed of the same ``one-zone'' gas clouds for which we will compute the line surface brightness. A one-zone cloud has a single volume density $\nH2$, column density $\NH2_$, gas kinetic temperature ($\Tkin$), CO abundance $[\mathrm{CO}/\mathrm{H_2}]$ and velocity gradient $\dvddr$. Note that although an equivalent cloud size $r$ is implied from the ratio of $\NH2_$ and $\nH2$, given that the calculation is in 1D, $r$ should not be taken as a physical cloud size. 
Also note that in our study we do not model the 3D distribution of one-zone models, therefore any radiative coupling between one-zone models along the same line of sight can not be accounted for. This is likely a minor issue for star-forming disk galaxies given their thin disks (a few hundred pc; \citealt{Wilson2019}) and systematic rotation which separates molecular clouds in the velocity space for inclined disks, but the actual effects need to be studied by detailed numerical simulations (e.g., \citealt{Smith2020,Tress2020}). 

Here we use RADEX (\citealt{vanderTak2007}) to compute the 
1D non-LTE radiative transfer. For a given $\nH2$, we loop $\NH2_$ from $10^{21}$ to $10^{24}\,\mathrm{cm}^{-2}$, and $r$ is then determined by: 
\begin{equation}
N_{\mathrm{H_2}} = 2 \times r \times n_{\mathrm{H_2}} = 6 \times 10^{18} \times \frac{r}{\mathrm{pc}} \times \frac{n_{\mathrm{H_2}}}{\mathrm{cm^{-3}}} \ \ [\mathrm{cm^{-2}}]
\end{equation}

We also loop over $\Tkin$ values of 25, 50, and 100\,K, while we fix $[\mathrm{CO}/\mathrm{H_2}] = 5 \times 10^{-5}$, a reasonable guess for star-forming clouds (e.g., \citealt{Leung1976}; \citealt{vanDishoeck1987,vanDishoeck1988}; although it varies from cloud to cloud and depends on chemistry; e.g., \citealt{Sheffer2008}). 
Note that there is one additional free parameter to set, i.e., either the LVG velocity gradient $\dvddr$, or the line width FWHM $\Delta{V}$, or the velocity dispersion $\sigma_{V}$. They are related to each other by: 
\begin{equation}
\begin{split}
\Delta{V} = 2 \times r \times \dvddr \ \ [\mathrm{km\,s^{-1}}] \\
\sigma_{V} = \Delta{V} / (2 \sqrt{2 \ln 2}) \ \ [\mathrm{km\,s^{-1}}]
\end{split}
\end{equation}
To determine these quantities and effectively reduce the number of free parameters while being consistent with observations, we use an empirical correlation between $\NH2_$, $r$, $\sigma_{V}$ and the virial parameter $\alphavir$. $\alphavir$ describes the ratio of a cloud's kinetic energy and gravitational potential energy (e.g., \citealt{Bertoldi1992}), and can be written as $\frac{5 \sigma_{V}^2 r}{f G M}$, where 
$\sigma_{V}$ and $r$ are introduced above, 
$G$ is the gravitational constant, $M$ is the cloud mass, and $f$ is a factor to account for the lack of balance between kinetic and gravitational potential (see Eq.~6 of \citealt{SunJiayi2018}). 
Observations show that clouds are not always virialized, i.e., $\alphavir$ is not always unity. 
Based on $\sim60$~pc CO mapping of 11 galaxies in the PHANGS-ALMA sample, \citet{SunJiayi2018} reported the following correlation in their Eq.~13 (helium and other heavy elements are included; see also Eq.~2 in the review by \citealt{Heyer2015Review}): 
\begin{equation}
\begin{split}
\alpha_{\mathrm{vir}} 
&= 5.77 \times 
    \left(\frac{\sigma_{V}}{\mathrm{km\,s^{-1}}}\right)^2 
    \left(\frac{\Sigma_{\mathrm{H_2}}}{\mathrm{M_{\odot}\,pc^{-2}}}\right)^{-1}
    \left(\frac{r}{40\,\mathrm{pc}}\right)^{-1} \\
&= 5.77 \times 
    \left(\frac{\sigma_{V}}{\mathrm{km\,s^{-1}}}\right)^2 
    \left(\frac{N_{\mathrm{H_2}}}{1.55 \times 10^{20}\,\mathrm{cm^{-2}}}\right)^{-1}
    \left(\frac{r}{\mathrm{pc}}\right)^{-1} \\
\end{split}
\label{Equation_alphavir}
\end{equation}

They find $\alphavir\approx1.5\,\textnormal{--}\,3.0$ with a $1\sigma$ width of 0.4\,$\textnormal{--}$\,0.65\,dex. 
For simplicity and also with the idea of focusing primarily on the effect of gas density, we adopt a constant $\alphavir$ of 2.3. As shown in later sections, this is already sufficient to explain the observed CO line ratios/SLEDs by our modeling. But note that more comprehensive descriptions of $\alphavir$ can be achieved in simulations and can be compared with the results from this work to better understand how a changing $\alphavir$ could affect CO line ratio predictions. 

Fig.~\ref{Plot_single_LVG_R52} presents how $\R52$ changes with the gas densities of one-zone cloud models for four representative redshifts where the cosmic microwave background (CMB) temperatures are different. 
We repeat our calculations for three representative $\Tkin$ as labeled in each panel. The comparison shows that $\Tkin$ significantly affects the $\R52$ line ratio, especially at low densities and at low redshifts. 
Note that due to the constant $\alphavir$ assumption, for a given $\nH2$, Eq.~\ref{Equation_alphavir} implies that $\sigma_{V} \propto r$, and that $\dvddr$ is not varying with $\NH2_$. Thus the actual choices of $\NH2_$ (or $r$) for each single one-zone model will not affect the modeling of $\R52$ (and of the optical depth $\tau$).

In addition, our modeling is also able to produce reasonable line optical depths ($\tau$) and [C~\textsc{i}]/CO line ratios, as presented in Appendix~\ref{Section_Appendix_Gas_Modeling}.

\begin{figure*}[htb]
\centering%
\includegraphics[width=1.0\textwidth, trim=5mm 0 0 0]{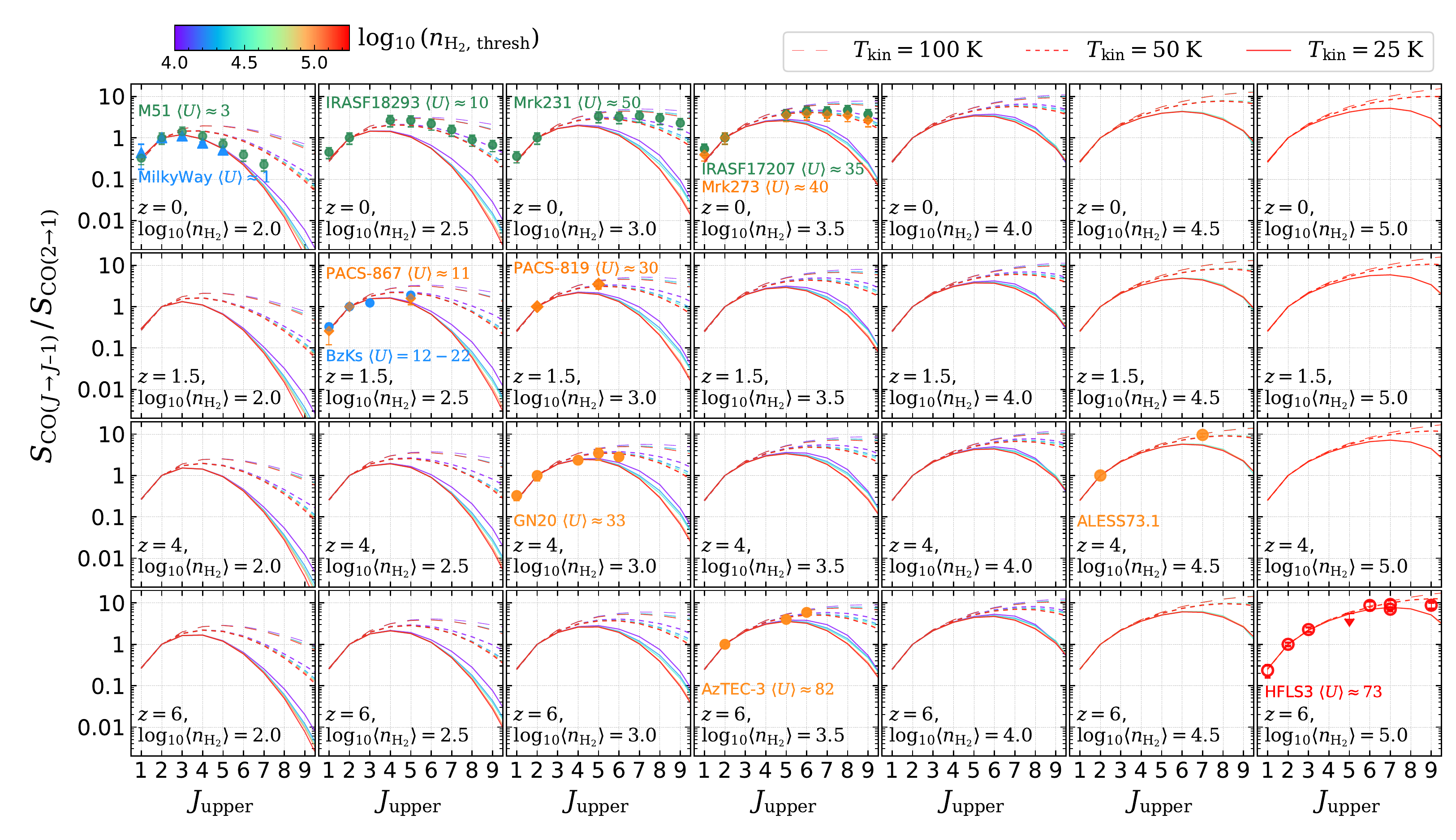}
\caption{%
Predicted CO SLEDs in Jansky units and normalized at CO(2-1). From top to bottom, CO SLEDs are at redshift $z=0$, $1.5$, $4$ and $6$, respectively. And from left to right, log-normal PDF's mean gas density $\lgnmean / \percmcubic$ changes from $2.0$ to $5.0$ in steps of 0.5. 
In each panel, solid, dashed and long-dashed lines represent $T_{\mathrm{kin}}=25$, $50$ and $100\,\mathrm{K}$ models, respectively. 
Line color coding indicates the threshold gas density of the power-law tail PDF, $\lgnthresh$. 
Colored data points are CO line fluxes in the following galaxies, with references in parentheses: 
the Milky Way (\citealt{Fixsen1999}), 
local spiral M51 (\citealt{Schirm2017}), 
local ULIRGs Mrk231, Mrk273, IRAS~F18293-3413 and IRAS~F17207-0014 (\citetalias{Liudz2015}; \citealt{Kamenetzky2014}), 
$z=1.5$ BzK galaxies (\citealt{Daddi2015}), 
$z=1.5$ starburst galaxies (\citealt{Silverman2015b} and this work), 
$z=4.055$ SMG GN20 (\citealt{Daddi2009GN20}; \citealt{Carilli2010GN20}; \citealt{Tan2014GN20}), 
$z=4.755$ SMG ALESS73.1 (\citealt{Coppin2010}; \citealt{ZhaoYinghe2020}), 
$z=5.3$ SMG AzTEC-3 (\citealt{Riechers2010AzTEC3}), 
and $z=6.3$ SMG HFLS3 (\citealt{Riechers2013HFLS3}). 
\label{Plot_modeled_line_ladder}
}

\vspace*{\floatsep}

\includegraphics[width=0.8\linewidth]{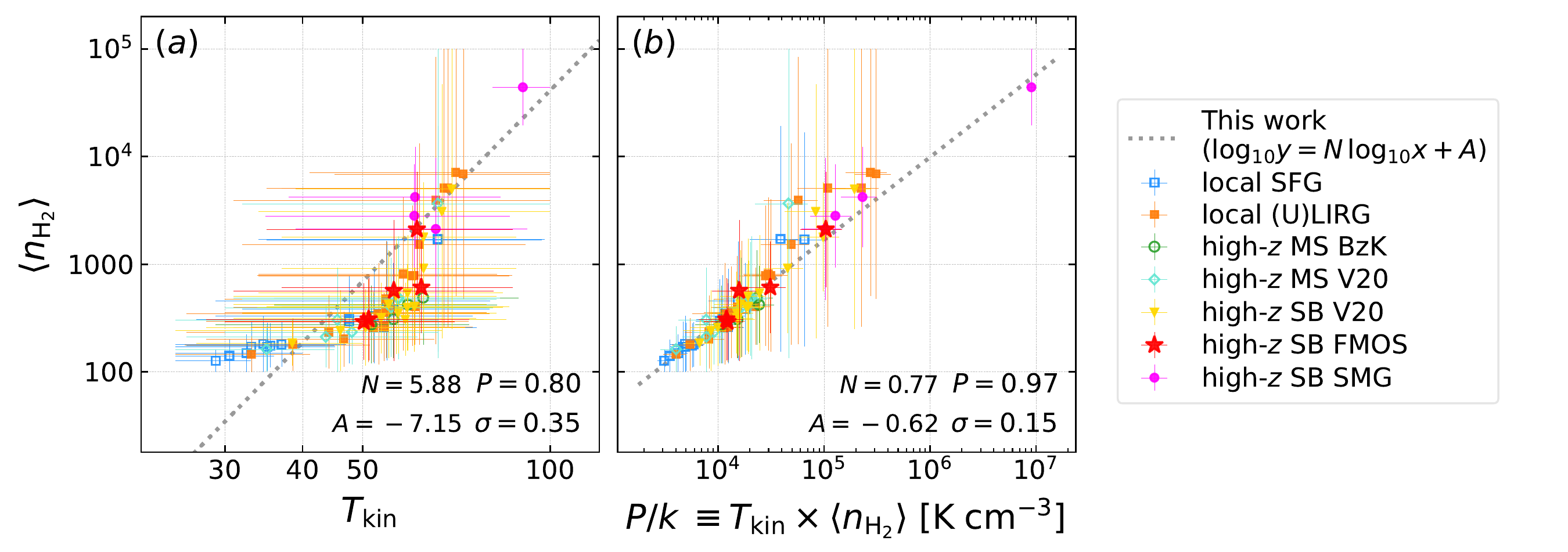}
\vspace{0.5ex}
\caption{%
Fitted mean gas density $\nmean$ versus fitted gas kinetic temperature $\Tkin$ (\textit{left panel}) and gas pressure $P/k$ (\textit{right panel}; $k$ is the Boltzmann constant) based on $\R52$ and its errors in our galaxy sample. This reflects the internal degeneracy between $\nmean$ and $\Tkin$ in our model grid. See fitting method in Sect.~\ref{Section_gas_modeling_composition}. 
\label{Plot_nH2_vs_Tkin}
}
\end{figure*}

\subsection{Converting $\R52$ to $\nmean$ and $\Tkin$ with the model grid}
\label{Section_gas_modeling_composition}

We compute the global line surface brightness by summing one-zone line surface brightnesses at each $\nH2$ bin according to the gas density PDF. 
With our assumptions, there are only four free parameters: the mean gas density of the log-normal PDF $\nmean$, the threshold density of the power-law tail $\nthresh$, gas kinetic temperature $\Tkin$, and redshift. Their grids are described in Sect.~\ref{Section_gas_density_modeling}. 

In Fig.~\ref{Plot_modeled_volume_density_nH2_R52}, we present the predicted $\R52$ as a function of the four free parameters. $\R52$ increases smoothly with $\nmean$ and $\Tkin$, while $\nthresh$ does not substantially alter the $\R52$ ratio, as indicated by the color coding. The minimum $\R52$ at the lowest density ($\lgnmean/\percmcubic \sim 2$) is nearly doubled from redshift 0 to 6 due to the increasing CMB temperature, but such a redshift effect is less prominent ($<\times1.5$) at both higher density ($\lgnmean/\percmcubic > 3$) and for higher $\Tkin$.

	In Fig.~\ref{Plot_modeled_line_ladder}, we further show the full CO SLEDs at $J_{\mathrm{u}}=1$ to $9$ from our model grid, and compare them with a subsample of galaxies with multiple CO transitions at various redshifts. These galaxies are displayed in panels where the $\nmean$ is closest to their $\R52$-derived $\nmean$ (see below). Our modeling can generally match these CO SLEDs given certain choices of $\nmean$ and $\Tkin$. Yet we caution that this is not a thorough comparison, and our model grid might not fit entirely well the CO SLED shape due to our simplifying assumptions of fixed Mach number and power-law tail slope or $\alphavir$. While this work only focuses on $\R52$ with the simplest assumptions, the model predictions seem overall already quite promising for the whole CO SLEDs, and can be further improved in future works.

	Based on the model grid, we describe below a method to determine the most probable $\nmean$, $\Tkin$ and $\nthresh$ ranges for a given $\R52$ and its error in galaxies with known redshift. This is done with a Monte Carlo approach. We first interpolate our 4D model grid to the exact redshift of each galaxy using \textsc{Python} \incode{scipy.interpolate.LinearNDInterpolator}, then resample the 3D model grid to a finer grid, perturb the $\R52$ given its error over a normal distribution for 300 realizations, and find the minimum $\chi^2$ best-fits for each realization. Finally, we combine best-fits to obtain posterior distributions of $\nmean$, $\Tkin$ and $\nthresh$, and determine their median, 16\% (L68) and 84\% (H68) percentiles. 
	This fitting method is coded in our \textsc{Python} package \incode{co-excitation-gas-modeling} that is made publicly available. 

	We note that although there is a single input observation ($\R52$) whereas there are three parameters to be determined ($\nmean$, $\Tkin$ and $\nthresh$), our method still produces reasonable results. In fact, our method is able to take into account the internal degeneracy between $\nmean$ and $\Tkin$ inside model grids, thus obtaining reasonable probability ranges. Fig.~\ref{Plot_nH2_vs_Tkin} shows the fitted $\nmean$ and $\Tkin$ for our galaxy sample, resulting in a non-linear trend between $\nmean$ and $\Tkin$. The galaxy-wide mean pressures of gas can also be calculated as $\Tkin \times \nmean$, and are found to agree with estimates in local galaxies \citep{Kamenetzky2014}. 

    In Figs.~\ref{Plot_nH2_vs_params} and \ref{Plot_Tkin_vs_params}, we present correlations between the $\R52$-fitted $\nmean$ and $\Tkin$, respectively, and various galaxy properties, similarly to what is presented in Fig.~\ref{Plot_R52_vs_params} for $\R52$. We discuss them in detail in the next sections (Sect.~\ref{Section_Umean_nH2}).

\vspace{0.5truecm}

\section{Results on ISM Physical Properties and Discussion}
\label{Section_Discussion}

\subsection{The underlying meaning of $\Umean$: a mass-to-light ratio for dust}
\label{Section_Discussion_Umean}

By definition, $\Umean$ is the mass-weighted ISRF intensity created by UV photons from stars in a galaxy. 
As indicated by the \citetalias{DL07} model and many of its applications, e.g., \citet{Draine2007SINGS,Draine2014}, \citet{Aniano2012,Aniano2020}, \citet{Dale2012,Dale2017}, \citet{Magdis2012SED,Magdis2017}, \citet{Ciesla2014} and \citet{Schreiber2018}, $\Umean$ is actually a mass-weighted, average mass-to-light ratio for the mixture of dust grains in a galaxy. It is driven by the young stars emitting most of the UV photons, but also reflects the mean distance between young stars and interstellar dust and the efficiency of UV photons heating the dust. 
For a given \citetalias{DL07} ISRF distribution power-law index ($=-2$) and $U_{\mathrm{max}}$ ($=10^{7}$; \citealt{Draine2014}), $\Umean$ is proportional to the ratio between $\LIR$ and $\Mdust$, with a coefficient $P_0 \approx 138$ from this work, where $P_0$ represents the power absorbed per unit dust mass in a radiation field $U=1$: 
\begin{equation}
\begin{split}
& L_{\mathrm{IR,\,8-1000{\mu}m}} = P_0 \cdot \Umean \cdot \Mdust \\[0.5ex]
& \textnormal{where} \ P_0 \approx 120 - 150 \ (\textnormal{mean} = 138)
\end{split}
\label{Equation_Umean_LIR_Mdust}
\end{equation}
Note that the $P_0$ factor is calibrated to be equal to 125 in \citet{Magdis2012SED} due to a slightly different $U_{\mathrm{max}}=10^{6}$, a small 10\% systematic difference. 

$\Umean$ is also positively linked to dust temperature, but it depends on how dust temperature is defined. For example, \cite{Draine2007SINGS} find that $T \approx 17 \cdot U^{1/6} \ \mathrm{[K]}$ for dust grains with sizes greater than 0.03\,$\mu$m whose blackbody radiation peaks around 160\,$\mu$m. \cite{Schreiber2018} calibrate the light-weighted dust temperature $T_{\mathrm{dust}}^{\mathrm{light}} = 20.0 \cdot U^{1/5.57} \ \mathrm{[K]}$ (and mass-weighted $T_{\mathrm{dust}}^{\mathrm{mass}} = 0.91 \cdot T_{\mathrm{dust}}^{\mathrm{light}}$) by fitting Wien's law to each elementary \cite{Galliano2011} template. 

Studies of $\Tdust$ and $\Umean$ have shown that dust (ISRF) is warmer (stronger) for increasing IR luminosity from local SFGs to (U)LIRGs (e.g., \citealt{Hwang2010}; \citealt{Symeonidis2013}; \citealt{HerreroIllana2019}), and increases with redshift for the majority of MS galaxies (e.g., \citealt{Magdis2012SED}; \citealt{Magnelli2014}; \citealt{Bethermin2015}; \citealt{Schreiber2018}). 
Some observations show colder dust temperatures in a few among the most extreme starburst systems (e.g., \citealt{Lisenfeld2000}; \citealt{Jin2019}; \citealt{Cortzen2020}). These are likely due to the presence of high dust opacity at shorter wavelengths which makes the dust SED apparently colder. Observations of SMGs also show colder dust temperatures in some of the less luminous ones. This phenomenon is likely driven by the fact that (sub-)mm selection favors cold-dust galaxies whose SEDs peak closer to (sub-)mm wavelengths (e.g., \citealt{Chapman2005}; 
\citealt{Kovacs2006}; \citealt{Symeonidis2009,Symeonidis2011}; \citealt{Magnelli2010}; \citealt{Hwang2010}; \citealt{Magdis2010}). 

There is also an interesting finding that for extreme SB galaxies with $\SFR/\SFR_{\mathrm{MS}}>4$, their $\Umean$ seem to not evolve with redshift (e.g., \citealt{Bethermin2015}). While $\Umean$ in MS galaxies does evolve with redshift, and extrapolation suggests that $\Umean$ in MS galaxies might become stronger than those in extreme SB galaxies at $z>2.5$, which seems at odds with the expectation. Yet this finding might also be limited by sample size and selection method, templates used for SED fitting, as well as the dust optically thin assumption in \citetalias{DL07} templates (e.g., \citealt{Jin2019,Cortzen2020}). 

Combining with the results from this work, the $\Tdust$ or $\Umean$ trends are easier to understand when correlating them with molecular gas mean density and temperature. We propose a picture in which the general increase of dust temperature and ISRF is mainly due to the increase in cold molecular gas temperature, either due to higher CMB temperature at higher redshifts or more intense star formation and feedback. While the mean molecular gas density has a weaker, non-linear trend driving $\Umean$ in most galaxies, merger-driven compaction could strongly increase gas density hence boost $\nmean$, $\Tkin$ and $\Umean$ in a small number of SB galaxies. Such an increase in gas density creates more contrast at lower redshifts due to the general decrease of the cosmic molecular gas density and CMB temperature. This could explain why $\Umean$ is more different between MS and SB galaxies at lower redshifts.

\subsection{Density or Temperature Regulated Star Formation? The $\Umean$--$\nmean$ and $\Umean$--$\Tkin$ Correlations}
\label{Section_Umean_nH2}

\begin{figure*}[htb]
\includegraphics[width=\textwidth]{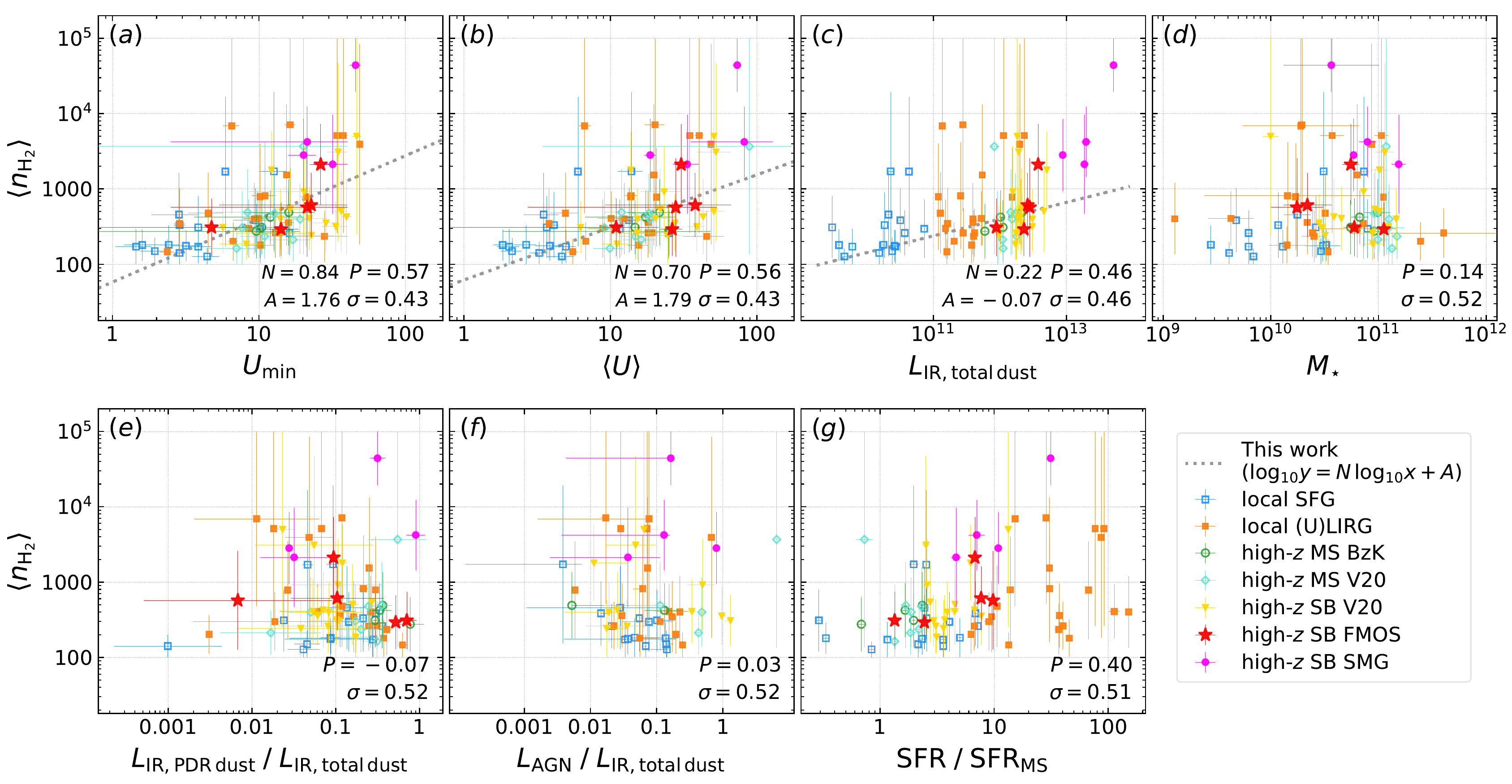}
\vspace{-1ex}
\caption{%
Fitted $\nH2$ versus galaxy properties same as in Fig.~\ref{Plot_R52_vs_params}. Data points' $\nH2$ and errorbars are the median and 1-sigma ranges of the fitting using our model grid as presented in Sect.~\ref{Section_gas_modeling} to the observed $\R52$. 
\label{Plot_nH2_vs_params}
}

\vspace*{\floatsep}

\includegraphics[width=\textwidth]{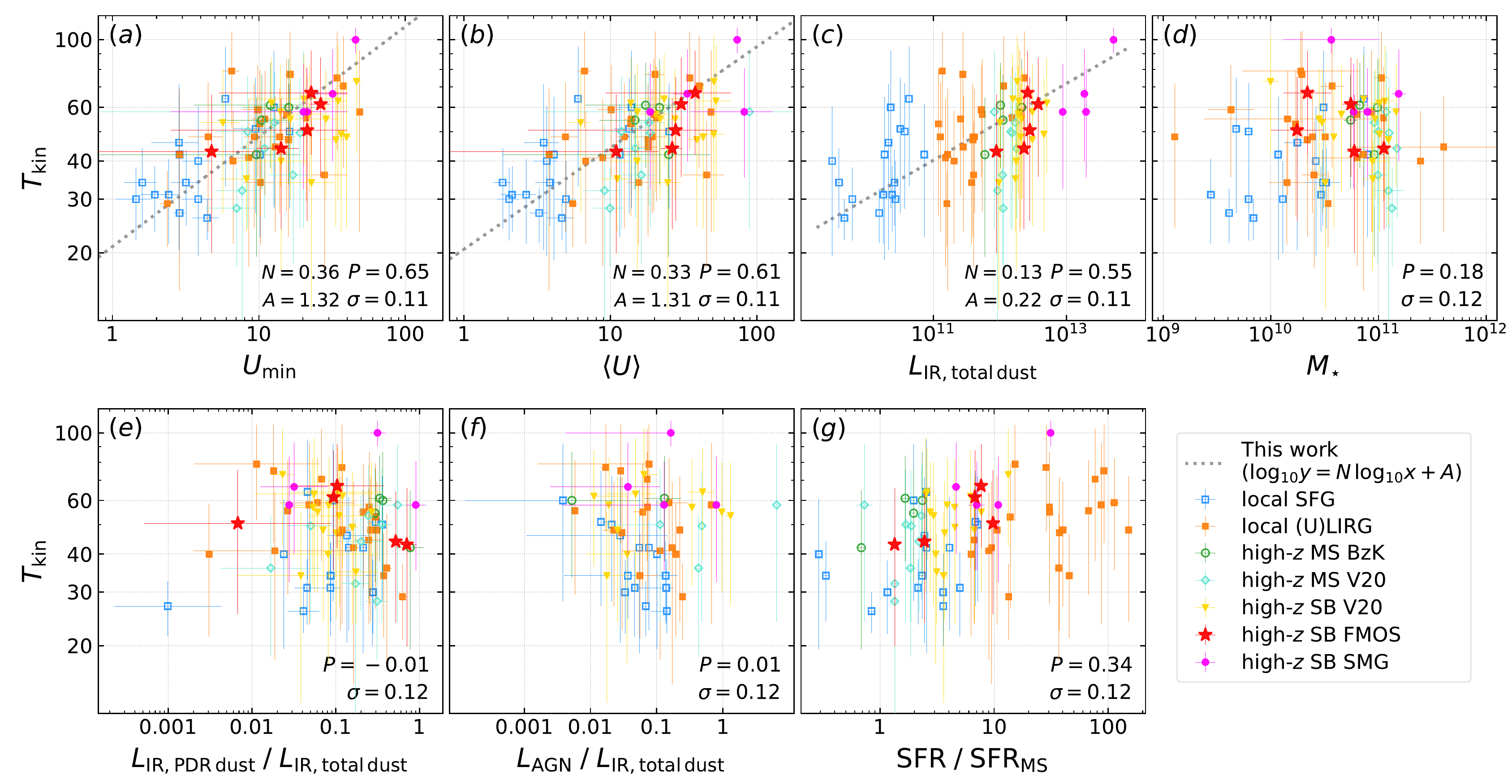}
\vspace{-1ex}
\caption{%
Fitted $\Tkin$ versus galaxy properties same as in Fig.~\ref{Plot_R52_vs_params}. $\Tkin$ is shown as median and errorbars representing the 1-sigma range of our model grid fitting to the observed $\R52$ as presented in Sect.~\ref{Section_gas_modeling}. 
\label{Plot_Tkin_vs_params}
}

\end{figure*}

Figs.~\ref{Plot_nH2_vs_params} and \ref{Plot_Tkin_vs_params} show that both $\nmean$ and $\Tkin$ positively correlate with $U$ and $\LIR$ but not with other properties like stellar mass or AGN fraction in our sample. 
Yet $\nmean$ correlates with $U$ or $\LIR$ in a non-linear way. Except for high-$z$ SMGs and a few local galaxies with large error bars coming from their large $\R52$ uncertainties, most galaxies are constrained within a narrow range of $\nmean \sim 10^{2}-10^{3} \; \percmcubic$. 
Despite the large scatter in the data, we observe a trend with $\nH2 \propto \Umean^{0.70}$ which seems to hold only within the intermediate $\Umean$ range ($\Umean \sim 5 - 20$). 

Meanwhile, $\Tkin$ has a tighter correlation ($\sigma\sim0.11$) with $U$ and $\LIR$. We find relations $\Tkin \propto \Umean^{0.33}$ and $\Tkin \propto \LIR^{0.13}$. 
Note that by calculating the [C\textsc{i}] $^3P_2$--$^3P_1$ and $^3P_1$--$^3P_0$ excitation temperatures as a probe of gas kinetic temperature under thermalized condition, \cite{Jiao2017,Jiao2019} and \cite{Valentino2020ApJ...890...24V} also found positive correlation between the gas kinetic temperature and dust temperature which is proportional to $\Umean^{0.16}$.
There is also a weak trend that $\Tkin$ increases with $\SFR/\SFR_{\mathrm{MS}}$ (Pearson correlation coefficient $P=0.34$), and the trend between $\nH2$ and $\SFR/\SFR_{\mathrm{MS}}$ is also marginal ($P=0.40$).

Given these results, it is very reasonable that both mean gas density and temperature increase from less to more intensively star-forming galaxies. 
Yet based on the datasets in this work, it is difficult to statistically decouple $\nmean$ and $\Tkin$ and hence to measure well the shapes of $\Umean$--$\nmean$ and $\Umean$--$\Tkin$ correlations. 
However, the non-linear or broken $\Umean$--$\nmean$ correlation and the more smooth $\Umean$--$\Tkin$ might imply two scenarios, one for ``normal'' star-forming galaxies, and one for merger-driven starbursts.
``Normal'' galaxies may have a smooth density- and temperature-regulated star formation, whereas strong gas compression in major merger events can induce extraordinarily high $\nmean$ with moderate $\Tkin$ and $\Umean$
(e.g., \citealt{Tacconi2008,Engel2010,Riechers2010AzTEC3,Cooray2014HFLS3,Larson2016,Calabro2019}). 
Further insights will require higher quality, multiple transition CO SLEDs, as we briefly discuss below (Sect.~\ref{Section_Limitations_and_outlook}).

\subsection{Implication for star formation law slopes}
\label{Section_Discussion_SF_Law}

The star formation (SF) law is known as the correlation between gas mass (or surface density) and star formation rate (or surface density), and can be expressed as:
\begin{equation}
\begin{split}
\SFR &= A \cdot \MH2_^{N} \quad \textnormal{or} \\[0.5ex]
\Sigma_{\SFR} &= A \cdot \Sigma_{\mathrm{gas}}^{N}
\end{split}
\end{equation}
where $A$ is the normalization and $N$ is the slope. 
After the initial idea presented by \cite{Schmidt1959}, \cite{Kennicutt1998ApJ} first systematically measured the SF law to be $\Sigma_{\SFR} \propto \Sigma_{\mathrm{gas}}^{1.4\pm0.15}$ based on observations of nearby spiral and starburst galaxies, where $\Sigma_{\mathrm{gas}}$ is the mass surface density of atomic plus molecular gas, and $\Sigma_{\SFR}$ is the SFR surface density traced by $\mathrm{H}\alpha$ and/or $\LIR$. 
This Kennicutt-Schmidt law with $N \approx 1.4$ has been extensively studied in galaxies with $\Sigma_{\mathrm{gas}}\sim1-10^{5}\;\Msun\,\mathrm{pc}^{-2}$ and is widely used in numerical simulations (see reviews by \citealt{Kennicutt2012Review}; \citealt{Carilli2013Review}). 

However, the actual slope $N$ of the SF law has been long debated. High-resolution (sub-kpc scale) observations in nearby spiral galaxies revealed that atomic gas does not correlate with SFR, whereas only molecular gas traces SFR, and $N$ is close to unity in these galaxies (e.g., \citealt{Wong2002,Leroy2008,Leroy2013,Bigiel2008,Schruba2011}). 
Meanwhile, from local SFGs to (U)LIRGs, observations suggest that $N$ is super-linear, ranging from $\sim 1$ to $\sim 2$ (e.g., \citealt{Kennicutt1998ApJ}; \citealt{Yao2003}; \citealt{Gao2004}; \citealt{Shetty2013,Shetty2014a,Shetty2014b}; \citealt{delosReyes2019}; \citealt{Wilson2019}). 
Furthermore, \cite{Daddi2010SFL} and \cite{Genzel2010} found that high redshift MS and SB galaxies follow two parallel sequences in the SF law ($M_{\mathrm{H_2}}$--$\SFR$) diagram, each with substantial breadth, and both with $N\sim1.1-1.2$ but with a 0.6~dex mean offset in normalization. 
Thus, why local SFG regions show a linear SF law, while high-$z$ SB galaxies have a much higher $\mathrm{SFE} \equiv \SFR / M_{\mathrm{H_2}}$ is still to be understood.

Here we decompose the SF law into $\Umean$ and $\nmean$ to gain some insights. First, it is known that the dust obscured SFR can be traced by the IR luminosity (e.g., \citealt{Kennicutt1998ApJ,Kennicutt2012Review}) as
$\SFR = \LIR / C_{\mathrm{IR}}$, where $C_{\mathrm{IR}} \sim 10^{10} \; [\Lsun \, (\Msun \mathrm{yr}^{-1})^{-1}]$ assuming a \citet{Chabrier2003} IMF. 
Second, as mentioned in the previous section, 
$ \Umean = P_0^{-1} \cdot \LIR / \Mdust $.
Third, we use the gas-to-dust ratio $\deltaGDR \equiv \Mgas / \Mdust$ to link gas to dust mass. This ratio varies with metallicity (e.g., \citealt{Israel1997}; \citealt{Leroy2007,Leroy2011}; \citealt{Sandstrom2013}; \citealt{Bolatto2013Review}; \citealt{RemyRuyer2014,RemyRuyer2015}), and also note that the definition of gas in $\deltaGDR$ is atomic plus molecular gas. We include an additional molecular hydrogen fraction $\fH2_ \equiv \MH2_ / \Mgas$ to the gas-to-dust ratio, having finally $\Mdust = \MH2_ \cdot (\fH2_ \deltaGDR)^{-1}$. 
Fourth, we consider MS galaxies to be disks with radius $r$ and height $h$, and assume that $\nmean$ is the global mean gas density, thus the molecular gas mass can be expressed as the product of the volume and the mean molecular gas density: 
$ \MH2_ = \pi \cdot \nmean \cdot r^2 \cdot h $.
And fifth, we ignore atomic gas and only considers molecular gas SF law. 

Then, we rewrite the SF law equation as:
\begin{equation}
\begin{split}
& \SFR = A \cdot \MH2_^{N} \\[0.5ex]
& \implies \Umean \cdot \Mdust = A \cdot P_0^{-1} \cdot C_{\mathrm{IR}} \cdot \MH2_^{N} \\[0.5ex]
& \implies \Umean \cdot (\fH2_ \, \deltaGDR)^{-1} = A \cdot P_0^{-1} \cdot C_{\mathrm{IR}} \cdot \MH2_^{N-1} \\[0.5ex]
& \implies \Umean \cdot (\fH2_ \, \deltaGDR)^{-1} = \\
& \qquad \qquad \qquad  A \cdot P_0^{-1} \cdot C_{\mathrm{IR}} \cdot (\pi \cdot \nmean \cdot r^2 \cdot h)^{N-1} \\[0.5ex]
\end{split}
\label{Equation_SFLaw_mol_1}
\end{equation}
Taking the logarithm of both sides, and assuming that $ \mathrm{log} \Umean $, $ \mathrm{log} (\fH2_ \deltaGDR) $ and $ \mathrm{log} (r^2 h) $ are functions of $ \mathrm{log} \nmean $, we have:
\begin{equation}
\begin{split}
& {\log \Umean} - {\log (\fH2_ \deltaGDR)} = \\[0.3ex]
& \quad \log (A \, P_0^{-1} \, C_{\mathrm{IR}}) \, + \, (N-1) \left[ \log \nmean + \log (\pi r^2 h) \right] \\[0.6ex]
& \implies N = \frac{ %
	\frac{\mathrm{d} \log \Umean}{\mathrm{d} \log \nmean} - %
	\frac{\mathrm{d} \log (\fH2_ \deltaGDR)}{\mathrm{d} \log \nmean} %
}{ %
	1 + %
	\frac{\mathrm{d} \log (r^2 h)}{\mathrm{d} \log \nmean} %
} + 1 \\
\end{split}
\label{Equation_SFLaw_mol}
\end{equation}

Therefore, the SF law slope $N$ depends on how $\Umean$, $\fH2_ \deltaGDR$ (metallicity) and $r^2 h$ (galaxy size) change with $\nmean$, which can further be described by the differentials $\frac{\mathrm{d} \log \Umean}{\mathrm{d} \log \nmean}$, $\frac{\mathrm{d} \log (\fH2_ \deltaGDR)}{\mathrm{d} \log \nmean}$ and $\frac{\mathrm{d} \log (r^2 h)}{\mathrm{d} \log \nmean}$, respectively. These differentials strongly depend on galaxy samples. When studying sub-kpc regions in local SFGs, if the ISRF, metallicity and galaxy size are similar among these SFGs, $N$ is close to 1. While when studying a sample including both SFGs and (U)LIRGs, $\Umean$ increases by a factor of a few tens with $\nmean$ changing from $10^{2}$ to $10^{4}\;\percmcubic$, and $r$ decreases by a factor of a few with $\nmean$ as (U)LIRGs are usually smaller and more compact (while the scale height $h$ seems constant, e.g., \citealt{Wilson2019}). As for the $\fH2_ \deltaGDR$ term, because $\fH2_$ increases with metallicity while $\deltaGDR$ decreases with it, their product $\fH2_ \deltaGDR$ likely does not change much. Therefore, $N$ can be much higher than 1. The overall effect is that the SF law does not have a single slope, yet the overall $N$ is about 1--2.

\subsection{Limitations and outlook}
\label{Section_Limitations_and_outlook}

We discuss three limitations of this work: the overall quality of current datasets, the assumptions in the gas modeling, and the contamination from AGN. First, CO line ratio or SLED studies require two or more CO line observations. These observations have different observing conditions, beam sizes, flux calibrations, etc., thus uncertainties are very likely underestimated even when the S/N of the line measurements are formally large (e.g., $>3$). For example, for our local SFG subsample, CO(5-4) data are from the \textit{Herschel} FTS with a certain beam size of $\sim40''$, which does not match the mapping area of CO(2-1) from ground-based telescope. The correction from the FTS beam to the entire galaxy can have a factor of two difference%
, which is reflected in the scatter of our data points although not fully reflected in their errorbars. The absolute flux calibration uncertainty of the observations in the literature can also be as high as $\sim30\%$, which is much poorer than current IRAM 30m and ALMA (total power) observations ($<10\%$). This also increases the scatter in our plots and necessarily makes observed correlations less significant. As for high-redshift galaxies, we use a $\SNR$ of 3 in both two CO lines to select our sample, which usually only reflects the quality of line measurements while it does not include the absolute flux calibration uncertainty. Their dust SEDs are also much more poorly covered, thus their $\Umean$ have fairly large uncertainties. Future ALMA Band 3 to 8 mapping of CO lines from $J_{\mathrm{u}}=1$ to $4$ in local galaxies, and VLA plus ALMA observations for suitable galaxies at high redshift with high-quality CO and continuum data will be the key to both spatially understand and statistically verify correlations between $\Umean$, $\nmean$ and $\Tkin$, as well as to unveil any evolutionary trend with redshift. 

Second, our assumptions in the gas modeling are also simplistic, in order to reflect only the effects of density and temperature on CO excitation. The constant $\alphavir$ assumption does not reflect the real situation in galaxies, e.g., as shown in \cite{SunJiayi2018}. Doubling the $\alphavir$ value from what we use in this work will result in a 20\% lower $\R52$ at $\log_{10} (\nmean/\mathrm{cm^{-3}}) = 3$, $\Tkin = 25\,\mathrm{K}$ and $z=0$. The constant $\Tkin$ assumption for all one-zone clouds in a galaxy is also a simplified ``toy model''-like condition. Adopting more realistic assumptions from observations (e.g., \citealt{SunJiayi2020}) or from hydrodynamic+chemistry simulations (e.g., \citealt{Smith2014a,Smith2014b,Smith2016,Smith2020}; \citealt{Tress2020}) in our gas modeling would naturally be the next step.

Third, it is known that some galaxies host AGNs which significantly contribute to optical or mid-IR SEDs as well as affect the CO excitation. Our SED fitting has already included a mid-IR AGN component that can dominate rest-frame $5-50\,\mu\mathrm{m}$ emission. This substantially improves the fitting $\chi^2$ for a number of galaxies showing mid-IR power-law SED feature, which, however, also brings in larger uncertainties in $\Umean$ as reflected in the errorbars in our plots. The used AGN SED templates could also slightly affect our results, although this effect should be well captured by the quoted uncertainties. Additional mid-IR photometry from future space telescopes like the James Webb Space Telescope (JWST) and the Origins Space Telescope (OST) will be key to solve this degeneracy and provide accurate AGN/ISRF decomposition. 
Meanwhile, an AGN can also boost highly-excited CO lines within X-ray dominated regions (XDRs) as shown by $J_{\mathrm{u}} \gtrsim 9$ CO studies (e.g., \citealt{vanderWerf2010,Rangwala2011,HaileyDunsheath2012,Spinoglio2012,Meijerink2013,PereiraSantaella2013,Rosenberg2014}). Decomposition of such AGN-dominated CO SLEDs usually requires three components, but the XDR component starts to dominate the CO SLED only at $J_{\mathrm{u}} \gtrsim 9$. Thus for this work, at CO(5-4) AGN likely contributes less than 10\% (e.g., see Fig.~2 of \citealt{vanderWerf2010}).

\vspace{0.5truecm}

\section{Summary}
\label{Section_Summary}

In this work, we compiled a comprehensive sample of galaxies from local to high redshift with CO(2-1) and CO(5-4) detections and well-sampled IR SEDs. 
This includes our new IRAM PdBI CO(5-4) observations of six $z\sim1.5$ COSMOS starburst galaxies. 
With this large sample, we measure their mean ISRF intensity $\Umean$ from dust SED fitting (Sect.~\ref{Section_SED_Fitting}), and their mean molecular gas density $\nmean$ converted from $\R52=S_{\mathrm{CO(5\textnormal{-}4)}}/S_{\mathrm{CO(2\textnormal{-}1)}}$ line ratios based on our density-PDF gas modeling (Sect.~\ref{Section_gas_modeling}). 
Our results can be summarised as following. 

\begin{itemize}
\item %
We confirm the tight $\Umean$--$\R52$ correlation first reported by \cite{Daddi2015}, and find that $\Umean$, $U_{\mathrm{min}}$ and $\LIR$ all strongly correlate with $\R52$, while stellar mass, AGN fraction, and the SFR offset to the MS all show weaker or no correlation with $\R52$ (Fig.~\ref{Plot_R52_vs_params}). 

\item %
We conduct density-PDF gas modeling to connect the mean molecular gas density $\nmean$ and kinetic temperature $\Tkin$ to the observable CO line ratio $\R52$. Based on which, we provide a Monte Carlo method (and a \textsc{Python} package \incode{co-excitation-gas-modeling}) to compute $\nmean$ and $\Tkin$'s probability ranges using our model grid for any given $J_{\mathrm{u}}=1-10$ CO line ratio (and for CO SLED as the next step; see, e.g., Fig.~\ref{Plot_modeled_line_ladder}). 

\item %
We find that both $\nmean$ and $\Tkin$ increase with $\Umean$, with $\Tkin$ having a tighter correlation with $\Umean$.

\item %
Based on these correlations, we propose a scenario in which the ISRF in the majority of galaxies is more directly regulated by the gas temperature and non-linearly by the gas density. A fraction of SB galaxies have gas densities larger by more than one order of magnitude with respect to MS galaxies and are possibly in a merger-driven compaction stage (Sects.~\ref{Section_Umean_nH2}~and~\ref{Section_Discussion_Umean}). 

\item %
We link the $\Umean$--$\nmean$ correlation to the Kennicutt-Schmidt SF law, and discuss how the SF law slope $N$ can be inferred from the $\Umean$--$\nmean$ correlation slope and other galaxy properties versus $\nmean$ correlations. We find that $N\sim1-2$ can be inferred from the trends of how $\Umean$ and galaxy size change with $\nmean$ in different galaxy samples (Sects.~\ref{Section_Discussion_SF_Law}).

\end{itemize}

Our study demonstrates that ISRF and molecular gas are tightly linked to each other, and density-PDF gas modeling is a promising tool for probing detailed ISM physical quantities, i.e., molecular gas density and temperature, from observables like CO line ratios/SLEDs.

\vspace{8ex}

\noindent%
\textit{%
\textbf{Data availability: }
Our \textsl{\michi2{}} SED fitting code is publicly available at \url{https://ascl.net/code/v/2533}. 
Our \textsc{Python} package \incode{co-excitation-gas-modeling} for computing $\nmean$ and $\Tkin$ from CO line ratios is publicly available at: \url{https://pypi.org/project/co-excitation-gas-modeling}. 
And our SED fitting figures as shown in Fig.~\ref{Figure_SED_fitting} and full Table~\ref{Table_All_in_one} are publicly available at: \url{https://doi.org/10.5281/zenodo.3958271}. 
}

\acknowledgments

We thank the anonymous referee for helpful comments. 
DL, ES and TS acknowledge funding from the European Research Council (ERC) under the European Union's Horizon 2020 research and innovation programme (grant agreement No. 694343).
GEM acknowledges the Villum Fonden research grant 13160 ``Gas to stars, stars to dust: tracing star formation across cosmic time'' and the Cosmic Dawn Center of Excellence funded by the Danish National Research Foundation under then grant No. 140.
YG's research is supported by National Key Basic Research and Development Program of China (grant No. 2017YFA0402700), National Natural Science Foundation of China (grant Nos. 11861131007, 11420101002), and Chinese Academy of Sciences Key Research Program of Frontier Sciences (grant No. QYZDJSSW-SLH008). 
SJ acknowledges financial support from the Spanish Ministry of Science, Innovation and Universities (MICIU) under grant AYA2017-84061-P, co-financed by FEDER (European Regional Development Funds). 
AP gratefully acknowledges financial support from STFC through grants ST/T000244/1 and ST/P000541/1. 
We thank A. Weiss and C. Wilson for helpful discussions. 
This work used observations carried out under project number W14DS with the IRAM Plateau de Bure Interferometer (PdBI). IRAM is supported by INSU/CNRS (France), MPG (Germany) and IGN (Spain). 
This work used observations carried out under project 17A-233 with the National Radio Astronomy Observatory's Karl G. Jansky Very Large Array (VLA). The National Radio Astronomy Observatory is a facility of the National Science Foundation operated under cooperative agreement by Associated Universities, Inc.

\clearpage

\appendix

\section{IRAM PdBI CO Observations of $z\sim1.5$ FMOS COSMOS Galaxies}
\label{Section_Appendix_PdBI_Observation}

\begin{figure*}[htb]
\centering
\includegraphics[width=\textwidth]{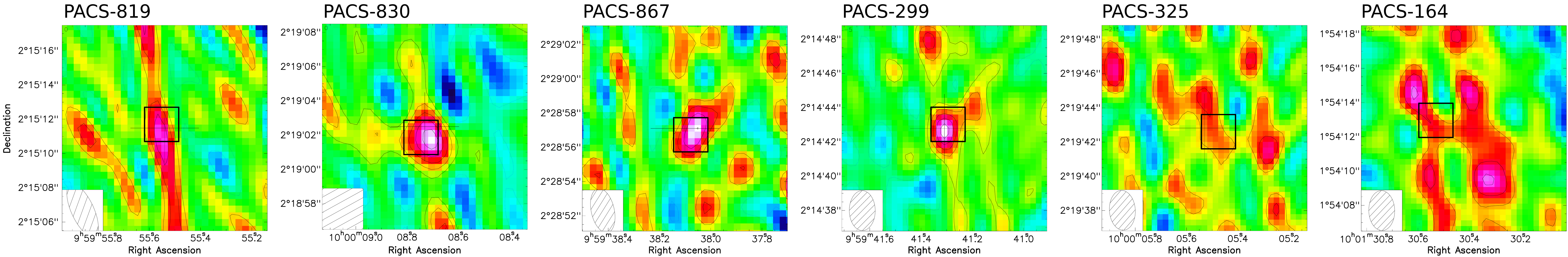}
\vspace{2ex}
\caption{%
CO(5-4) line maps for PACS-819, 830, 867, 299, 325 and 164, respectively. In the last two panels, PACS-325 and 164 are undetected. The field of view is $12''\times12''$ in all panels. Contours have a spacing of $1\,\sigma$ noise in each panel. The cross-hair indicates the phase center and the box indicates the ALMA CO(2-1) emission peak position, which is also the position where we extract the CO(5-4) line fluxes. 
\label{Figure_CO54_linemaps}
}
\end{figure*}

We present the sample table and CO(5-4) imaging of our PdBI observations in Table~\ref{Table_Appendix_A} and Fig.~\ref{Figure_CO54_linemaps}. The observations are described in Sect.~\ref{Section_Sample_FMOS_COSMOS}.

\begin{table*}[htb]
\centering
\caption{%
CO observation results. 
\label{Table_Appendix_A}}
\vspace{-2.5ex}
\begin{tabular}{ccccccccc}
\hline
\hline
Source & R.A.$_\mathrm{CO}$ & Dec.$_{\mathrm{CO}}$ & z$_{\mathrm{CO}}$ & $\Delta{V}_{\mathrm{CO}}$ & CO Size & $S_\mathrm{\mathrm{CO}(1-0)}$ & $S_\mathrm{\mathrm{CO}(2-1)}$ & $S_\mathrm{\mathrm{CO}(5-4)}$ \\
& & & & [$\mathrm{km\,s^{-1}}$] & [''] & [$\mathrm{Jy\,km\,s^{-1}}$] & [$\mathrm{Jy\,km\,s^{-1}}$] & [$\mathrm{Jy\,km\,s^{-1}}$] \\
(1) & (2) & (3) & (4) & (5) & (6) & (7) & (8) & (9) \\
\hline
PACS-819 & 09:59:55.552 & 02:15:11.70 & 1.4451 & 592 &   0.335 &           \nodata & 1.10 $\pm$ 0.07 & 3.850 $\pm$ 0.922 \\
PACS-830 & 10:00:08.746 & 02:19:01.87 & 1.4631 & 436 &   0.973 &           \nodata & 1.18 $\pm$ 0.10 & 1.876 $\pm$ 0.387 \\
PACS-867 & 09:59:38.078 & 02:28:56.73 & 1.5656 & 472 & \nodata & 0.119 $\pm$ 0.064 & 0.46 $\pm$ 0.04 & 0.731 $\pm$ 0.218 \\
PACS-299 & 09:59:41.295 & 02:14:43.03 & 1.6483 & 590 & \nodata &  $<$ 0.210 $^{a}$ & 0.67 $\pm$ 0.08 & 1.758 $\pm$ 0.325 \\
PACS-325 & 10:00:05.475 & 02:19:42.61 & 1.6538 & 764 & \nodata &           \nodata & 0.28 $\pm$ 0.06 &  $<$ 0.942 $^{a}$ \\
PACS-164 & 10:01:30.530 & 01:54:12.96 & 1.6481 & 894 & \nodata &  $<$ 0.222 $^{a}$ & 0.61 $\pm$ 0.11 & 1.175 $\pm$ 0.465 \\
\hline
\end{tabular}
\begin{minipage}{0.85\textwidth}
\vspace{2pt}
Columns (1--6) and (8) are the ALMA CO(2-1) properties reported by \cite{Silverman2015a}. 
Column (7) and (9) show the results from this work for VLA CO(1-0) and IRAM PdBI CO(5-4), respectively. \\
$^{a}$ $3\,\sigma$ upper limits. 
\end{minipage}
\end{table*}


\section{Some Notes on CO Observations of Individual Nearby Galaxies in the Literature}
\label{Section_Appendix_CO}

{\bf CenA}: We excluded this galaxy because its CO(2-1) and CO(5-4) data only cover the center of the galaxy, and significant correction is needed for recovering the entire galaxy. For example, \cite{Kamenetzky2014} applied a correction factor of $1/0.48$, where $0.48$ is the beam-aperture-to-entire-galaxy fraction denoted as ``BeamFrac'' and reported in the full Table~\ref{Table_All_in_one} (available online), to convert the CO(2-1) observed at the galaxy center with a beam of $22''$ (\citealt{Eckart1990}) to a beam of $43''$ for their study. They derived this factor based on SPIRE $250\,\mu$m image aperture photometry. Based on PACS $70-160\,\mu$m (as presented in \citetalias{Liudz2015}), we obtain correction factors of $1/0.088$ and $1/0.185$ from a $22''$ and $43''$ beam to the entire galaxy, respectively. Thus the $22''$-to-$43''$ correction factors in two works fully agree ($0.088/0.185 \approx 0.48$). 
Despite the good agreement in beam related correction among these works, we caution that using far-infrared data to correct CO(2-1) is very uncertain as low-$J$ CO lines do not linearly correlate with far-infrared emission (\citetalias{Liudz2015}).

{\bf M83}: We excluded this galaxy in this work as well. \cite{Israel2001} reported a CO(2-1) line flux of $261 \pm 15 \; \Kkms_$ ($5501 \pm 316 \; \Jykms$) within a $22''$ beam (with SEST 15m) at the M83 galaxy center, \cite{Lundgren2004} reported $98.1 \pm 0.8 \; \Kkms_$ ($2068 \pm 17 \; \Jykms$) within a $22''$ aperture (with JCMT 15m) at the same center, and \cite{Bayet2006} reported $67.4 \pm 2.2 \; \Kkms_$ ($2721 \pm 88 \; \Jykms$) within a $30.5''$ beam (with CSO 10.4m) also at the center position. \cite{Kamenetzky2014} adopted the \cite{Bayet2006} line flux and applied a factor of $1/0.76$ correction to obtain the line flux within a $43''$ beam. This correction factor agrees with \citetalias{Liudz2015}. However, if we want to obtain the entire flux for M83, we will need to correct the $43''$ flux by a factor of $1/0.232$ based on the Herschel PACS aperture photometry in \citetalias{Liudz2015}. We caution that the uncertain in such a correction is large, and the CO(2-1) line fluxes at the galaxy center in the literature are already inconsistent by a factor of two. 

{\bf NGC0253}: At the galaxy center position, the reported CO(2-1) line fluxes are: 
$6637 \pm 996 \; \Jykms$ within a $12''$ beam (\citealt{Bradford2003}), 
$10684 \pm 1602 \; \Jykms$ within a $15''$ beam (\citealt{Bradford2003}), 
$17757 \pm 3551 \; \Jykms$ within a $21''$ beam (\citealt{Bayet2004}), 
$24428 \pm 2686 \; \Jykms$ within a $23''$ beam (\citealt{Bayet2004}), 
$33800 \pm 3200 \; \Jykms$ within a $43.5''$ beam (\citealt{Kamenetzky2014}), 
and $34300 \pm 3600 \; \Jykms$ within a $43.5''$ beam (\citealt{Kamenetzky2014}; corrected from the original beam in \citealt{Harrison1999}). 
The beam-to-entire-galaxy fraction, ``BeamFrac'', is 0.518 from $43.5''$ to the entire galaxy based on \citetalias{Liudz2015}. 
These fluxes are roughly consistent, and the ``BeamFrac''-based correction factor is only a factor of two, thus we take the last $43.5''$-beam flux and obtain $66216 \pm 15010 \; \Jykms$ as the CO(2-1) flux for the entire NGC0253, where we added a 0.2~dex uncertainty to the $43.5''$-beam flux error. We caution that even with the additional uncertainty, the flux error might still underestimate the true uncertainty, which includes original flux calibration and measurement error in \citet{Harrison1999}, correction from original beam to $43.5''$ by \citet{Kamenetzky2014}, and from $43.5''$ beam to entire galaxy by 
\citetalias{Liudz2015}.

{\bf NGC0891}: We excluded this galaxy in this work. \cite{Braine1992} reported a CO(2-1) line flux of $86 \pm 6 \; \Kkms_$ ($1974 \pm 138 \; \Jykms$) within a convolved $23''$ beam at the galaxy center position. \cite{Baan2008} converted the same line brightness temperature from \cite{Braine1992} to a flux of $381 \pm 26 \; \Jykms$, which, however, is lower than our converted value in parentheses, and is possibly mistaking the original $12''$ beam for calculation while the brightness temperature that \cite{Braine1992} reported has been convolved to $23''$ beam as mentioned in their Table~1 caption. 
We note that the correction factor from $12''$ or $23''$ to the entire NGC0891 is as large as $\sim10$, e.g., the ``BeamFrac'' from $16.9''$ to entire galaxy is 0.112 as measured by \citetalias{Liudz2015}. Thus it is too uncertain to consider this galaxy in this work. 

As for CO(1-0), \cite{Braine1992} reported a line flux of $96 \pm 5 \; \Kkms_$ ($551 \pm 28 \; \Jykms$) within a $23''$ beam at the galaxy center position. This can be corrected to the entire galaxy scale as $3908 \pm 204 \; \Jykms$ based on \citetalias{Liudz2015}. 
\cite{Gao2004,Gao2004ApJS} reported a line flux of $35.5 \pm 5 \; \Kkms_$ ($963 \pm 136 \; \Jykms$) within a $50''$ beam (with FCRAO 14m), and a global scale integrated flux of $3733.7 \; \Jykms$. 
They are consistent within errors. 

{\bf NGC1068}: \cite{Braine1992} reported a CO(2-1) line flux of $240 \pm 10 \; \Kkms_$ ($5488 \pm 229 \; \Jykms$) within a convolved $23''$ beam at the galaxy center position. 
\cite{Baan2008} converted the same line brightness temperature from \cite{Braine1992} to a flux of $1967.2 \pm 80 \; \Jykms$, which is also inconsistent with our converted value (in parentheses) and possibly due to the mistaking of the original $12''$ beam in their calculation. 
\cite{Papadopoulos2012} reported a CO(2-1) line flux of $11300 \pm 2200 \; \Jykms$ within the inner $40''$ of NGC1068 (originally from \citealt{Papadopoulos1999}). 
\cite{Kamenetzky2011} reported a CO(2-1) line flux of $8366 \pm 19 \; \Jykms$ within a beam of $30''$ (with CSO 10.4m), which is then corrected to $43''$-beam flux of $11700 \pm 1100 \; \Jykms$ by \cite{Kamenetzky2014}. 
\cite{Kamenetzky2014} also cited \cite{Baan2008}'s flux and reported a $43''$-beam flux of $12600 \pm 2500 \; \Jykms$ converted from a $12''$ beam. But note that \cite{Baan2008} might have mistaken a $12''$ beam for the calculation. 
If we directly correct the \cite{Braine1992} $23''$-beam flux to a $43''$ beam, it is $8669 \; \Jykms$, which, however, is 30\% smaller than that in \cite{Kamenetzky2014}. 
Meanwhile, if we correct \cite{Kamenetzky2011}'s flux from $30''$-beam to $43''$-beam, it is $10542\; \Jykms$, consistent with both \cite{Kamenetzky2014} and \cite{Papadopoulos2012}. 
Given the difference is only about 30\%, in this work we adopt the average of these fluxes, 
i.e., $10170 \; \Jykms$ for a $43''$-beam, or $15551 \; \Jykms$ corrected to the entire galaxy scale (based on \citetalias{Liudz2015} BeamFrac). 

For CO(1-0), we perform our own photometry using the Nobeyama 45m COAtlas Survey data (\citealt{Kuno2007}) and obtain a flux of $5228 \; \Jykms$. 
This is 40\% higher than the global scale line flux of $3651.1 \; \Jykms$ measured by \cite{Gao2004ApJS} using FCRAO mapping observations, but more close to the line flux of $4240 \; \Jykms$ within a $43''$-beam reported by \cite{Kamenetzky2014} which is citing \cite{Baan2008} and originally also from \cite{Gao2004ApJS}. 

{\bf NGC1365}: NGC1365 were observed at two positions by Herschel SPIRE FTS, one at North-East (NGC1365-NE) and one at South-West (NGC1365-SW). They have similar CO(5-4) within 10\% but the IR luminosity within each aperture differ by 25\%. This means our aperture-based beam-to-entire-galaxy correction has at least 25\% uncertainty (same in the independent analysis of the similar method by \citealt{Kamenetzky2014}). For CO(2-1) we use the same \cite{Sandqvist1995} SEST 15m ($24''$ beam) data as in \cite{Kamenetzky2014}, and correct it to the entire galaxy scale to match our corrected CO(5-4).

{\bf NGC1614}: CO(2-1) is from \cite{Aalto1995}, observed with SEST 15m ($22''$ beam; $\eta_{\mathrm{mb}}=0.5$, $\int T_{\mathrm{mb}} \mathrm{d} v = 56 \pm 2 \; \Kkms_$ or line flux $1180 \pm 42 \; \Jykms$). We correct from $22''$ beam to the entire galaxy with a BeamFrac of $0.792$ (\citetalias{Liudz2015}). 
Meanwhile, note that \cite{Wilson2008} reported an interferometric integrated CO(2-1) flux of $670 \pm 7 \; \Jykms$ (synthesized beam $3.7'' \times 3.3''$). The discrepancy of about 50\% is likely due to the missing flux of the interferometry (see \citealt{Wilson2008}). 

{\bf NGC2369}: CO(2-1) is from \cite{Aalto1995}, observed with SEST 15m ($22''$ beam; $\eta_{\mathrm{mb}}=0.5$, $\int T_{\mathrm{mb}} \mathrm{d} v = 74 \pm 2.4 \; \Kkms_$ or line flux $1560 \pm 51 \; \Jykms$). 
Meanwhile, note that \cite{Baan2008} reported $959.4 \pm 14.3 \; \Jykms$ which is originally from \cite{Garay1993} also with SEST 15m ($\int T_{\mathrm{mb}} \mathrm{d} v = 46.8 \pm 0.7 \; \Kkms_$; with $\eta_{\mathrm{mb}}=0.54$). The reason for this factor of two discrepancy is unclear. Here we take their average ($1259.7 \; \Jykms$) and correct from the $22''$ beam to the entire galaxy with a BeamFrac of $0.808$ (\citetalias{Liudz2015}). 

{\bf NGC2623}: \cite{Wilson2008} reported an interferometric integrated CO(2-1) flux of $267 \pm 8 \; \Jykms$ observed with SMA. \cite{Papadopoulos2012} cited this flux in their study, and discussed that this flux is unlikely affected by missing flux.

{\bf NGC3256}: \cite{Aalto1995} reported a CO(2-1) flux of $\int T_{\mathrm{mb}} \mathrm{d} v = 314 \pm 8 \; \Kkms_$ ($6619 \pm 169 \; \Jykms$) observed with SEST 15m ($22''$ beam; $\eta_{\mathrm{mb}}=0.5$). 
Meanwhile, \cite{Baan2008} reported $2980.7 \pm 14.3 \; \Jykms$ which is originally from \cite{Garay1993} also observed with SEST 15m ($\int T_{\mathrm{mb}} \mathrm{d} v = 145.5 \pm 0.7 \; \Kkms_$; with $\eta_{\mathrm{mb}}=0.7$). Similar to NGC2369, the reason for the factor of two to three discrepancy is unclear. 
We take their average ($4799.85 \; \Jykms$) and correct from the $22''$ beam to the entire galaxy with a BeamFrac of $0.744$ (\citetalias{Liudz2015}). 

{\bf NGC3351}: We obtain CO(2-1) and CO(1-0) line fluxes for the entire galaxy with our own photometry as $2681 \; \Jykms$ and $1138 \; \Jykms$, respectively, to the HERACLES data and the Nobeyama 45m COAtlas Survey (\citealt{Kuno2007}) data. Uncertainties contributed by the noise in the moment-0 maps are about 6\% of the measured fluxes. 
Note that \cite{Braine1992} observed a CO(2-1) and CO(1-0) flux of about $642 \; \Jykms$ and $97 \; \Jykms$, respectively, convolved to a $23''$ beam. \cite{Leroy2009} reported a CO(2-1) luminosity of $0.78 \times 10^{5} \; \mathrm{K\,km\,s^{-1}\,arcsec^{2}}$, or a line flux of $2808 \; \Jykms$, for the entire galaxy, consistent with ours. 
\cite{Usero2015} reported a CO(1-0) flux of about $210 \; \Jykms$ within a $21.3''$ beam at the central position.

{\bf NGC3627}: Similar to NGC3351, we obtain the CO(2-1) and CO(1-0) line fluxes for the whole galaxy via our photometry using the HERACLES and the NRO45m COAtlas data, respectively. We measured $9219 \; \Jykms$ and $7366 \; \Jykms$, respectively. Note that \cite{Gao2004ApJS} reported a global CO(1-0) flux of $4477 \; \Jykms$, which is about 40\% lower than ours. 

{\bf NGC4321}: Similar to NGC3351 and NGC3627, the global CO(2-1) and CO(1-0) line fluxes are obtained as $9088 \; \Jykms$ and $2251 \; \Jykms$, from the HERACLES and the NRO45m COAtlas data, respectively. 
Note that \cite{Braine1992} observed a CO(1-0) flux of $445 \; \Jykms$ within a $23''$ beam, which can be corrected to a consistent entire galaxy flux of $2280 \; \Jykms$ by a BeamFrac of $0.195$ (\citetalias{Liudz2015}). While \cite{Komugi2008} observed a CO(1-0) flux of $174 \; \Jykms$ within a $16''$ beam, which is somehow lower than others.

{\bf NGC4945}: \cite{Wang2004NGC4945} observed the central position of NGC4945 with SEST 15m and obtained a CO(2-1) flux of $\int T_{\mathrm{mb}} \mathrm{d} v = 920.9 \pm 0.6 \; \Kkms_$ ($19412 \pm 12.6 \; \Jykms$, for point source response in a $22''$ beam). 
\cite{Baan2008} cited the same \cite{Wang2004NGC4945} CO(2-1) flux as $18878.5 \pm 12.3 \; \Jykms$, which is consistent with our conversion. 
\cite{Curran2001NGC4945} also observed the central position of NGC4945 with SEST 15m. They reported a CO(2-1) flux of $\int T_{\mathrm{mb}} \mathrm{d} v = 740 \pm 40 \; \Kkms_$, about 20\% lower than that of \cite{Wang2004NGC4945}. 
As discussed in \cite{Wang2004NGC4945}, the reason for the discrepancy is unclear, but this shows that the uncertainty in the CO(2-1) flux at the galaxy center is at least 20\%. 
We take the average ($17505 \; \Jykms$) in this work, and estimate the entire galaxy CO(2-1) flux to be $31770 \; \Jykms$ based on a BeamFrac of $0.551$ (\citetalias{Liudz2015}) from the $22''$ beam.

{\bf NGC6946}: The global CO(2-1) and CO(1-0) line fluxes are obtained as $36296 \; \Jykms$ and $11454 \; \Jykms$, from the HERACLES and the NRO45m COAtlas data, respectively. 
This is in good agreement with the global scale CO(1-0) flux of $11400.5 \; \Jykms$ reported by \cite{Gao2004ApJS} using NRAO 12m mapping data.

\section{Comparison of SED Fitting Codes}
\label{Section_Appendix_SED}

We performed additional \textsc{MAGPHYS} (\citealt{daCunha2008,daCunha2015,Battisti2019}) and \textsc{CIGALE} (\citealt{Burgarella2005,Noll2009,Ciesla2014,Ciesla2015,Boquien2019,Yang2020}) SED fitting to verify our \michi2{} SED fitting results. We use the updated \textsc{MAGPHYS} version with high-$z$ extension (\url{http://www.iap.fr/magphys/download.html}), and \textsc{CIGALE} version 2020.0 (June 29th, 2020) (\url{https://cigale.lam.fr/download/}). We modified the \textsc{MAGPHYS} FORTRAN source code to allow for longer photometry filter names and larger filter number. \textsc{MAGPHYS} and \textsc{CIGALE} require a list of preset filters, which we choose the following list: 
GALEX FUV and NUV, KPNO MOSAIC1 $u$, CFHT MegaCam $u$ band, SDSS $ugriz$, Subaru SuprimeCam $BVriz$, GTC $griz$, VISTA VIRCAM $Y,\,J,\,H,\,K_s$, \textit{HST} ACS F435W/F606W/F755W/F814W and WFC3 F125W/F140W/F160W, \textit{Spitzer} IRAC ch1/2/3/4, IRS PUI 16\,$\mu$m and MIPS 24\,$\mu$m, \textit{Herschel} PACS 70/100/160 and SPIRE 250/350/500\,$\mu$m, SCUBA2 450/850\,$\mu$m, VLA 3/1.4\,GHz, and pseudo 880/1100/1200/2000\,$\mu$m filters. 
Other photometry data like sub-mm interferometry data (e.g., from ALMA) and some optical data are ignored. Note that in our \michi2{} fitting these bands without a known filter curve are automatically used with a pseudo delta-function filter curve. 

The current \textsc{MAGPHYS} code does not include the fitting of a mid-IR AGN SED component, although such an extension has been used non-publicly in some studies (\citealt{Chang2015}). \textsc{MAGPHYS} has preset stellar libraries, dust attenuation laws and dust libraries, therefore no need to adjust any parameters. Except that we run \textsc{MAGPHYS} only for $z>0.03$ galaxies, as \textsc{MAGPHYS} computes the luminosity and mass properties with the luminosity distance, which does not match the physical distance at a very low-$z$. 

\textsc{CIGALE} has the capability of including a mid-IR AGN component, so does our \michi2{} code. The current version of \textsc{CIGALE} uses AGN emission models computed from physical modeling of AGN torus by \cite{Fritz2006}. It has much more freedom than the observationally-derived AGN templates by \cite{Mullaney2011} used by \michi2{}. However, this can also easily over-fit the data when there are only a few broadband photometry data point at mid-IR $\sim8-100\,\mu\mathrm{m}$. 
For our fitting with \textsc{CIGALE}, we fix several AGN parameters based on the fitting results of starburst galaxies in \cite{Fritz2006}: \incode{r_ratio = 60}, \incode{beta = -1.0}, \incode{gamma = 6.0}, \incode{opening_angle = 140.0}, and let following parameters to vary: \incode{tau = 1.0,3.0,6.0}, \incode{psy = 0.001, 10.100, 20.100, 30.100}, and \incode{fracAGN = 0.0,0.2,0.4,0.6}. 

For the stellar component in our \textsc{CIGALE} fitting of high-redshift galaxies, we use a constant SFH as in our \michi2{} fitting. This is achieved by adding \incode{sfhperiodic} into the \textsc{CIGALE} \incode{sed_modules}, and setting \incode{type_bursts = 2}, \incode{delta_bursts = 200}, \incode{tau_bursts = 200}. To allow the fitting of a range of stellar ages, we set \incode{age = 200,300,400,500,600,700,800,900,1000,2000} for the \incode{bc03} SED module. And we adopt the \cite{Calzetti2000} dust attenuation law as in our \michi2{} fitting by adding \incode{dustatt_modified_starburst} to the SED modules, and setting \incode{E_BV_lines} to 0.0 to 2.0 in steps of 0.2. 
Meanwhile, for local galaxies ($z<0.03$) in our sample whose stellar ages are generally older, a constant SFH stellar component can not fit the stellar SED well. Thus, we adopt the exponentially declining SFH \incode{sfh2exp} in \textsc{CIGALE}, and set \incode{tau_main = 200,500,1000,2000,4000} and \incode{age = 200,500,1000,2000,4000,6000,8000,10000}. We turn off the burst model by setting \incode{f_burst = 0.0}. 

The dust template used in \textsc{CIGALE} fitting is also the same as used in our \michi2{} fitting, i.e., the \cite{Draine2014} updated DL07 templates. $\Umin$ (\incode{umin}) is set to vary from 1.0 to 50, and $\fPDR$ (\incode{gamma}) 0.0 to 1.0. 
We also set \incode{lim_flag = True} to allow \textsc{CIGALE} analyzing the photometry upper limits (to achieve this we need to flip the sign of the flux errors for the photometry with a $\SNR<3$). Then, each galaxy has about 696960 models fitted. In comparison, in \textsc{MAGPHYS} in general 13933 optical models and 24999 IR dust models are fitted for each galaxy. 

\begin{figure}[H]
\centering
\includegraphics[width=0.450\textwidth]{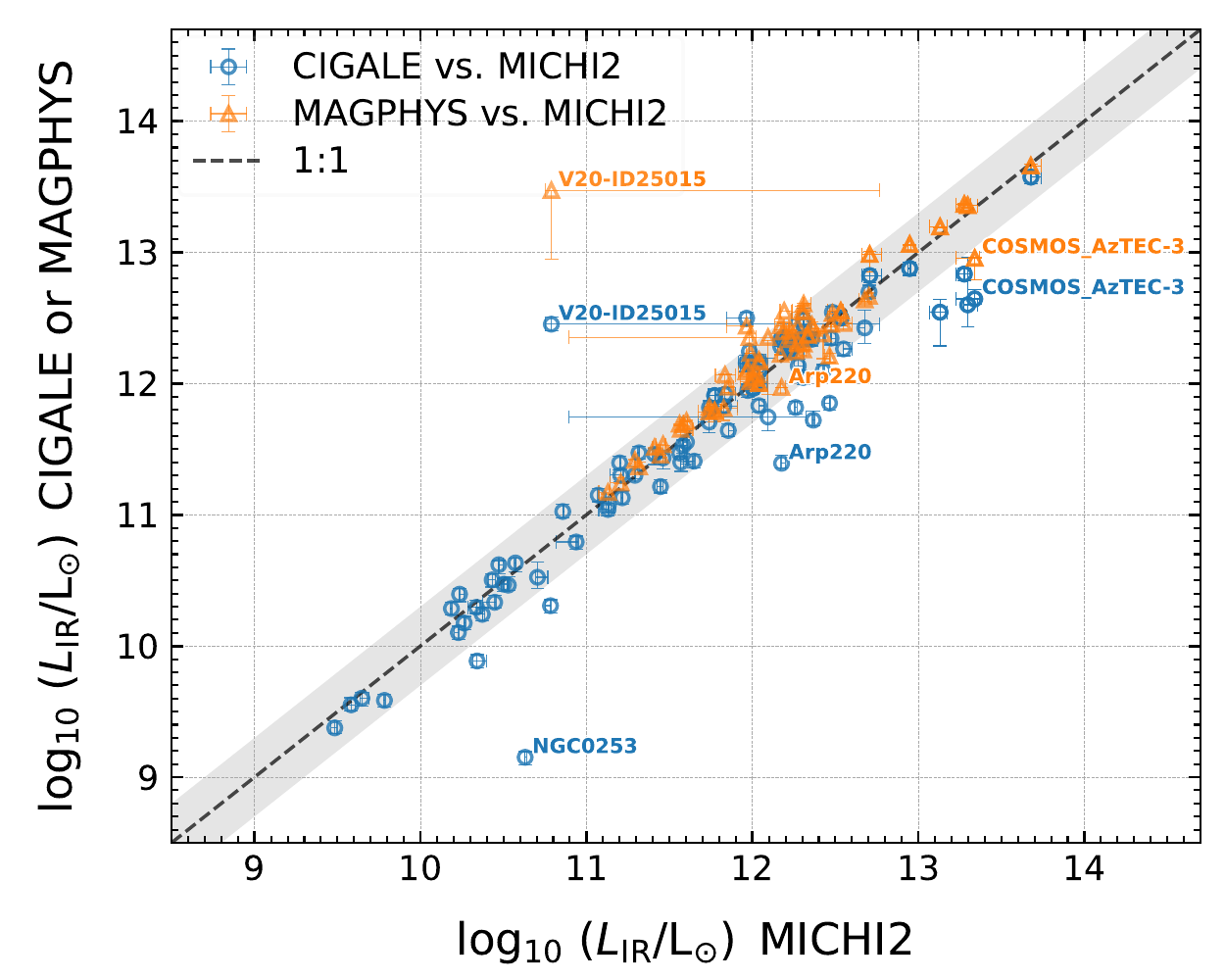}
\includegraphics[width=0.450\textwidth]{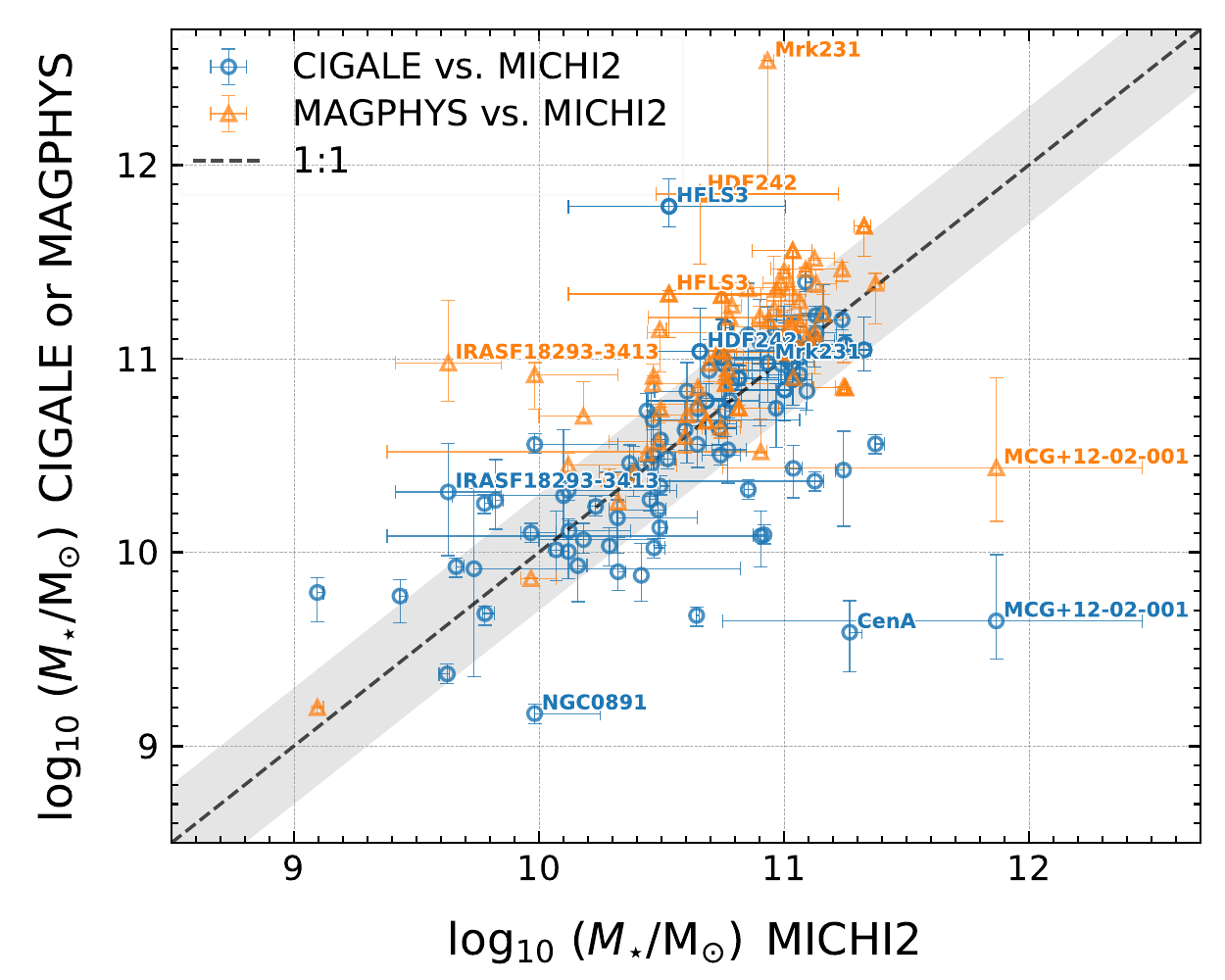}
\caption{Comparison of the fitted 8--1000\,$\mu$m dust luminosities (\textit{left panel}) and stellar masses (\textit{right panel}) from three SED fitting codes: \michi2{}, \textsc{CIGALE} and \textsc{MAGPHYS}. X-axes in both panels indicate the fitted parameters from the \michi2{}, whereas Y-axes indicate those from either \textsc{CIGALE} (blue circles) or \textsc{MAGPHYS} (orange triangles). The dashed line is a one-to-one relation, and the grey shading indicates a $\pm 0.3\,\mathrm{dex}$ range. Error bars show the fitted 16\%- and 84\%-percentiles and symbols center at the minimum-$\chi^2$/highest-probability values. 
}
\label{Plot_SED_fitting_parameter_comparison_lgMstar_lgLIR}
\end{figure}

\begin{figure}[H]
\centering
\includegraphics[width=0.450\textwidth]{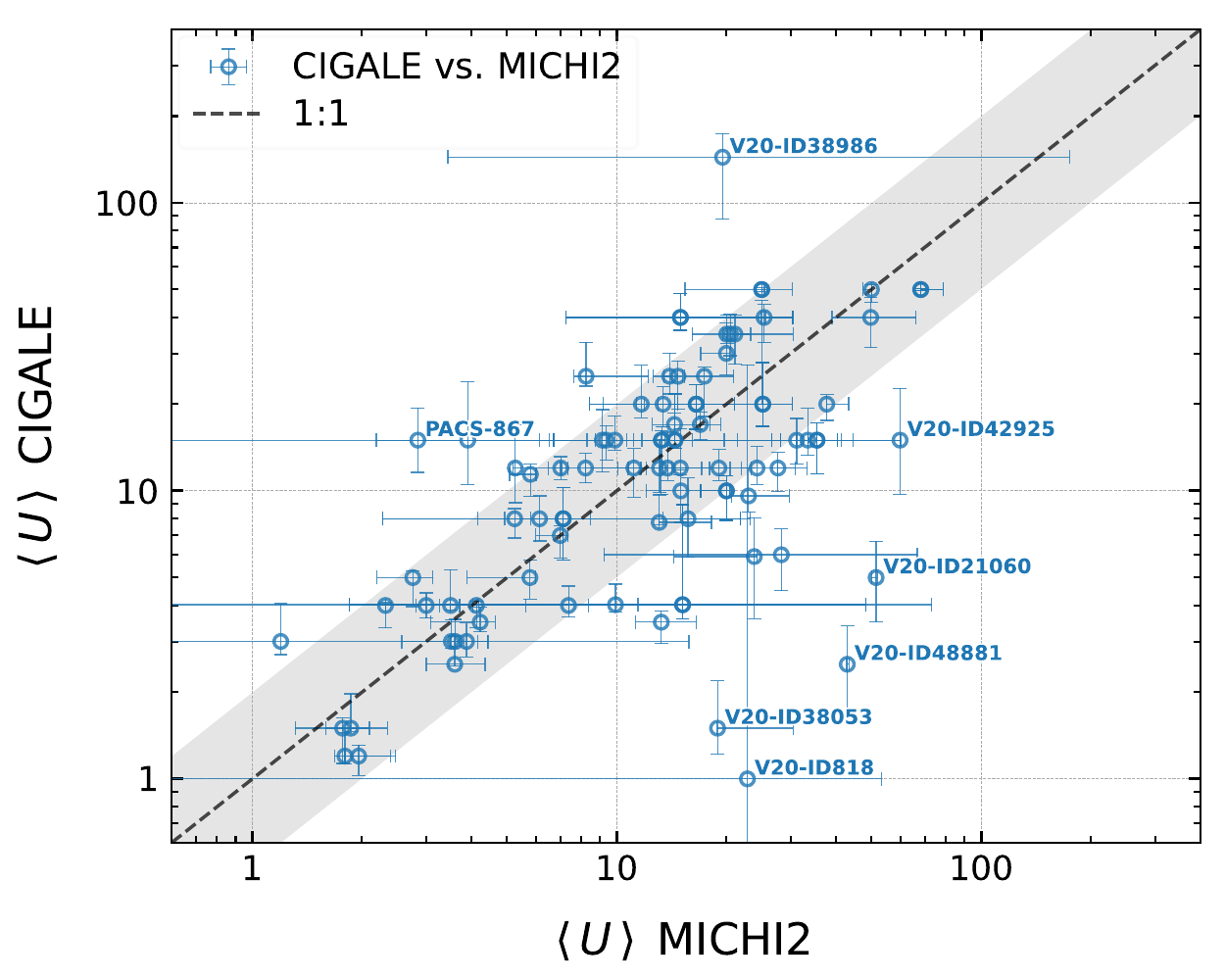}
\includegraphics[width=0.450\textwidth]{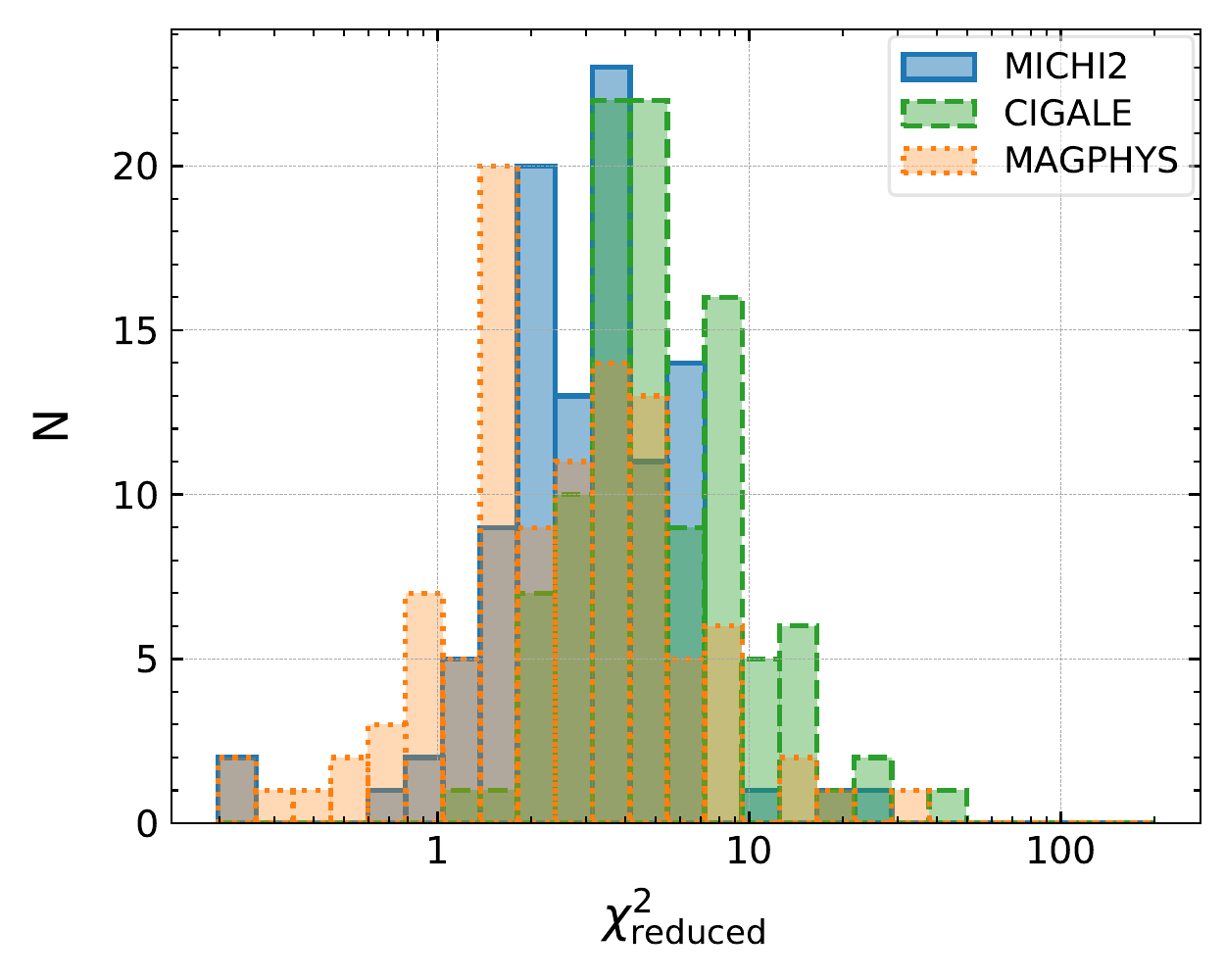}
\caption{%
\textit{Left panel} shows the comparison of the fitted $\Umean$ from \michi2{} and \textsc{CIGALE}. Symbols are similar to Fig.~\ref{Plot_SED_fitting_parameter_comparison_lgMstar_lgLIR}, except that data points with reduced-$\chi^2 > 10$ are excluded. 
\textit{Right panel} shows the histograms of the reduced-$\chi^2$ from \michi2{}, \textsc{CIGALE} and \textsc{MAGPHYS} SED fittings. Given that our fittings span from UV/optical to mm/radio wavelengths, a reduced-$\chi^2$ larger than unity is not unexpected. A value of a few still indicates a reasonable fitting in our cases, but $>10$ usually means poor or failed fitting. 
}
\label{Plot_SED_fitting_parameter_comparison_Umean}
\end{figure}

In Fig.~\ref{Plot_SED_fitting_parameter_comparison_lgMstar_lgLIR}, we compare the fitted dust 8--1000\,$\mu$m luminosities and stellar masses from the three SED fitting codes. In the left panel of Fig.~\ref{Plot_SED_fitting_parameter_comparison_Umean} we compare the fitted $\Umean$ from \michi2{} and \textsc{CIGALE}, as \textsc{MAGPHYS} does not have the same \cite{DL07} library. Note that not all fittings show a reasonable $\chi^2$, as can be seen in the right panel of Fig.~\ref{Plot_SED_fitting_parameter_comparison_Umean}, where the histograms of reduced-$\chi^2$ are shown for the three fitting codes. \textsc{CIGALE} fittings in general have a higher reduced-$\chi^2$, which means poorer fitting than \michi2{}, whereas \textsc{MAGPHYS} produces slightly better fittings than \michi2{}. However, both \textsc{MAGPHYS} and \textsc{CIGALE} have a number of very poor/failed fitting cases which have reduced-$\chi^2 \gtrsim 10$ and are the outlier data points in Figs.~\ref{Plot_SED_fitting_parameter_comparison_lgMstar_lgLIR}. The threshold reduced-$\chi^2 \sim 8-10$ is empirically estimated after visually examining the SED fitting results%
\,\footnote{All SED fitting figures are available at \url{https://doi.org/10.5281/zenodo.3958271}.}%
. There are only two sources exhibit reduced-$\chi^2 > 10$ in \michi2{} fitting, and their IR-to-mm are actually well fitted, leaving the stellar part poorly constrained (CenA and NGC0253). 
In comparison, there are 12 poor/failed cases in \textsc{CIGALE} fitting, 
and 4 in \textsc{MAGPHYS} fitting. 
The main reason for these poor/failed cases is likely the energy balance forced in \textsc{MAGPHYS} and \textsc{CIGALE}. In these cases, the stellar part of the SED fitting gives a dust attenuation that can not fully balance the far-IR/mm emission, and this is also mentioned in other studies of extremely dust-obscured high-redshift galaxies (e.g., \citealt{Simpson2017,Casey2017,Miettinen2017b,Miettinen2017d}). 
Except for these poor/failed fittings, the fitted IR luminosities and stellar masses are reasonably well agreed within about 0.3\,dex. 

In the comparison of $\Umean$ in Fig.~\ref{Plot_SED_fitting_parameter_comparison_Umean}, we excluded the 12 sources with reduced-$\chi^2 > 10$ in \textsc{CIGALE} fitting. Although most sources have consistent $\Umean$, a small number of sources do not have consistent $\Umean$, and they mostly come from the V20 subsample. This is mainly because they have very poor IR photometry except for one or two sub-mm interferometry photometry. But we chose to skip these interferometry photometry in our \textsc{CIGALE} fitting tests due to the filter setting. Adding a fake filter in \textsc{CIGALE} and re-run the fitting for each of these sources is required in order to fit the sub-mm interferometry photometry, then more consistent results are expected. Therefore, these comparisons show that for sources with good reduced-$\chi^2$ and photometry data, \michi2{} and \textsc{CIGALE} have similar constraints on $\Umean$.

\section{Gas Modeling Prediction on Line Optical Depth and [CI]/CO Line Ratio}
\label{Section_Appendix_Gas_Modeling}

\begin{figure*}[h]
\includegraphics[width=0.99\textwidth]{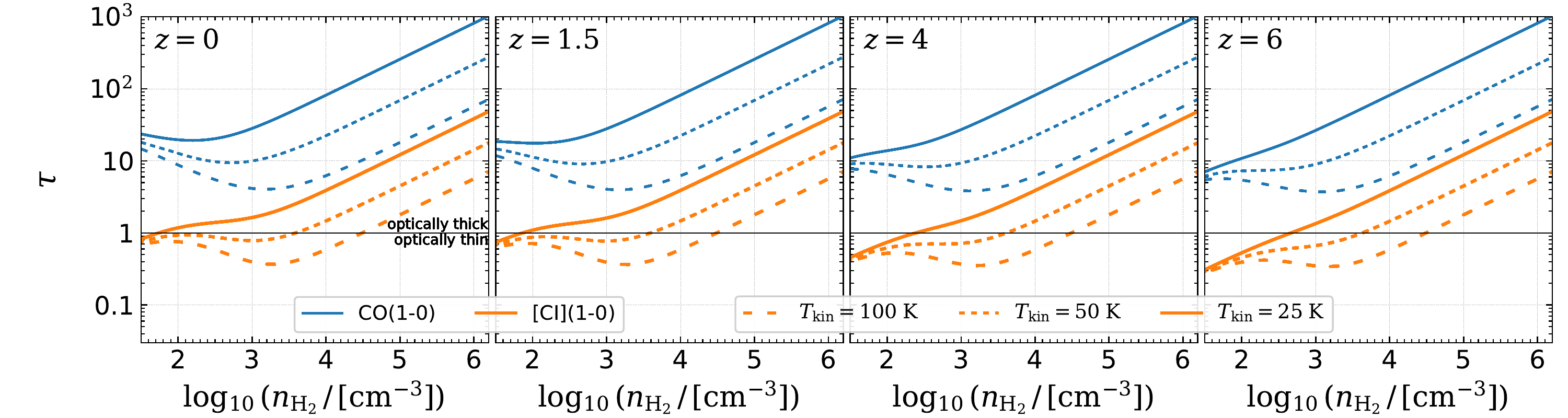}
\vspace{1ex}
\caption{%
Optical depths ($\tau$) of CO(1-0) and [CI](1-0) as a function of the gas density of each single LVG (one-zone) model in four redshift panels. Blue lines are CO(1-0) and orange lines are [CI](1-0). Line styles (solid, dashed and long-dashed) represent different gas kinetic temperatures as labeled. These show that the derived optical depths from our models (Sect.~\ref{Section_gas_modeling}) roughly agree with observations which usually show optically thin ($\tau\sim1$) [CI](1-0) and optically thick CO(1-0). 
\label{Plot_single_LVG_tau}
}
\end{figure*}

\begin{figure*}
\includegraphics[width=0.99\textwidth]{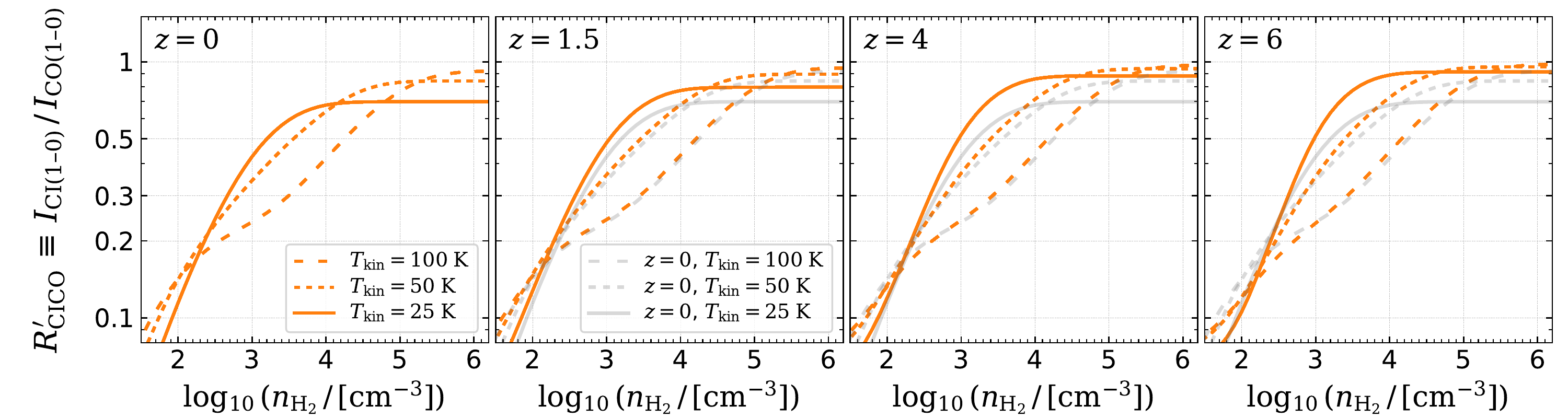}
\vspace{1ex}
\caption{%
Ratio between [CI](1-0) and CO(1-0) line surface brightness from our single LVG one-zone models. Lines and symbols are similar to those in Fig.~\ref{Plot_single_LVG_R52}, but note that the ratio in Fig.~\ref{Plot_single_LVG_R52} is flux ratio while here we show the surface brightness ratio ($R^{\prime}_{\mathrm{CI/CO}} \equiv L^{\prime}_{\mathrm{[CI](1-0)}} / L^{\prime}_{\mathrm{CO(1-0)}}$). 
\label{Plot_single_LVG_RCICO}
}
\end{figure*}

\begin{figure*}
\includegraphics[width=0.99\textwidth]{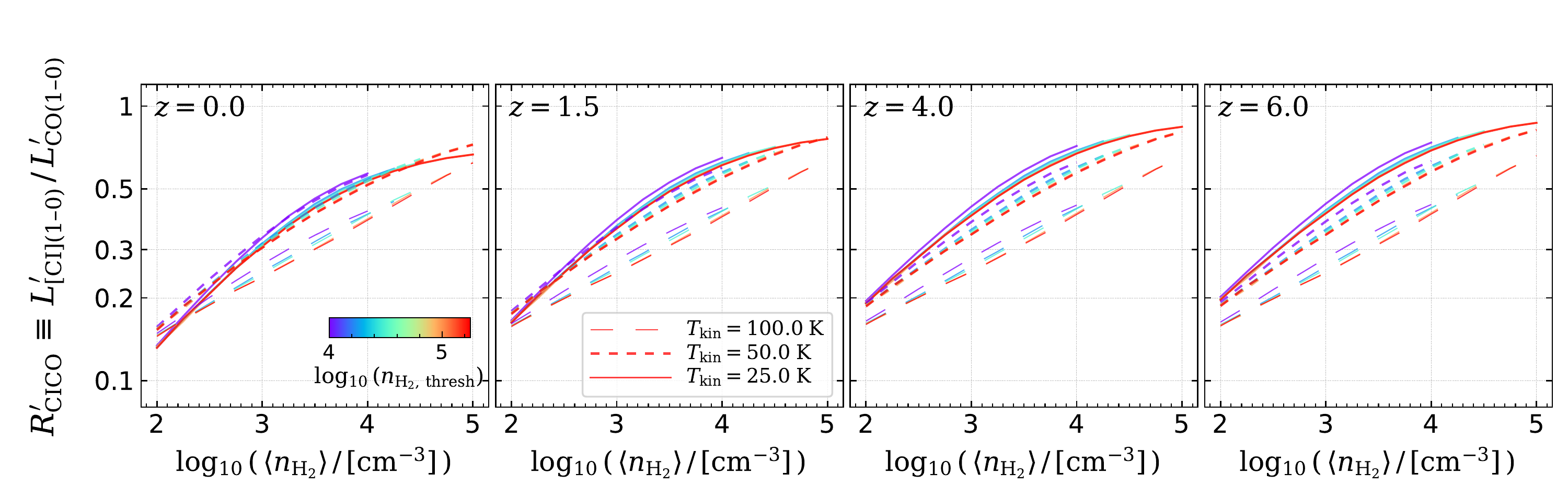}
\vspace{1ex}
\caption{%
Ratio between [CI](1-0) and CO(1-0) line surface brightness from our density-PDF gas modeling. Lines and symbols are similar to those in Fig.~\ref{Plot_modeled_volume_density_nH2_R52} (see also the note about the different ratio definition in Fig.~\ref{Plot_single_LVG_RCICO} caption). 
\label{Plot_modeled_volume_density_nH2_RCICO}
}
\end{figure*}

We present the predicted [C~\textsc{i}]($^{3}P_1-^{3}P_0$) (hereafter [C~\textsc{i}](1-0)) and CO(1-0) line optical depths and the [C~\textsc{i}](1-0)/CO(1-0) line ratio in surface brightness unit ($R^{\prime}_{\mathrm{CICO}}$) in Fig.~\ref{Plot_single_LVG_tau}, Fig.~\ref{Plot_single_LVG_RCICO} and Fig.~\ref{Plot_modeled_volume_density_nH2_RCICO}. 
The optical depths shown in Fig.~\ref{Plot_single_LVG_tau} agree with normal conditions where CO(1-0) is optically thick while [C~\textsc{i}](1-0) is roughly optically thin or has $\tau\sim1$. 

Fig.~\ref{Plot_single_LVG_RCICO} and Fig.~\ref{Plot_modeled_volume_density_nH2_RCICO} show $R^{\prime}_{\mathrm{CICO}}$ as a function of one-zone cloud molecular gas density and mean molecular gas density of the composite PDF, respectively. Similar as in Fig.~\ref{Plot_single_LVG_R52} and Fig.~\ref{Plot_modeled_volume_density_nH2_R52}, we show our prediction at four redshifts, $z=0$, $1.5$, $4$ and $6$, and with three representative $\Tkin=25$, $50$, and $100\;\mathrm{K}$. 
$R^{\prime}_{\mathrm{CICO}}$ increases with $\nH2$ or $\nmean$, but strongly decreases with $\Tkin$ at intermediate $\nH2\sim10^{3-4}\;\percmcubic$. Future study of this line ratio with our gas modeling will shed light on how to better constrain $\Tkin$ and $\nmean$.

\clearpage

\DeclareRobustCommand{\disambiguate}[3]{#1}
\bibliography{Biblio.bib}

\end{document}